\documentclass[usenatbib]{mnras}
\usepackage{newtxtext,newtxmath}
\usepackage[T1]{fontenc}
\usepackage{CJK}
\usepackage{graphicx}
\usepackage{amsmath}
\usepackage{natbib}

\newcommand{\begit}{\begin{itemize}}
\newcommand{\enit}{\end{itemize}}
\newcommand{\begen}{\begin{enumerate}}
\newcommand{\enen}{\end{enumerate}}
\newcommand{\pp}{\partial}    

\newcommand{\beq}{\begin{equation}}
\newcommand{\eeq}{\end{equation}}
\newcommand{\beqa}{\begin{eqnarray}} 
\newcommand{\eeqa}{\end{eqnarray}}

\newcommand       \be           {\begin{equation}}
\newcommand       \ee           {\end{equation}}
\newcommand       \bea          {\begin{eqnarray}}
\newcommand       \eea          {\end{eqnarray}}
\newcommand       \kms		{\,{\rm km \,\, s}^{-1}}

\newcommand       \mspy 	{\,{\rm M_\odot \, yr^{-1}}}

\newcommand       \kpc		{\,{\rm kpc }}

\newcommand 	     \geff		{{\rm \gamma_{eff}}}
\newcommand 	     \Veff		{{\rm V_{g,eff}}}
\newcommand 	     \Veffsq		{{\rm V^2_{g,eff}}}
\newcommand 	     \qc		{{ q_{\rm cool}}}

\title[Galactic Winds Driven by Streaming Cosmic Rays]{The Physics of Galactic Winds Driven by Cosmic Rays II:  Isothermal Streaming Solutions}

\author[E. Quataert, Y.-F. Jiang, T. A. Thompson]{Eliot Quataert$^{1}$,  Yan-Fei Jiang(\begin{CJK*}{UTF8}{gbsn}姜燕飞\end{CJK*})$^{2}$, and Todd A. Thompson$^{3,4}$   \\
 $^{1}$Department of Astrophysical Sciences, Princeton University, Princeton, NJ 08544, USA\\ 
 $^{2}$ Center for Computational Astrophysics, Flatiron Institute, 162 Fifth Avenue, New York, NY, 10010, USA \\
$^{3}$ Department of Astronomy, The Ohio State University, 140 West 18th Avenue, Columbus, OH 43210, USA \\
$^{4}$ Center for Cosmology and Astro-Particle Physics (CCAPP), The Ohio State University, 191 West Woodruff Ave., Columbus, OH 43210, USA \\}

\date{\vspace{0pt}Accepted XXX. Received YYY; in original form ZZZ}

\pubyear{2021}

\begin{document}
\label{firstpage}
\pagerange{\pageref{firstpage}--\pageref{lastpage}}
\maketitle


\begin{abstract}
We use analytic calculations and time-dependent spherically-symmetric
simulations to study the properties of isothermal galactic winds driven by cosmic-rays (CRs) streaming at the Alfv\'en velocity.  The
simulations produce time-dependent flows permeated by strong shocks; we identify a new linear instability of sound waves that sources these shocks.  The shocks substantially modify the wind dynamics, invalidating previous steady state models: the CR pressure $p_c$ has a staircase-like structure with $dp_c/dr \simeq 0$ in most of the volume, and the time-averaged CR energetics are in many cases better approximated by $p_c \propto \rho^{1/2}$, rather than the canonical $p_c \propto \rho^{2/3}$.  Accounting for this change in CR energetics, we analytically derive new expressions for the mass-loss
rate, momentum flux, wind speed, and wind kinetic power in galactic winds driven by CR streaming. We show that streaming CRs are ineffective at directly driving cold gas out of galaxies, though CR-driven winds in hotter ISM phases may entrain cool gas. For the same physical conditions, diffusive CR transport (Paper I) yields mass-loss rates that are a few-100 times larger than streaming transport, and asymptotic wind powers that are a factor of $\simeq 4$ larger. We discuss the implications of our results for galactic wind theory and observations; strong shocks driven by CR-streaming-induced instabilities produce gas with a wide range of densities and temperatures, consistent with the multiphase nature of observed winds.   We also quantify the applicability of the isothermal gas approximation for modeling streaming CRs and highlight the need for calculations with more realistic thermodynamics.
\end{abstract}


\section{Introduction}
\label{section:introduction}

A significant fraction of the mechanical energy supplied by supernovae to the interstellar medium (ISM) goes into cosmic-ray (CR) protons (e.g., \citealt{Blandford1987}).   Those cosmic-rays may in turn play a number of important roles in galaxy formation.    Cosmic-rays regulate the ionization state of the dense interstellar medium (e.g., \citealt{Dalgarno2006}), can contribute to driving galactic winds from star-forming galaxies (e.g., \citealt{Ipavich1975}), and may be an important source of heating in low-density phases of the interstellar, circumgalactic, and intracluster medium (e.g., \citealt{Guo2008, Wiener2013b}).   
Despite their potential importance in galaxy formation, our understanding of the impacts of CRs is still relatively rudimentary.  The primary theoretical and observational challenge is that the physical processes regulating CR transport are still not fully understood (see, e.g., \citealt{Amato2017} for a review in the context of the Milky Way).   Empirically, we know that CRs have a short mean-free path and thus do not leave galaxies on a light-crossing time, despite their relativistic energies.    Theoretically, CRs scatter off of small-scale magnetic fluctuations that can either be the small-scale tail of a turbulent cascade (e.g., \citealt{Yan2002}) or fluctuations generated by the CRs themselves (e.g., \citealt{Kulsrud1969}).  If CRs are not efficiently scattered by ambient turbulence, as is plausibly the case in many physical conditions, any net drift of the cosmic-rays exceeding the Alfv\'en speed $v_A$ will excite the gyro-resonant streaming instability \citep{Lerche1967}; short-wavelength Alfv\'en waves can then grow to the point that they can scatter the cosmic-rays and limit the resulting CR streaming speed (e.g., \citealt{Bai2019}).   

The mechanism of CR transport can have a large effect on their broader astrophysical impacts.   For example, if CRs stream relative to the thermal gas, they inevitably heat the gas (mediated via the streaming-excited waves) at a rate  $|{\bf v_A \cdot \nabla} p_c|$, where $p_c$ is the CR pressure (\citealt{Wentzel1971}). This heating is absent in the case of pure diffusive transport. Previous numerical work has also shown that the properties of galactic winds driven by CRs change significantly depending on the mechanism of CR transport  (e.g., \citealt{Wiener2017}), as does the impact of CRs on the circumgalactic medium (e.g., \citealt{Butsky2018,Hopkins2020b}).

In this paper and a companion (\citealt{Quataert2021}; hereafter Paper I) we study the physical properties of galactic winds driven by CRs and their dependence on galaxy properties and the mechanism of cosmic-ray transport.    There is a  large body of previous analytic and numerical work on galactic winds driven by CRs with either diffusive or streaming transport (e.g., \citealt{Ipavich1975,Breitschwerdt1991,Everett2008,Booth2013,Recchia2016,Chan2019}; see the introduction to Paper I for a more comprehensive discussion of previous work).  Our work provides an analytic framework for understanding these previous results and provides estimates of wind properties suitable for use in cosmological simulations or semi-analytic models of galaxy formation. Paper I considered the case of cosmic-ray transport by diffusion. In this paper, we consider the case of CR transport by streaming at the Alfv\'en speed. We also directly compare these solutions to their counterparts with diffusion.  In both papers we assume that the gas is isothermal, as is plausible for photoionized gas or when cooling is rapid.   In addition to our analytic estimates, we carry out spherically symmetric time-dependent simulations of CR-driven winds using the two-moment CR transport scheme of \citet{JiangOh2018}.   In the case of diffusive CR transport in Paper I, the simulations largely validated the analytic estimates. As we show in this paper, the case of CR transport by streaming is more interesting:  the time-dependent simulations show that steady-state wind models with streaming are linearly unstable. This invalidates previous steady-state calculations because the winds are intrinsically time-dependent and have a time-averaged structure that is different from their steady-state counterparts. The simulations motivate a revised theory of CR-driven winds that accounts for the effects of this instability and the non-linear structures it produces. We also show that the assumption of  isothermal gas breaks down as a result of CR heating in important regimes of parameter space where earlier isothermal CR streaming wind models have been applied. 

The remainder of this paper is organized as follows.   In \S \ref{section:streaming}, we present analytic estimates of the mass-loss rate and terminal velocity in galactic winds driven by CRs streaming at the Alfv\'en speed using standard theoretical assumptions.  In \S \ref{sec:numerics}, we present time-dependent numerical simulations, which show that in most cases the solutions are highly time-dependent, with strong shocks permeating the flow.   These shocks significantly modify the time-averaged energetics of the CRs relative to standard CR wind models in the literature.  In \S \ref{sec:strmod}, we develop a modified CR-driven wind model that accounts for the energetics seen in the simulations.  \S \ref{sec:disc} synthesizes our numerical results compared to the analytics (\S \ref{sec:synthesis}), compares the properties of galactic winds driven by streaming CRs to those driven by diffusing CRs (\S \ref{sec:strvsdiff}; see Paper I), quantitatively assesses the validity of the isothermal gas approximation used throughout this work (\S \ref{sec:iso}), and discusses the observational implications of our results (\S \ref{sec:obs}).   In \S \ref{sec:summary}, we summarize our main results.   Appendix \ref{sec:appendixA} presents linear stability analyses relevant for interpreting our numerical simulations.   We identify two new (to the best of our knowledge) instabilities driven by CR streaming, which are the origin of the time dependence seen in the simulations.  Appendix \ref{sec:AppB} discusses a few aspects of the time-averaged CR energetics not addressed in the main text.    

\section{Analytic Approximations for Galactic Winds Driven by Streaming Cosmic Rays}

\label{section:streaming}

The equation for the CR energy density $E_c$ in the absence of CR sources and pionic losses, and including diffusion along magnetic field lines and streaming at the Alfv\'en velocity down the CR pressure gradient, is given by
\beq
\frac{\pp E_c}{\pp t}+\nabla\cdot {\bf F}_c=\left({\bf v}+{\bf v_{\rm s}}\right)\cdot \nabla p_{c},
\label{cr_energy}
\eeq
where $p_c=E_c/3$ is the CR pressure, ${\bf v_s} = -{\bf v_A}|\nabla p_c|/\nabla p_c$ with ${\bf v_A}={\bf B}/(4\pi\rho)^{1/2}$, and 
the ``equilibrium" CR flux\footnote{The two-moment CR model we solve numerically in \S \ref{sec:numerics} evolves ${\bf F_c}$ as an independent variable and the flux reduces to equation \ref{equilibrium_flux} only when time variations are sufficiently slow (see eq. \ref{eq:CR2mom}).} is 
\be
{\bf F}_c=4 p_c \left({\bf v}+{\bf v_{\rm s}}\right)-\kappa \,{\bf n}\left({\bf n}\cdot\nabla E_c\right).
\label{equilibrium_flux}
\eeq
Here $\kappa$ is the diffusion coefficient and ${\bf n}={\bf v_{\rm A}}/|\bf v_{\rm A}|$.    Equation \ref{cr_energy}, and indeed a scalar CR pressure in the momentum equation for the gas, is formally valid only on scales larger than the mean-free-path of the $\sim$ GeV energy CRs that dominate the total energy of the CR population.

In this paper we focus primarily on the properties of galactic winds in which CRs stream at the Alfv\'en velocity.  We also compare those results to the case of CR diffusion discussed in Paper I.  We consider the simplified model problem of  a spherical CR-driven wind in an isothermal gravitational potential. We further assume an isothermal gas of sound speed $c_i$, as is a priori plausible for rapidly cooling or photoionized gas. In \S \ref{sec:iso} we assess the isothermal gas assumption and show that it breaks down in important regimes of parameter space because of the inevitable presence of CR heating when CRs stream at the Alfv\'en speed.    The assumption of spherical symmetry also implies that the magnetic field is assumed to be a split-monopole configuration, and only shows up dynamically in setting the Alfv\'en speed (and hence the streaming speed) as a function of radius. 

In this section we derive steady-state analytic approximations to the wind solutions by considering hydrostatic equilibrium below the sonic point and assuming rapid streaming (high $v_A$); the latter simplification is analogous to the assumption of rapid diffusion made in the analytic estimates of Paper I.  In Section \ref{sec:numerics}, we then treat the same problem numerically by solving the time-dependent CR-driven wind problem using the numerical scheme of \cite{JiangOh2018}.  These numerical calculations invalidate the analytic solutions for many physical parameters, because the numerical solutions are time-dependent (due to instabilities discussed in \S \ref{sec:lin} \& Appendix \ref{sec:appendixA}).    Nevertheless, the standard streaming case presented here provides important pedagogical insights, makes contact with the previous literature (e.g., \citealt{Ipavich1975,Mao2018}), and sets the stage for \S \ref{sec:strmod}, in which we present a modification to the streaming wind solutions that better captures the dynamics found in our numerical simulations.   

\subsection{The Density Profile and Mass-loss Rate}

Under the approximation of time-steady spherical flow, subject to gas pressure $p$, CR pressure $p_c$, and an isothermal gravitational potential characterized by velocity dispersion $V_g$,  the equations describing mass and momentum conservation are
\be
\dot M_w = 4 \pi r^2 \rho v = {\rm const}
\label{eq:mdot}
\ee
\be
v \frac{dv}{dr} = -\frac{1}{\rho}\frac{dp}{dr} - \frac{1}{\rho}\frac{dp_c}{dr} - \frac{2 V_g^2}{r}.
\label{eq:mom}
\ee
Assuming $dp_c/dr < 0$, the steady-state CR streaming flux is given by 
\beq
{\bf F_c}=4 p_c({\bf v}+{\bf v_A}).
\eeq
Neglecting CR source terms and hadronic losses, the CR energy equation is given by
\beq
\nabla\cdot\left[4 \, \left({\bf v}+{\bf v}_{\rm A}\right) p_c \right]=\left({\bf v}+{\bf v_{\rm A}}\right)\cdot \nabla p_c.
\label{cr_energy2}
\eeq
Noting that in spherical symmetry
\beq
\nabla\cdot {\bf v_{\rm A}}=-\frac{1}{2}v_{\rm A}\frac{d\ln\rho}{dr},
\eeq
the CR energy equation becomes
\beq
\frac{dp_c}{dr}=\frac{4}{3}\frac{p_c}{\rho}\frac{d\rho}{dr}\left(\frac{v_{\rm A}/2+v}{v_{\rm A}+v}\right).
\label{cr_energy_simple}
\eeq
Combining this expression with equations \ref{eq:mdot} and \ref{eq:mom}, we obtain the wind equation for the fluid in the streaming limit:
\beq
\frac{1}{v} \frac{dv}{dr}=\frac{2}{r}\left(\frac{c_i^2+c_{\rm eff}^2-V_g^2}{v^2-c_i^2-c_{\rm eff}^2}\right) \equiv \frac{N_{st}}{D_{st}}
\label{wind_streaming}
\eeq
where we have defined the numerator $N_{st}$ and denominator $D_{st}$ of the wind equation, and where
\beq
c_{\rm eff}^2=\frac{4}{3}\frac{p_c}{\rho}\left(\frac{v_{\rm A}/2+v}{v_{\rm A}+v}\right),
\label{eq:ceff}
\eeq
is the effective CR sound speed. We will use several different definitions of the CR sound speed in this paper, depending on the exact context; see Table \ref{tab:cs}.

Equation (\ref{wind_streaming}) closely follows the relations derived in \cite{Ipavich1975}. Setting the numerator and denominator in the wind equation equal to zero at the sonic point $r_s$ implies that 
\beq
v(r_s)=V_g \ \ \ {\rm and \ \ \ } c_{\rm eff}^2(r_s)=V_g^2-c_i^2 \equiv \Veffsq
\label{sonic}
\eeq
where we have defined an effective gravitational speed $\Veff$ for future use.

\begin{table*}
	\caption{Summary of different definitions of cosmic-ray sound speed used in this paper.  In Paper I we exclusively used $c_c$.  The value of the CR sound speed at the base of the wind appears in many of our analytic estimates, and is denoted $c_{\rm eff,0}$ or $c_{c,0}$.}
\centerline{	\begin{tabular}{c|ccc}
Quantity & Definition & Context & Use in Paper \\ \hline
General $c_{\rm eff}$ & eq. \ref{eq:ceff} & full streaming wind equation & \S \ref{section:streaming} \\
$c_{\rm eff}$ for $\geff = 2/3$ & $\sqrt{\frac{2}{3} \frac{p_c}{\rho}}$ (eq. \ref{ceffstr}) & high $v_A$ limit of standard streaming analytics & \S \ref{section:streaming} Analytics \\
$c_{\rm eff}$ for general $\geff$ & $\sqrt{\geff \frac{p_c}{\rho}}$
(eq. \ref{eq:ceffgen}) & high $v_A$ limit of modified streaming analytics & \S \ref{sec:geffgen} Analytics \\

$c_{\rm eff}$ with $\geff=1/2$ & $\sqrt{\frac{1}{2} \frac{p_c}{\rho}}$ (eq. \ref{eq:ceffgen} with $\geff=1/2$) & high $v_A$ limit of modified streaming analytics & \S \ref{sec:geff0.5} Analytics \\
$c_{\rm eff}$ with $\geff=4/3$ & $\sqrt{\frac{4}{3} \frac{p_c}{\rho}}$ (eq. \ref{eq:omc}) & adiabatic CR sound speed & \S \ref{sec:lin2mom} Stability Calculation \\
$c_c$ & $\sqrt{\frac{p_c}{\rho}}$ & `Isothermal' CR Sound Speed & \S \ref{sec:numerics} Simulations and \S \ref{sec:lin1mom} Stability Calculation \\
\hline
\label{tab:cs}
\end{tabular}}
\end{table*}

We now seek an analytic expression for the mass loss rate in CR-driven winds in the streaming limit for comparison with the more complete numerical calculations presented in Section \ref{sec:numerics}. To do so, we employ the commonly-used strategy of assuming that the system can be approximated as in hydrostatic equilibrium with $v\rightarrow0$ in the region below the sonic point $r<r_s$. This approximation is justified deep in the atmosphere of the wind, but becomes increasingly suspect as the sonic point is approached. Nevertheless, as in discussions of the isothermal and polytropic Parker-type thermal winds, this approximation yields the scalings of how the mass-loss rate depends on physical parameters and an estimate for the normalization of the mass loss rate $\dot{M}_w$.

To make analytic progress we assume that in the hydrostatic portion of the solution, $v \ll v_A$.  We check below a posteriori when this assumption is valid (eq. \ref{eq:vAlarge}) and note that our numerical simulations in \S \ref{sec:numerics} do not employ the high $v_A$ assumption.   Taking the limit $v\ll v_{\rm A}$, equation (\ref{cr_energy_simple}) becomes
\beq
\frac{d\ln p_c}{dr}=\frac{2}{3}\frac{d\ln\rho}{dr} \Longrightarrow \left(\frac{p_c}{p_{c,0}}\right)=\left(\frac{\rho}{\rho_0}\right)^{2/3},
\label{simple_crenergy}
\eeq
where $p_{c,0}$ and $\rho_0$ represent parameters of the problem at the base of the outflow, or at any other reference point. Note that equation (\ref{simple_crenergy}) can also be written as 
\beq
c_{\rm eff}^2=c_{\rm eff,0}^2\left(\frac{\rho_0}{\rho}\right)^{1/3} = \frac{2}{3} \frac{p_{c,0}}{\rho_0}\left(\frac{\rho_0}{\rho}\right)^{1/3},
\label{eq:ceffstr}
\eeq
where $c_{\rm eff}^2=(2/3)p_c/\rho$ in the limit that $v\ll v_{\rm A}$. These expressions show that the CR sound speed increases as the gas density decreases and the wind accelerates, allowing the flow to eventually reach the sonic conditions given in equation \ref{sonic}.    If, however, $v \gg v_A$, $c_{\rm eff}^2 \simeq (4/3)p_c/\rho$ and equation \ref{cr_energy_simple} becomes $p_c \propto \rho^{4/3}$, i.e., the CR are adiabatic.  In this case $c_{\rm eff}^2 \propto \rho^{1/3}$ and the CR sound speed decreases with decreasing density. This shows that there cannot be an isothermal CR-driven outflow in the limit of small $v_A$, because the CR sound speed can never reach $V_g$ if $c_{\rm eff,0} \lesssim V_g$, i.e., if the CR are initially bound. The condition $v \lesssim v_A$ must thus apply over a significant range of radii in order for the outflow to accelerate and reach the escape speed.  Equations \ref{simple_crenergy} \& \ref{eq:ceffstr} are an approximation to this requirement that enables considerable analytic progress.  We will use these for the remainder of this section.

We now calculate the density profile in the hydrostatic portion of the flow assuming $p_c \propto \rho^{2/3}$ (eq. \ref{simple_crenergy}), via
\beq
\frac{c_i^2}{\rho}\frac{d\rho}{dr} + \frac{c_{\rm eff}^2}{\rho}\frac{d\rho}{dr}= \frac{c_i^2}{\rho}\frac{d\rho}{dr} + \frac{c_{\rm eff,0}^2}{\rho}\left(\frac{\rho_0}{\rho}\right)^{1/3}\frac{d\rho}{dr} = - \frac{2 V_g^2}{r},
\label{ceffstr}
\eeq
which has the solution
\beq
c_i^2\ln\left(\frac{\rho}{\rho_0}\right)-3c_{\rm eff,0}^2\left(\frac{\rho_0^{1/3}}{\rho^{1/3}}-1\right)=-2V_g^2\ln\left(\frac{r}{r_0}\right).
\label{density_he_streaming}
\eeq
This expression yields a simple implicit relation for the density profile that is easily solved numerically for $\rho(r)$.  Using the numerical solutions to equation \ref{density_he_streaming} we can also estimate the mass-loss rate in galactic winds driven by streaming CRs in the high $v_A$ limit as follows: using equation \ref{eq:ceffstr}, equation \ref{sonic} directly implies the density at the sonic point, namely 
\be
\rho(r_s) \simeq \rho_0 \left(\frac{c_{\rm eff,0}}{\Veff}\right)^{6}.
\label{eq:rhosonic}
\ee
The sonic radius $r_s$ is then determined numerically using the solution to equation \ref{density_he_streaming} as the radius where $\rho(r_s)$ is reached; we derive analytic approximations for $r_s$ below.   The mass-loss rate is then $\dot M_{\rm w} \simeq 4 \pi r_s^2 \rho({r_s}) V_g$ (eq.~\ref{sonic}).  

\begin{figure}
\centering
\includegraphics[width=84mm]{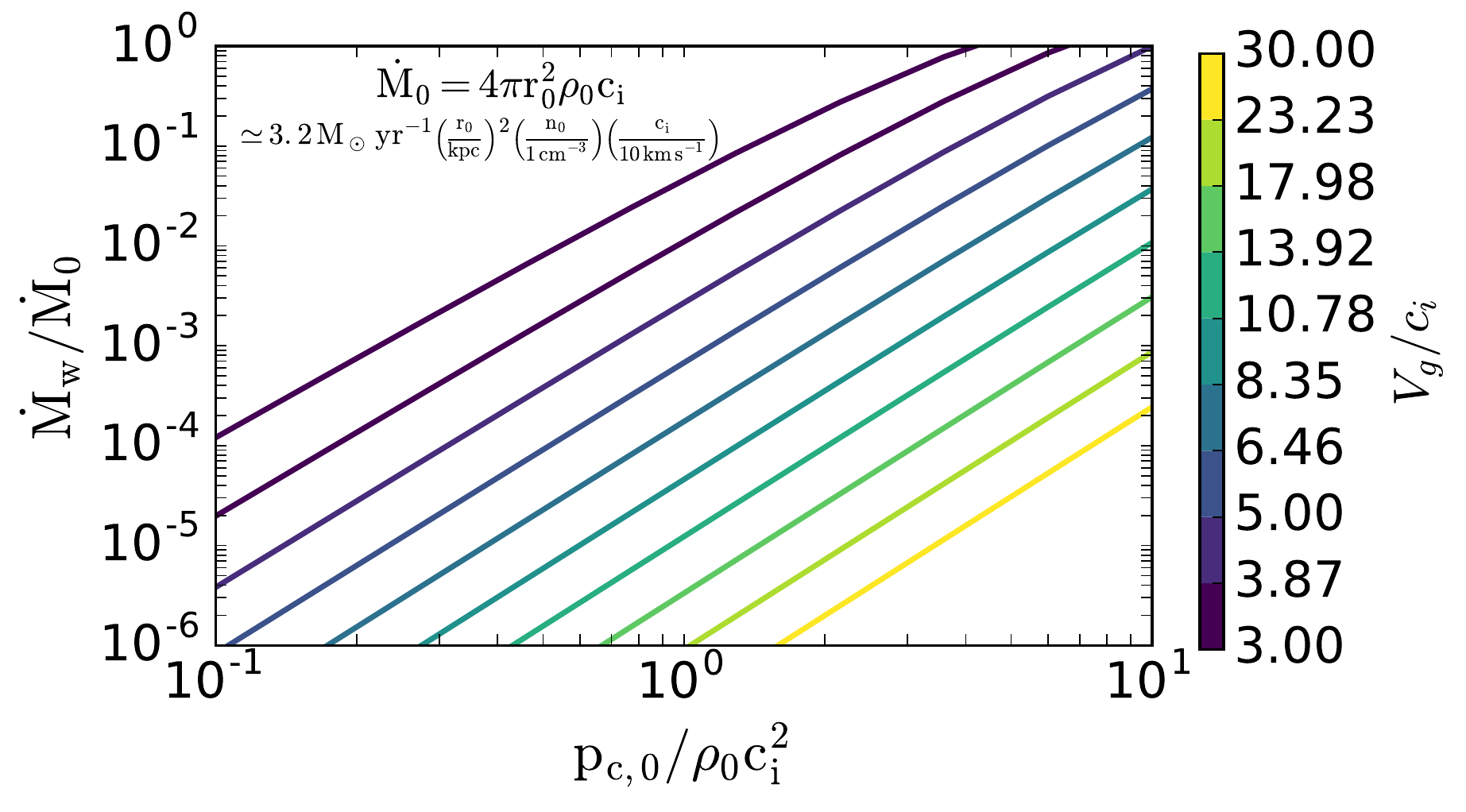}
\vspace{-0.2cm}
\caption{Analytic mass-loss rate for CR driven galactic winds in the limit of rapid streaming, as a function of the strength of gravity relative to the gas sound speed in the disk ($V_g/c_i$) and the base CR pressure (${\rm p_{c,0}/\rho_0 c_i^2}$).  The mass-loss rate is normalized by eq. \ref{eq:mdot0}. Labeled values of $V_g/c_i$ on the color bar are logarithmically distributed and correspond to the curves on the plot.}  
\label{fig:mdotstr}
\end{figure}

Figure \ref{fig:mdotstr} shows the mass-loss rate as a function of the two key dimensionless parameters in the problem, namely the strength of gravity $(V_g/c_i)$ and the base CR pressure ($p_{c,0}/\rho_0 c_i^2$).   As in Paper I, we express the mass-loss rate in units of
\be
\dot M_0 = 4 \pi r_0^2 \rho_0 c_i \simeq 3.2 \, \mspy \,  \left(\frac{r_0}{\kpc}\right)^2 \left(\frac{n_0}{1 \, {\rm cm^{-3}}}\right) \left(\frac{c_i}{10 \, \kms}\right)
\label{eq:mdot0}
\ee
where $n_0 \equiv \rho_0/m_p$.  Figure \ref{fig:mdotstr} shows that the mass-loss rate due to streaming CRs is a particularly strong function of the strength of gravity relative to the sound speed in the galactic disk; as we show explicitly below, the dominant dependence is actually on $V_g/c_{\rm eff,0}$, i.e., on the CR sound speed rather than the gas sound speed.  It is useful to also think of different values of $c_i$ and $c_{\rm eff,0}$ in Figure \ref{fig:mdotstr} as corresponding to different phases of the ISM. Figure \ref{fig:mdotstr} implies that most of the mass loss will be from the warmer phases of the ISM, which have larger $c_i$ and $c_{\rm eff,0}$, unless $p_{c,0}$ is much lower in those phases. 

We can also find analytic approximations to the numerical solutions in Figure \ref{fig:mdotstr}.  We first consider the limit where $c_i^2\ll 3 c_{\rm eff,0}^2$. The analytic solution to equation (\ref{density_he_streaming}) is then
\beq
\frac{\rho}{\rho_0}\simeq\left[1+\frac{2}{3}\frac{V_g^2}{c_{\rm eff,0}^2}\,\ln\left(\frac{r}{r_0}\right)\right]^{\,-3}.
\label{eq:rhostrpcr}
\eeq
Equation \ref{eq:rhostrpcr} is in fact a good approximation to the numerical solution of \ref{density_he_streaming} even for the case of $c_{\rm eff,0}=c_i$. Given equation \ref{eq:rhostrpcr}, the mass loss rate can be estimated using equation (\ref{eq:mdot}), but with all quantities evaluated at the sonic point. Comparing equation \ref{eq:rhosonic} and \ref{eq:rhostrpcr} we derive an approximation for the sonic point radius
\beq
\frac{r_s}{r_0}\simeq\exp\left[\frac{3}{2}\frac{c_{\rm eff,0}^2}{V_g^2}\left(\frac{\Veffsq}{c_{\rm eff,0}^2}-1\right)\right].
\eeq
The general expression for the wind mass loss rate in the streaming limit, assuming a hydrostatic density profile with $c_i^2\ll 3 c_{\rm eff,0}^2$, is then
\beq
\dot{M}_w \simeq 4\pi r_0^2\,\rho_0\,V_g \,\left(\frac{\,c_{\rm eff,0}}{\Veff}\right)^6\,\exp\left[\frac{3c_{\rm eff,0}^2}{V_g^2}\left(\frac{\Veffsq}{c_{\rm eff,0}^2}-1\right)\right].
\label{mdot_streaming}
\eeq
This expression for $\dot M_w$ is independent of the gas sound speed $c_i$.  The apparent dependence on $c_i$ in Figure \ref{fig:mdotstr} is primarily because $c_i$ is used to non-dimensionalize $\dot M_w$, $c_{\rm eff,0}$ and $V_g$; there is also a weak dependence on $c_i$ at low $c_{\rm eff,0}/c_i$ derived below (eq. \ref{mdotstrgam0.6weak}).  We retain this choice of dimensionless variables for consistency with Paper I.

For case of massive galaxies like the Milky Way, where one expects $V_g^2\gg c_{\rm eff,0}^2$ and $V_g^2\gg c_i^2$, the expression for the sonic point radius reduces to the remarkably simple form
\beq
\frac{r_s}{r_0}=e^{3/2},
\label{eq:rssimp}
\eeq
and equation (\ref{mdot_streaming}) becomes
\beq
\begin{split}
\dot{M}_w= & \, 4\pi r_0^2\,\rho_0\,V_g \, e^3\,\left(\frac{\,c_{\rm eff,0}}{V_g}\right)^6 \simeq 6 \times 10^{-4} \mspy \left(\frac{r_0}{1 \kpc}\right)^2 \\ & \times \, \left(\frac{n_0}{1 \, {\rm cm^{-3}}}\right)   \left(\frac{c_{\rm eff,0}}{10 \, \kms}\right)^6 \left(\frac{V_g}{100 \, \kms}\right)^{-5} 
\label{mdotstrgam0.6}
\end{split}
\eeq
For reference, if we scale for parameters appropriate to a galaxy like the Milky Way with mass-averaged ISM conditions $V_g\simeq150$\,km s$^{-1}$, $c_{\rm eff,0}\simeq10$\,km s$^{-1}$, $n_0\simeq 1$ cm$^{-3}$, and $r_0\sim5$\,kpc, this expression yields $\dot{M}_w\simeq 2 \times 10^{-3}$\,M$_\odot$ yr$^{-1}$, much less than the star formation rate, with the most important factor being the strong suppression $(c_{\rm eff,0}/V_g)^6\simeq10^{-7}$.  However, the mass-loss rate in equation \ref{mdotstrgam0.6} can also be expressed as $\dot M_w \propto p_{c,0} c_{\rm eff,0}^4/V_g^5$.  Assuming that the hot ISM has a similar CR pressure, its mass-loss rate will be much larger than that of the cold ISM, with, e.g., $\dot M_w \simeq 1 \mspy$ for $c_{\rm eff,0} \simeq 50 \kms$, comparable to the star formation rate.  Most of the mass loss is thus likely to originate from the hotter phases of the ISM.   Cold gas can still be unbound by such an outflow if it is entrained in the hot CR wind. Equation \ref{mdotstrgam0.6} and Figure \ref{fig:mdotstr} show, however, that {\em direct} CR driving of winds from $\sim 10^4$ K gas is very inefficient.

An instructive expression for the wind mass-loss rate  can be obtained by comparing the mass-loss rate estimated here to the star formation rate.  To do so, we note that feedback-regulated models of star formation in galaxies predict that (e.g., \citealt{Thompson2005,Ostriker2011}) $\dot M_* \approx \pi r_0^2 \dot \Sigma_*$ with $\dot \Sigma_* \simeq 2 \sqrt{2} \pi G \Sigma_g^2 \phi /(p_*/m_*)$ where $p_*/m_* \approx 3000 \kms$ is the momentum per unit mass associated with stellar feedback, which supports the disk against its own self-gravity, and $\phi$ quantifies the stellar and dark matter contribution to the gravity of the disk.   Equation \ref{mdotstrgam0.6} for the mass-loss rate in the streaming limit takes the form
\be
\begin{split}
\frac{\dot M_{\rm w}}{\dot M_*} \simeq & \, 0.06 \left(\frac{c_{\rm eff,0}}{10 \, \kms}\right)^4 \left(\frac{p_*/m_*}{3000 \, \kms}\right) \\ & \times \left(\frac{V_g}{100 \, \kms}\right)^{-5}  \frac{p_{c,0}}{\pi G \Sigma_g^2 \phi}
\end{split}
\label{eq:mdotstr2}
\ee
For Milky-way like conditions in which the CR pressure is comparable to that needed for hydrostatic equilibrium in the galactic disk, the wind mass-loss rate due to streaming CRs from the `mass-average' ISM (with $c_{\rm eff,0} \simeq 10 \kms$) is much less than the star formation rate.   As noted previously, however, volume filling lower density gas with larger $c_{\rm eff,0}$ will have a significantly higher mass-loss rate, plausibly comparable to the star formation rate. For example, taking $V_g =150$\,km s$^{-1}$ with $c_{\rm eff,\,0}=30$\,km s$^{-1}$ and $50$\,km s$^{-1}$ gives $\dot{M}_w/\dot{M}_*\simeq0.2$ and $\simeq2$, respectively, if $p_{c,0} \simeq 0.3 \pi G \Sigma_g^2 \phi$ in the warmer phases of the ISM.

We now consider analytic approximations for the mass-loss rate in the limit of weak CR pressure compared to gas pressure at the base in the galactic disk, i.e., $c_{\rm eff,0} \ll c_i$.  In this limit the gas density profile is initially set by gas pressure, with
\beq
\frac{\rho}{\rho_0}=\left(\frac{r_0}{r}\right)^{2V_g^2/c_i^2} \ \ \ \ (r < r_{tr})
\label{eq:weakCR}
\eeq
However, as the density drops, the CR pressure increases in importance relative to the gas pressure.   So long as $V_g \gtrsim c_i$,\footnote{This is the only interesting regime for a wind solution since otherwise the gas is effectively unbound in the galactic disk even without CRs.} the sonic point condition (eq. \ref{sonic}) requires that the pressure be CR dominated at the sonic point.   The transition between gas pressure and CR pressure support happens at a radius 
\be
r_{tr} \simeq r_0 \left(\frac{\rho_0 c^2_i}{p_{c,0}}\right)^\frac{3c_i^2}{2 V_g^2}.
\label{eq:rtr}
\ee 
The density profile exterior to this transition radius is then like equation \ref{eq:rhostrpcr} but with a different boundary condition set by continuity at $r_{tr}$.   This yields
\be
\frac{\rho}{\rho_0}\simeq \left(\frac{p_{c,0}}{\rho_0 c^2_i}\right)^{3}\left[1+ \frac{V_g^2}{c_i^2}\,\ln\left(\frac{r}{r_{tr}}\right)\right]^{\,-3} \ \ \ \ (r > r_{tr})
\label{eq:rhostrweak}
\ee
The sonic point is then located at $r_s \simeq e^{3/2} r_{tr}$ and the mass-loss rate is
\be
\dot{M}_w\simeq4\pi r_0^2\,\rho_0\,V_g \, e^3 \,
\left(\frac{\rho_0 c^2_i}{p_{c,0}}\right)^\frac{3 \, c_i^2}{V_g^2} \left(\frac{c_{\rm eff,0}}{V_g}\right)^6.
\label{mdotstrgam0.6weak}
\ee
Note that equation \ref{mdotstrgam0.6weak} (valid for $c_{\rm eff,0} < c_i$) differs from equation \ref{mdotstrgam0.6} (valid for $c_{\rm eff,0} > c_i$) only by a factor of $(\rho_0 c^2_i/p_{c,0})^{3 c_i^2/V_g^2}$, which is not that different from one for massive galaxies with $V_g \gg c_i$; the latter is the condition that $r_{tr} \simeq r_0$ (eq. \ref{eq:rtr}).  Physically, the sonic point conditions uniquely determine the density at the sonic point (eq. \ref{eq:rhosonic}).    So long as $r_{tr} \simeq r_0$, then the radius of the sonic point can be estimated entirely neglecting the gas-dominated region near the base of the wind, and is still given by equation \ref{eq:rssimp}.

\subsection{Validity of the High Alfv\'en Velocity Approximation}
\label{sec:highvA}
The analytic estimate of the mass-loss rate in Figure \ref{fig:mdotstr} assumes that $v_A \gg v$.  Since the mass-loss rate is set by conditions at the sonic point, our estimate is applicable when $v_A(r_s) \gtrsim v(r_s) = V_g$ so that equation \ref{simple_crenergy} is valid out to the sonic point.  Using equations \ref{sonic} and \ref{eq:ceffstr} this constraint can be rewritten as
\be
v_{A,0} \gtrsim v_{A,crit} \equiv {c_{\rm eff,0}} \, \frac{V_g \, c^2_{\rm eff,0}}{V^3_{\rm g, eff}}\left(\frac{r_s}{r_0}\right)^2 \simeq \frac{e^3 c_{\rm eff,0}^3}{V_g^2} \ \ {\rm High \ v_A \ Limit} 
\label{eq:vAlarge} 
\ee
where the last expression is appropriate for  massive galaxies with $V_g^2\gg c_{\rm eff,0}^2,c_i^2$.  For sufficiently small $v_{A,0} \ll v_{A,crit}$ there cannot be an unbound wind because the CRs are effectively adiabatic (see the discussion after eq. \ref{eq:ceffstr}).    A plausible conjecture is in fact that if $v_{A,0} \lesssim v_{A,crit}$ the solution will not be able to pass through the sonic point and will thus formally be a `breeze' rather than a wind. We shall see that this is borne out by our numerical solutions in \S \ref{sec:vA}.

Comparing equation \ref{eq:mdotstr2} and equation \ref{eq:vAlarge} we see that for a given base Alfv\'en speed $v_{A,0}$, a larger base CR sound speed $c_{\rm eff,0}$ will both increase the mass-loss rate and make the high $v_A$ approximation used to derive equation \ref{eq:mdotstr2} suspect.   One might thus anticipate that the largest mass-loss rate for a given set of base conditions would be obtained for conditions that just satisfy $v_A \sim v_{A,crit}$.   We shall see below in \S \ref{sec:maxmdot} that this is indeed correct (eq. \ref{eq:mdotmax}).

\subsection{Energetics, Momentum Flux, \& Terminal Velocity}

There is no conserved total energy flux for our model problem of isothermal gas with streaming CRs.    The reason is that the gas is heated by the CRs (the standard streaming term $\sim v_A dp_c/dr$) but this energy is assumed to be instantaneously radiated away to keep the gas isothermal. Instead, the steady state CR energy equation takes the form
\be
\frac{1}{r^2}\frac{d}{d r}\left(r^2 F_c \right) = \left(v + v_A\right)\frac{d p_c}{dr} 
\label{CRenstr}
\ee
An effective energy equation for the gas can be derived using the gas momentum equation, which can be rewritten as
\be
\frac{d}{dr}\left(\frac{1}{2} v^2 + c_i^2 \ln \rho + 2 V_g^2 \ln r \right) = - \frac{1}{\rho}\frac{d p_c}{dr} 
\label{gasenstr}
\ee
Multiplying by $r^2 v$ and combining with equation \ref{CRenstr} yields
\be
\frac{1}{r^2} \, \frac{d}{dr}\left(\dot M \left[\frac{1}{2} v^2 + c_i^2 \ln \rho + 2 V_g^2 \ln r \right] + \dot E_c \right) =  4 \pi v_A\frac{d p_c}{dr} 
\label{gasenstr3}
\ee
where $\dot E_c = 4 \pi r^2 F_c$ is the CR `luminosity/power.'   Because $d p_c/dr < 0$, the right-hand-side of equation \ref{gasenstr3} is negative and so the total power carried by the wind decreases with increasing radius.

To the extent that $p_c \propto \rho^{2/3}$ (eq. \ref{cr_energy_simple}), the CR pressure gradient term in equation \ref{gasenstr} can be rewritten as a CR enthalpy, leading to a conserved Bernoulli-like constant (e.g., \citealt{Mao2018})
\be
\frac{d}{dr}\left(\frac{1}{2} v^2 + c_i^2 \ln \rho + 2 V_g^2 \ln r - 3 c_{\rm eff,0}^2  \left[\frac{\rho}{\rho_0}\right]^{-1/3}\right)  = 0.
\label{gasenstr2}
\ee

We can roughly estimate the terminal speed of CR-driven winds in the streaming limit using equation \ref{gasenstr2} as follows.   The velocity of the gas will increase with increasing radius so long as $v_A > v$.   This is because the CR enthalpy $\sim c_{\rm eff}^2 \propto \rho^{-1/3}$ increases to arbitrarily large values with decreasing density so long as $p_c \propto \rho^{2/3}$. The acceleration of the flow ceases, however, when $v_A < v$ because then the gas becomes adiabatic and $p_c \propto \rho^{4/3}$, i.e., $c_{\rm eff}^2 \propto \rho^{1/3}$ (eq. \ref{eq:ceff}), which decreases with decreasing density.  There is thus a key radius in the flow where $v \sim v_A$ which determines where the acceleration ceases.   We call this the Alfv\'en radius\footnote{The Alfv\'en radius in magnetocentrifugal winds is conceptually different from that defined here.} $r_A$ and estimate the terminal speed using $v_\infty \simeq v_A(r_A)$.   We can estimate $r_A$ and $v_\infty$ by equating the Bernoulli-like constant in equation \ref{gasenstr2} at the sonic point $r_s$ and $r_A$ and using equations \ref{sonic}.   In doing so we neglect the gravity term which makes only a small logarithmic correction to the result (and for any potential with a finite escape speed the correction is even less important).   This yields
\be
v_\infty^2 \simeq V_g^2 + 6 \Veffsq\left(\left[\frac{\rho(r_A)}{\rho(r_s)}\right]^{-1/3} - 1 \right)
\label{eq:vinfstr}
\ee
Given $\rho(r_s)$ from the sonic conditions (eq. \ref{eq:rhosonic}), equation \ref{eq:vinfstr} has 3 unknowns:  $v_\infty$, $r_A$, $\rho(r_A)$.  The two additional equations are the definition $v_A(r_A) = v_\infty$, which yields $\rho(r_A) = B_0^2/(4\pi v_\infty^2)\, (r_0/r_A)^4$ and conservation of mass ($\dot M$ = const) between $r_s$ and $r_A$, which yields $\rho(r_A)r_A^2 v_\infty \simeq r_s^2 \rho(r_s) V_g$.    These two equations combine to yield
\be
\frac{\rho(r_A)}{\rho(r_s)} = \left(\frac{r_s}{r_0}\right)^4 \left(\frac{V_g \, c_{\rm eff,0}}{\Veff \, v_{A,0}}\right)^2 \left(\frac{c_{\rm eff,0}}{\Veff}\right)^4 = \left(\frac{v_{A,crit}}{v_{A,0}}\right)^2
\label{eq:rhorat}
\ee
where the second equality uses equation \ref{eq:vAlarge}.  Recall that $v_{A,0} > v_{A,crit}$ is required for the validity of our high $v_A$ approximation, in which case $v_\infty \gtrsim V_g$ from equation \ref{eq:vinfstr}.  Equation \ref{eq:rhorat} can be substituted into equation \ref{eq:vinfstr} to estimate $v_\infty$.  The expression is particularly simple in the high $v_A$ limit of massive galaxies with $\Veff \simeq V_g \gg c_{\rm eff,0} \gtrsim c_i$ and $r_s \simeq e^{3/2} r_0$ (eq. \ref{eq:rssimp}):
\beq
\begin{split}
    v_\infty \simeq &  \ \sqrt{6} \frac{V_g}{e} \left(\frac{V_{A,0}}{c_{\rm eff,0}}\right)^{1/3} \left(\frac{V_g}{c_{\rm eff,0}}\right)^{2/3}  \simeq 400 \kms  \, \times \\ &  \, \left(\frac{V_g}{100 \kms}\right)^{5/3} \left(\frac{v_{A,0}}{10 \kms}\right)^{1/3} \left(\frac{c_{\rm eff,0}}{10 \kms}\right)^{-1}
    \label{eq:vinfstr2}
    \end{split}
\eeq
where  we again note that equation \ref{eq:vinfstr2} requires that equation \ref{eq:vAlarge} is satisfied.   When $v_{A,0} \lesssim v_{A,crit}$, either there is no wind (for very small $v_{A,0}$) or $v_\infty \lesssim V_g$.   As discussed in the context of equations \ref{mdotstrgam0.6} and \ref{eq:mdotstr2}, the largest mass-loss rates generally arise in phases of the ISM with larger $c_{\rm eff,0}$.   Equation \ref{eq:vinfstr2} shows that a consequence of this larger mass-loading is that the velocity of the outflow is significantly smaller at larger $c_{\rm eff,0}$, probably at most $\sim V_g$.

For the case of massive galaxies with $v_{A,0} > v_{A,crit}$, equations \ref{mdotstrgam0.6} and \ref{eq:vinfstr2} can be combined to yield the total energy carried by the wind.   We express this wind `luminosity' in terms of the CR `luminosity' at the base of the wind, $\dot E_c(r_0) = 16 \pi r_0^2 p_{c,0} v_{A,0}$, yielding
\be
\frac{0.5 \dot M_w v_\infty^2}{\dot E_c(r_0)} \simeq \frac{e}{2} \frac{c_{\rm eff,0}^2}{v_{A,0}^{1/3} V_g^{5/3}} \simeq \frac{1}{2} \left( \frac{v_{A,crit}}{v_{A,0}}\right)^{1/3} \left(\frac{c_{\rm eff,0}}{V_g}\right) \ll 1
\label{eq:Edotstr}
\ee
where the second equality uses the definition of $v_{A,crit}$ valid for massive galaxies. The final inequality in equation \ref{eq:Edotstr} follows from requiring  $c_{\rm eff,0} < V_g$ and $v_{A,0} > v_{A,crit}$ (the latter for the validity of our analytics) and implies that in the streaming limit, the total energy carried by the gas at large radii is less than that supplied to CRs at the base of the wind.  This is because for an isothermal gas equation of state, streaming losses $\sim v_A dp_c/dr$ remove energy from the CRs and are assumed to be rapidly radiated away by the gas.  

Finally, we can derive an expression for the asymptotic momentum flux in the wind $\dot p_w = \dot M_w v_\infty$.   This quantity can be compared with the total momentum rate carried by photons from star formation $\dot{p}_{ph}=L_{\rm bol}/c=\epsilon_{ph}\dot{M}_* c$, where $\epsilon_{ph,\,-3.3}=\epsilon_{ph}/5\times10^{-4}$ for steady-state star formation and a standard IMF.  Using equations \ref{eq:mdotstr2} and \ref{eq:vinfstr2}, we then have that 
\be
\begin{split}
\frac{\dot p_w}{\dot p_{ph}} & \simeq \frac{\dot M_w}{\dot M_*} \frac{v_\infty}{\epsilon_{ph} c} \\ & \simeq  0.16 \, \epsilon_{ph,\,-3.3}^{-1} \, \left(\frac{c_{\rm eff,0}}{10 \, \kms}\right)^3 \left(\frac{p_*/m_*}{3000 \, \kms}\right) \\ & \times \left(\frac{V_g}{100 \, \kms}\right)^{-10/3} \left(\frac{v_{A,0}}{10 \, \kms}\right)^{1/3}  \frac{p_{c,0}}{\pi G \Sigma_g^2}
\end{split}
\label{eq:pdotw}
\ee
Equation \ref{eq:pdotw} shows that the momentum flux in isothermal CR-driven winds is in general less than or at most comparable to that in the stellar radiation field.  For comparison, IMF-averaged SNe-ejecta have momentum fluxes $\sim L/c$ and the momentum can increase by a factor of $\sim 10$ or more in the Sedov-Taylor phase  (\citealt{Ostriker2011}).   

\subsection{Maximum Mass-Loss Rate}
\label{sec:maxmdot}
There is an upper limit to the mass-loss rate associated with CR-driven winds that is set by energy conservation.   This upper limit can be rigorously derived for diffusive CR transport because in that regime there is a conserved wind power that is independent of radius (eq. 34 of Paper I).   This is not the case for isothermal winds with CR streaming because streaming losses remove energy from the cosmic-rays (and that energy is assumed to be radiated away).   It is nonetheless instructive to provide a rough estimate of the maximum mass-loss rate for a given set of base conditions.   

The cosmic-ray power supplied at the base of the wind is $\dot E_c(r_0) = 16 \pi r_0^2 p_{c,0} v_{A,0}$.    Assume that a fraction $\zeta$ of this energy goes into lifting gas out to larger radii (the remaining energy is lost radiatively).  The maximum mass-loss rate is realized when the asymptotic kinetic energy vanishes, so that $\dot M_{\rm max} \simeq 2 \zeta \dot E_{c}(r_0)/v^2_{\rm esc}(r_0)$ where $v_{\rm esc}(r_0)$ is the escape speed from the base of the wind.\footnote{$v_{\rm esc}(r_0)$ is not defined for a $\ln(r)$ potential, but it is $\sim 2 V_g$ for a more realistic potential which deviates from isothermal at larger radii (and/or for a computational domain of reasonable size even with a $\ln(r)$ potential).}   

Equation \ref{eq:Edotstr} provides an estimate of the energy in the wind at large radii, i.e., $\zeta$.  However, this result only holds when the asymptotic kinetic energy is finite, which is not the case as $\dot M_w \rightarrow \dot M_{max}$.  We do not have an analogous approximation for $\zeta$ valid as $\dot M_w \rightarrow \dot M_{max}$.   To account for this we use equation \ref{eq:Edotstr} but multiply the result by a dimensionless factor $\zeta_0$.   Doing so, it is straightforward to combine equation \ref{mdotstrgam0.6}, \ref{eq:vAlarge}, \& \ref{eq:Edotstr} to find
\be
\frac{\dot M_w}{\dot M_{\rm max}} \simeq \zeta_0^{-1} \, \left(\frac{v_{A, crit}}{v_{A,0}}\right)^{2/3}
\label{eq:mdotmax}
\ee
Equation \ref{eq:mdotmax} shows that in the high base Alfv\'en speed limit, the predicted mass-loss rate is below $\dot M_{max}$, consistent with the finite terminal speed predicted by equation \ref{eq:vinfstr2}.    For lower base Alfv\'en speeds our analytic estimates are less applicable, but extrapolation of our results suggests that $\dot M_w \simeq \dot M_{\rm max}$ will be realized for $v_{A,0} \lesssim v_{A,crit}$.   We shall see in \S \ref{sec:numerics} that this extrapolation is borne out: our numerical solutions for low $v_{A,0}$ produce outflows but these outflows have $\dot M_w \simeq \dot M_{\rm max}$ and never reach the critical point at which $v = V_g$ (Fig. \ref{fig:vA}); they are formally `breezes' rather than transonic winds.

We can also express the maximum mass-loss rate $\dot M_{\rm max}$ in a form that is easier to compare to observational constraints.    To do so, we write the CR energy injection rate at the base of the outflow as 
\be
\dot E_c = \epsilon_c \dot M_* c^2,
\label{eq:edotc}
\ee
where $\dot{M}_*$ is the star formation rate and $\epsilon_c \equiv 10^{-6.3} \epsilon_{c,-6.3}$ is set by the fraction of SNe energy that goes into CRs:  for $10^{51}$\,ergs per SNe and 1 SNe per 100 $M_\odot$ of stars formed, $\epsilon_c = 10^{-6.3}$ if $10 \%$ of the SNe energy goes into primary CRs.  Our expression for the maximum mass-loss rate can thus be written as 
\be
\frac{\dot M_{\rm max}}{\dot M_*} \simeq \, 0.2 \, \zeta_0 \, \epsilon_{c,-6.3} \, \left(\frac{100 \, \kms}{V_g}\right)^{7/3} \, \left(\frac{v_{A,0}}{10 \, \kms}\right)^{1/3} 
\label{eq:mdotmaxvsmdotstarfin}
\ee
where we have assumed $v^2_{\rm esc}(r_0) \simeq 4 V_g^2$ and have maximized the CR energy flux by taking $v_{A,0} \simeq v_{A,crit}$ and hence  $c_{\rm eff,0}/V_g \simeq e^{-1}(v_{A,0}/V_g)^{1/3}$ (eq. \ref{eq:vAlarge}) and $\zeta \simeq 0.18 \zeta_0 \, (v_{A,0}/V_g)^{1/3}$ (eq. \ref{eq:Edotstr}).  

\begin{table*}
\caption{Parameters and properties of our numerical simulations.  Columns are diffusion coefficient $\kappa$, base Alfv\'en speed $v_{A,0}$, velocity $V_g$ of the isothermal gravitational potential, base CR pressure $p_{c,0}$, outer radius of domain $r_{out}$, reduced speed of light $V_m$,   mass-loss rate in units of $\dot M_0 = 4 \pi r_0^2 \rho_0 c_i$ (eq. \ref{eq:mdot0}), velocity of the gas at the outer radius $v_(r_{out})$, and ratio of the wind power at the outer radius to the CR power at the base $\dot E_w(r_{out})/\dot E_c(r_0)$.  Solutions labeled as 'Breeze' are not transonic and never reach velocities comparable to $V_g$.  Numerical resolution is $\delta r/r = 5.25 \times 10^{-4}$ unless noted otherwise.}
\begin{tabular}{c|cc|cccccccccc}
Transport  & $\kappa$ & $v_{A,0}$ & $V_g$ & $p_{c,0}$ & $r_{out}$ & $V_m$ & $\dot M_{sim}$ & $v(r_{out})$ & $\frac{\dot E_w(r_{out})}{\dot E_c(r_0)}$ & Notes \\ [1pt]
 & ($r_0 c_i$) & ($c_i$) & ($c_i$) & ($\rho_0 c_i^2$) & ($r_0$) & ($c_i$) & ($\dot M_0$) & ($V_g$) & -- \\ [3pt] \hline
 {\bf Streaming} \\ \hline
&  -- & 1 & 10 & 1 & 10 & 3000 & $5 \times 10^{-4}$ & 1.9 & 0.16\\
 &  -- & 1 & 10 & 1 & 10  & $3 \times 10^4$ & $4.6 \times 10^{-4}$ & 2.2 & 0.15 & High $V_m$ \\
  & --  & 3 & 10 & 1 & 5 & 3000 & $9 \times 10^{-4}$ & 2.2 & 0.17  \\
     &  --  & 10 & 10 & 1 & 5 & 6000 & $1.2 \times 10^{-3}$ & 2.8 & 0.16 \\
        &  --  & 10 & 10 & 1 & 5 &  3000 & $2.1 \times 10^{-3}$ & 2.9 & 0.19 & $\delta r/r=2.46\times10^{-5}$  \\
&  --  & 10 & 10 & 1 & 10  & $2 \times 10^4$ & $1.1 \times 10^{-3}$ & 4.1 & 0.14 & High $V_m$$^{a}$       \\
    &  --  & 1 & 10 & 0.3 & 5  & 3000 & $4.1 \times 10^{-5}$ & 2.4 & 0.1 \\
          &  --  & 0.1 & 6 & 1 & 55  & 3000 & $1.3 \times 10^{-4}$ & 0.56 & 0.12 & Breeze \\
      &  --  & 0.3 & 6 & 1 & 55  & 3000 & $4.3 \times 10^{-4}$ & 0.65 & 0.14 & Breeze \\
      &  --  & 1 & 6 & 1 & 15  & 3000 & $2 \times 10^{-3}$ & 0.77 & 0.19 \\
      &  --  & 3 & 6 & 1 & 15  & 3000 & $4.2 \times 10^{-3}$ & 2.6 & 0.21 \\
            &  --  & 10 & 6 & 1 & 15  & 3000  & $6 \times 10^{-3}$ & 4.1 & 0.19  \\
      &  --  & 1 & 3 & 1& 15 & 3000  & $0.014$  & 0.75 & 0.29  \\
            &  --  & 3 & 3 & 1 & 15 & 3000 & $0.04$ & 1.1 & 0.33 \\
            &  --  & 10 & 3 & 1 & 5 & 3000 & $0.065$ & 3.0 & 0.3
        \\[3pt] \hline
  {\bf Streaming \& Diffusion} \\ \hline &  0.03 & 1 & 10 & 1 & 5 & 3000 & $2.6 \times 10^{-3}$ &   0.93 & 0.33 \\
  & 0.33  & 1 & 10 & 1 & 5 & 3000 & 0.012 & 1.0 & 0.5 \\
     &  3.3  & 1 & 10 & 1 & 5 & 3000 & 0.033 & 4.3 & 0.72\\[3pt] \hline
     {\bf Diffusion} \\ \hline
     &  $3.3$ & 0 & 10 & 1 & 15  & 3000  & 0.03 & 4.6 & 1 \\
\label{tab:compare}
\end{tabular}\\
$^{a}$ Differences in velocity and $\dot E_w/\dot E_c$ relative to $V_m=6000$ are primarily due to the larger box:  $v(r=5) \simeq 2.6 V_g$ \& $\dot E_w(r=5)/\dot E_c(r_0) \simeq 0.19$. \\
\end{table*}

\section{Numerical Simulations}
\label{sec:numerics}
\subsection{Equations}
\label{sec:methods}

We solve the cosmic ray hydrodynamic equations in one dimensional spherical polar coordinates using the two moment approach developed by \cite{JiangOh2018} . This algorithm has been implemented in the magneto-hydrodynamic code {\sf Athena++} \citep{Stone2020}.
As in \S \ref{section:streaming}, we use an isothermal equation of state  for the gas with isothermal sound speed $c_i$ and take the gravitational potential to be $\phi=2V_g^2\log r$. 

The magnetic field is assumed to be a split-monopole, with 
\begin{eqnarray}
B(r)=B_0(r_0/r)^2,
\end{eqnarray} 
where $B_0$ is the magnetic field at the bottom boundary of our simulation box with radius $r_0$. We do not evolve the magnetic field in our 1D calculations. It is only used to calculate the Alfv\'en velocity, which is needed for cosmic ray streaming. 

The full set of equations for gas density $\rho$, flow velocity $v$, CR energy density $E_c$, CR pressure $p_c=E_c/3$ and flux $F_c$ in 1D spherical polar coordinates are
\begin{eqnarray}
\frac{\partial \rho}{\partial t}+\frac{1}{r^2}\frac{\partial}{\partial r}\left( r^2\rho v\right)&=&0,\nonumber\\
\frac{\partial \left(\rho v\right)}{\partial t}
+\frac{1}{r^2}\frac{\partial }{\partial r}\left(r^2 \rho v^2\right)&=&-\rho\frac{\partial \phi}{\partial r}-c_i^2\frac{\partial \rho}{\partial r} +\sigma_c\left[F_c-v(E_c+p_c)\right],\nonumber\\
\frac{\partial E_c}{\partial t}+\frac{1}{r^2}\frac{\partial\left( r^2F_c\right)}{\partial r}&=&
-(v+v_s)\sigma_c [F_c-v(E_c+p_c)],\nonumber\\
\frac{1}{V_m^2}\frac{\partial F_c}{\partial t}+\frac{\partial p_c}{\partial r}&=&-\sigma_c\left[F_c-v(E_c+p_c)\right].
\label{eq:CR2mom}
\end{eqnarray}
The streaming velocity is $v_s=-v_A \nabla p_c/|\nabla p_c|$, where the Alfv\'en velocity is calculated based on the assumed magnetic field and gas density $\rho$ as $v_A=B/(4\pi \rho)^{1/2}$. The reduced speed of light is $V_m$, which is chosen to be much larger than $v$ and $v_A$ in the whole simulation box. We will primarily carry out simulations with cosmic ray transport by  streaming;  in a few cases, diffusion will be included as well.  For the case of streaming, 
\begin{eqnarray}
\sigma_c=\frac{|\nabla p_c|}{v_A(E_c+p_c)}.
\label{eq:kapst}
\end{eqnarray} 
while for cases with both diffusion\footnote{As in the radiation transfer literature, \citet{JiangOh2018} write the diffusive flux as $F_c = -\sigma_c^{-1} \nabla p_c$. In the rest of this paper, and in Paper I, we follow the CR literature and define the diffusive flux as $F_c = -\kappa \nabla E_c$ (eq. \ref{cr_energy}).  This accounts for the factor of 3 relating $\kappa$ and $\sigma_c^{-1}$ in equation \ref{eq:kapboth}.} and streaming
\begin{eqnarray}
\sigma_c^{-1} = 3 \kappa + \frac{v_A(E_c+p_c)}{|\nabla p_c|}
\label{eq:kapboth}
\end{eqnarray} 
Substituting the right-hand-side of the fourth of equations \ref{eq:CR2mom} into the 3rd term on the right-hand-side of the 2nd of equations \ref{eq:CR2mom} produces the usual $\partial p_c/\partial r$ cosmic-ray pressure gradient in the momentum equation, along with another term related to the time variation of the CR flux (this latter term is small in our simulations, as we have verified post facto).   Note also that multiplying the fourth of equations \ref{eq:CR2mom} by 
 equation \ref{eq:kapboth} yields equation \ref{equilibrium_flux}, along with another term related to the time variation of the CR flux.   The time dependent term in the CR flux equation {\em is} dynamically important in our simulations in some regions as we will describe more below.  Physically this means that the assumption that the CR energy flux is given by equation \ref{equilibrium_flux} is not well-satisfied in many locations in our simulations.      

A few additional comments about the physical content of the two-moment model are in order.   In the presence of CR diffusion, so that $\kappa \ne 0$, $\sigma_c$ is always finite.  This is not the case for transport by streaming alone with $\kappa \rightarrow 0$.  In that case, $\sigma_c \rightarrow 0$ when $\nabla p_c \rightarrow 0$ and the two-moment CR equations reduce to a hyperbolic system in which the intrinsic transport speed is the speed of light $V_m$.  Physically, this captures the fact that if $\nabla p_c = 0$ there is nothing driving the streaming instability and thus nothing to limit the CR transport to the Alfv\'en speed.  This is also precisely the regime in the which the time-dependent term in the fourth of equations \ref{eq:CR2mom} is important.  A transport speed of $V_m$ does not, however, imply $F_c \sim V_m p_c$.   Rather, the relation between CR flux and energy density is set by boundary conditions.    This is analogous to the fact that for photons,  $F \sim  p c$ in the optically thin limit for a single source but not if one is measuring the flux in between two equal sources of photons, in which case $F \sim 0$.     

In what follows we refer to the CRs as ``coupled" to the gas when there is a finite CR pressure gradient such that the CR energy flux due to streaming is close to the one-moment value of $4 p_c (v_A + v)$.   Formally, this holds when the time dependent term in the CR flux equation (eq. \ref{eq:CR2mom}) can be neglected. 

\subsection{Initial and Boundary Conditions}
For each simulation, we pick gas density $\rho_0$, magnetic field $B_0$ (parameterized by $v_{A,0}=v_A(r_0)$) and cosmic ray pressure $p_{c,0}$ at the bottom boundary $r_0$ and then initialize the gas density and cosmic ray energy density at each radius $r$ as
\begin{eqnarray}
\rho(r)&=&\rho_0 \left(r/r_0\right)^{-V_g^2/c_i^2},\nonumber\\
E_c(r)&=&3p_c(r_0)\left(r/r_0\right)^{-V_g^2/c_i^2}.
\end{eqnarray}
We apply floor values of $10^{-4}$ for $\rho/\rho_0$ and $p_c(r)/p_c(r_0)$ in the initial condition. The flow velocity and cosmic ray flux are set to be zero in the whole simulation domain initially.

For the bottom boundary condition, we fix the cosmic ray energy density $E_c$ and gas density $\rho$ to the desired values at $r_0$. Then we determine the gradients of $E_c$ and $\rho$ in the ghost zones by assuming steady state. Flow velocity in the ghost zones is set such that mass flux $\rho v r^2$ is continuous from the last active zone to the ghost zones. The cosmic ray flux in the ghost zones is set to be $(v+v_A)(E_c+p_c)$ for the streaming simulations.  

For the boundary condition at the top of the simulation domain, we keep the gradient of $\rho$ and $E_c$ continuous across the boundary but find that it is necessary to set the cosmic ray flux to be $V_m E_c/3^{1/2}$ (so that $F_c/E_c > v_A$) at the top boundary for the streaming case to get an outflow solution (for the diffusion-only case it is sufficient to simply keep the gradient of $F_c$ continuous across the outer boundary). The flow velocity at the top boundary is  set by requiring the mass flux $\rho v r^2$ to be continuous.  The boundary conditions for the simulations with both streaming and diffusion are the same as those with streaming alone.

\subsection{Simulation Suite}

\begin{table}
	\caption{Summary of Units for Numerical Simulations. $c_i$ is the gas isothermal sound speed, $\rho_0$ is the gas density at the base of the wind (radius $r = r_0$), and $p_{c,0}$ is the base CR pressure.}
\centerline{	\begin{tabular}{c|cc}
Quantity & Symbol & Units \\ \hline
Radial Velocity & $v$ & $c_i$ \\
`Isothermal' CR Sound Speed & $c_c \equiv \left(\frac{p_c}{\rho}\right)^{1/2}$ & $c_i$ \\
Alfv\'en Speed & $v_A$ & $c_i$ \\
Gravitational Velocity  & $V_g$ & $c_i$ \\
Density & $\rho$ & $\rho_0$ \\
CR Pressure & $p_c$ & $p_{c,0}$ \\
CR Flux & $F_c$ & $c_i p_{c,0}$ \\
CR Diffusion Coefficient & $\kappa$ & $c_i r_0$ \\
\hline
\label{tab:units}
\end{tabular}}
\end{table}

Table \ref{tab:compare} summarizes our simulations.   The key physical parameters of each simulation are $V_g/c_i$ and $p_{c,0}/\rho_0 c_i^2$, as in the analytics of \S \ref{section:streaming}, as well as the base Alfv\'en speed $v_{A,0}$ (in units of $c_i$) and, for a few of the simulations, the diffusion coefficient $\kappa$ (in units of $r_0 c_i$). The key numerical parameters are the resolution, reduced speed of light $V_m$, and box size.   Table \ref{tab:compare} also includes one of our diffusion only simulations from Paper I, for comparison to the analogous simulations with both streaming and diffusion and streaming alone.   In \S \ref{sec:strvsdiff} we present a detailed comparison of the streaming results of this paper and the diffusion results from Paper I.

The units for the results of our numerical simulations are summarized in Table \ref{tab:units}: gas density is in units of the base density $\rho_0$, speeds are in units of the gas isothermal sound speed $c_i$, CR pressure is in units of the base CR pressure $p_{c,0}$, and CR fluxes are in units of $p_{c,0} c_i$.

\subsection{Streaming-Only Simulations}
\label{sec:streamnum}

\begin{figure*}
\centering
\includegraphics[width=175mm]{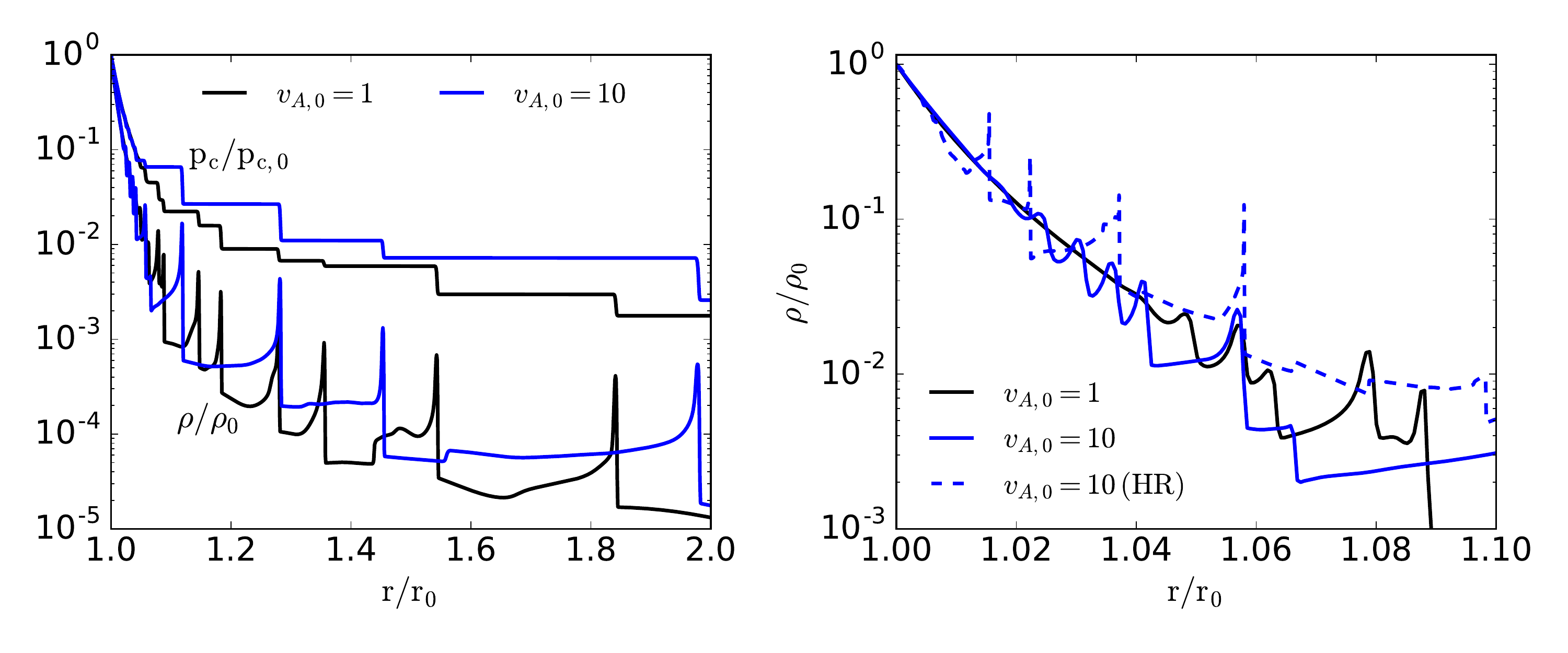}
\vspace{-0.2cm}
\caption{{\em Left:} Instantaneous CR pressure and density profiles in several of our $V_g=10$ simulations.  The isothermal gas sound speed $c_i=1$ in our units so gas density is also gas pressure.   The flow is permeated by strong shocks, which are sourced by a linear instability near the base of the wind (\S \ref{sec:lin}).   The density increase in the vicinity of the shock triggers the `bottleneck' effect which locally flattens the CR pressure profile leading to the staircase like structure in $p_c$.  Figure \ref{fig:shocks} shows a zoom-in on the structure of one of the shocks.  {\em Right:} Density profile near the base of the wind where the instability and shocks originate; they grow more quickly at higher $v_{A,0}$ and higher resolution (HR); see \S \ref{sec:lin} and Appendix \ref{sec:appendixA}.}
\label{fig:rho_stream}
\end{figure*}

In this section we summarize the properties of our numerical models of galactic winds with cosmic-ray transport via streaming at the Alfv\'en speed.   The models explored are summarized in Table \ref{tab:compare}.   We also compare  the numerical solutions to the analytic solutions summarized in \S \ref{section:streaming} (and, in a few cases, to the modified analytic models in \S \ref{sec:strmod} that are motivated by the numerical solutions).   We begin by analyzing simulations with $V_g/c_i = 10$ and $v_{A,0}/c_i =1, 10$ because these most dramatically highlight some of the new features revealed by our time-dependent simulations.  We then consider different values of $V_g/c_i$ in \S \ref{sec:Vg} and a wider range of $v_{A,0}$ in \S \ref{sec:vA}. 

\begin{figure*}
\centering
\includegraphics[width=174mm]{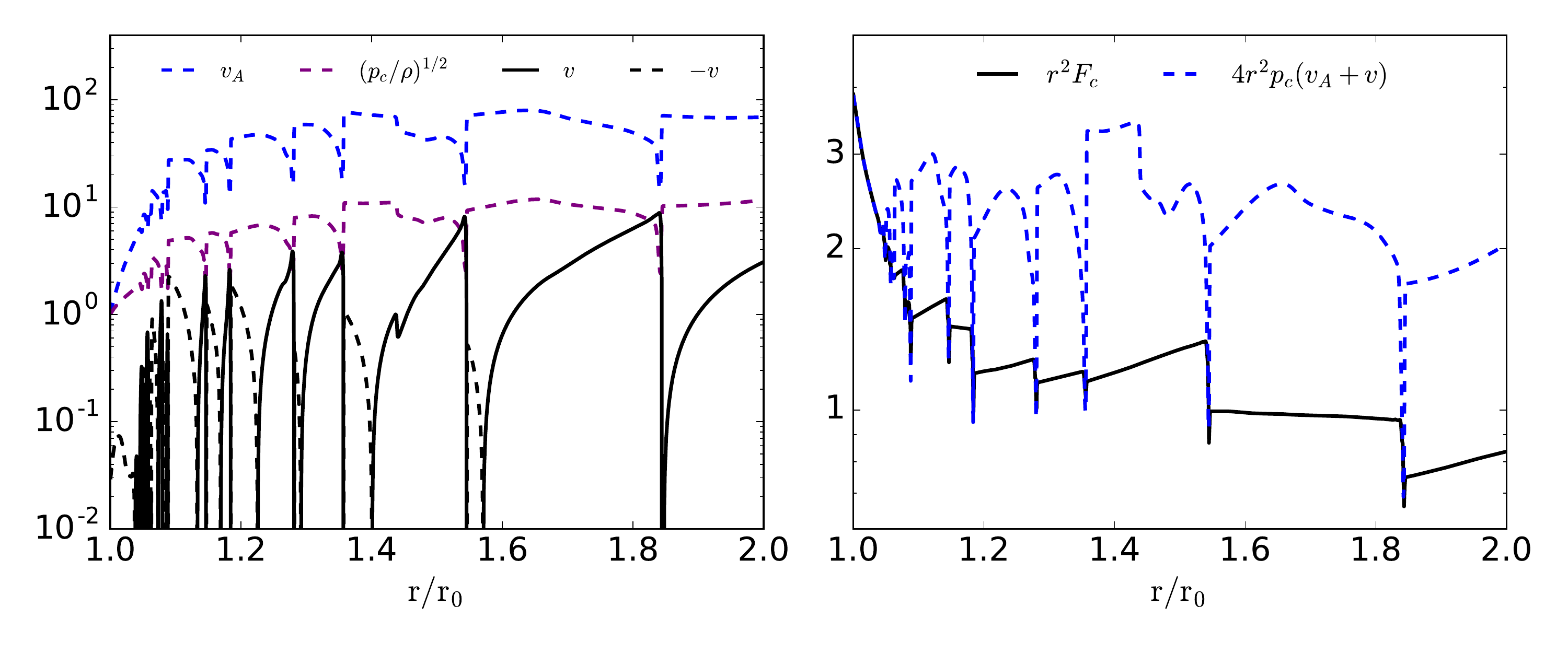}
\vspace{-0.2cm}
\caption{{\em Left:}  Alfv\'en velocity, flow velocity, and `isothermal' CR sound speed for our $V_g=10, v_{A,0}=1$ simulation; gas sound speed $c_i=1$.  Note the hierarchy of velocities at large radii with $v_A > \sqrt{p_c/\rho} > |v| > c_i$.  {\em Right:} CR flux $F_c$ evolved in our two moment formalism and `equilibrium' CR flux $4p_c(v+v_A)$ assumed in one-moment CR models. The equilibrium CR flux is a good approximation near the the base of the wind but not exterior to that once the strong shocks and CR bottleneck develop.}
\label{fig:vel_flux_stream}
\end{figure*}

A key feature of nearly all of the streaming simulations in this paper is that they are time dependent due to an instability that develops near the base of the wind that steepens into shocks.   This is shown in Figure \ref{fig:rho_stream}, which plots the CR pressure and gas density as a function of radius once the simulations have evolved to a statistical steady state.\footnote{The full domain is larger than that shown in Figures \ref{fig:rho_stream}-\ref{fig:vel_flux_stream} (see Table \ref{tab:compare}), but we focus on $r/r_0=1-2$ because otherwise the number of shocks in the Figures is  large and obscures legibility.}   Recall that in our units the isothermal sound speed is 1 so that the gas density in Figure \ref{fig:rho_stream} and subsequent Figures is also the gas pressure.  The right panel of Figure \ref{fig:rho_stream} zooms in further to the inner region near the base of the wind where the instability and shocks originate.  We derive the physical origin of the instability in Appendix \ref{sec:appendixA} and summarize the results below in \S \ref{sec:lin}.  As a point of contrast, we note that the diffusion simulations in Paper I nearly all reached a laminar steady state with no analogous instability or shocks.

\begin{figure}
\centering
\includegraphics[width=84mm]{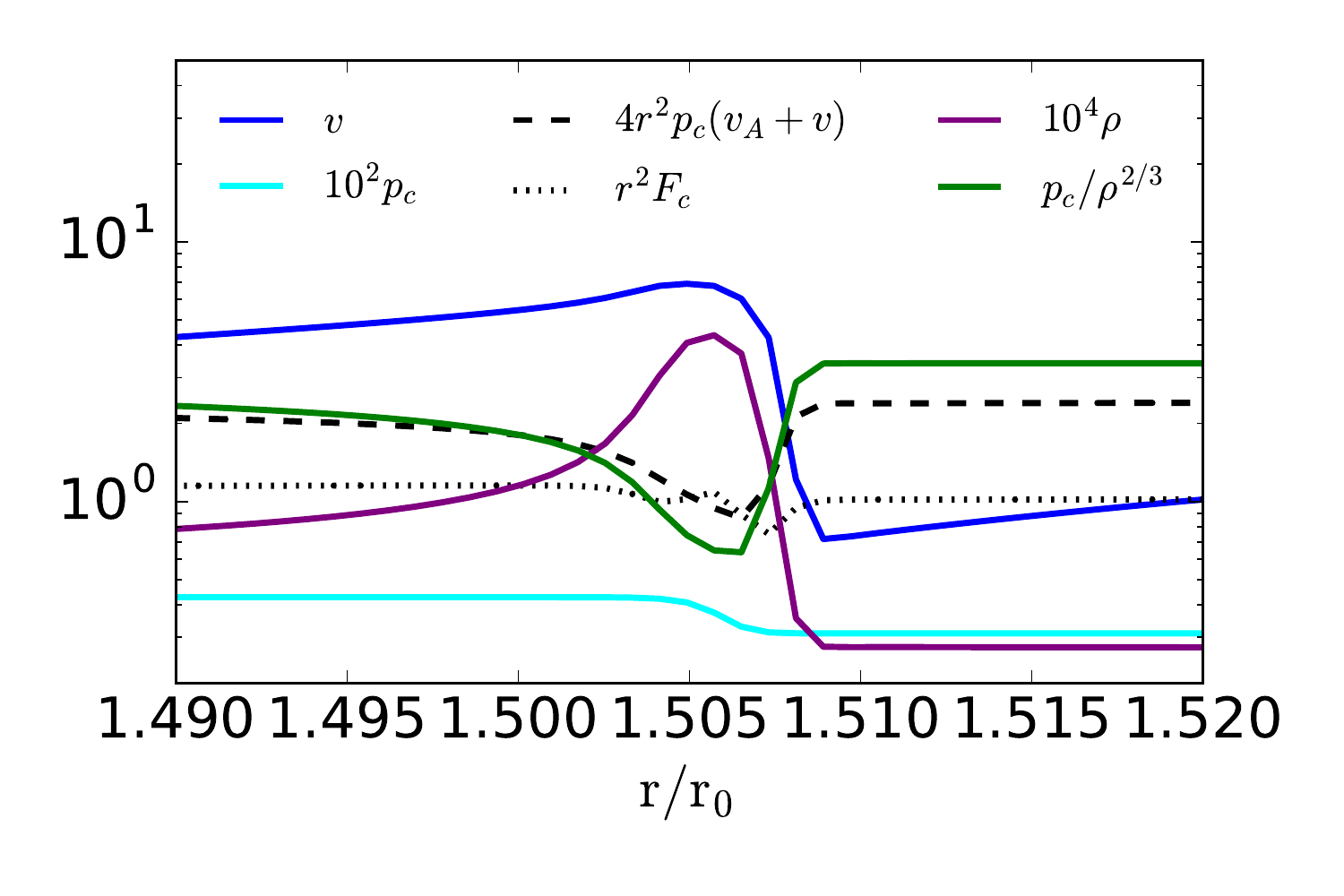}
\vspace{-0.2cm}
\caption{Zoom-in on the structure of one of the shocks in the $V_g=10, v_{A,0}=1$ simulation.  The CRs are well coupled, i.e., $dp_c/dr \ne 0$ and $F_c \approx 4 p_c (v_A + v)$ only in a small subregion of the shock where the density is larger than average and $v_A$ is smaller than average.  This minimizes the overall magnitude of CR streaming losses ($v_A dp_c/dr$) in the wind.  The large density compression is roughly consistent with that expected for an isothermal shock (recall that $c_i=1$ in our units so this is a high Mach number shock).}
\label{fig:shocks}
\end{figure}

Additional features of the time-dependent streaming solution for $v_{A,0}=1$ are shown in Figures \ref{fig:vel_flux_stream} \& \ref{fig:shocks}:   the left panel of Figure \ref{fig:vel_flux_stream} shows the gas velocity, Alfv\'en velocity, and the `isothermal' CR sound speed $(p_c/\rho)^{1/2}$; while the right panel shows the true CR flux from our time dependent solution compared to the steady state flux assumed in  one-moment CR transport models.  Figure \ref{fig:shocks} zooms in and shows the detailed structure of one of the shocks.

The most striking features of Figure \ref{fig:rho_stream} are (1) the outflows are permeated by strong shocks in which the gas density increases by at least a factor of $\sim 10$, and (2) the CR pressure profile is not continuous, but is instead largely flat with significant changes only near the shocks.   These two features are related because of the bottleneck effect:   if CRs are well-coupled to the gas by scattering due to the streaming instability, they stream at the Alfv\'en speed relative to the gas and the solution to the steady state CR energy equation for $v_A \gg v$ is  $p_c \propto \rho^{2/3}$ (see \S \ref{section:streaming}).  In regions where $\rho$ increases outwards, which occupy a large fraction of the volume in our wind solutions (Fig. \ref{fig:rho_stream}),  $p_c$ thus needs to increase outwards if the CRs are well-coupled.   This conclusion is, however, inconsistent with the fact that CRs stream down the CR pressure gradient.  As a result, the only physical solution when the gas density increases outwards is that the CR pressure is flat.  The absence of a CR pressure gradient  implies that there is no driving of the streaming instability and hence no scattering to pin the CR streaming speed at $v_A$.
A corollary of the fact that $p_c \sim$ constant over much of the domain is that the CR flux deviates from the canonical streaming value of $F_c = 4 p_c (v + v_A)$.  The reason is that when $p_c \sim$ constant, the steady state CR energy equation only requires $r^2 F_c \sim$ constant, with no direct constraint on its value.   Figure \ref{fig:vel_flux_stream} (right panel) shows that in general the CR flux is a factor of few less than the standard streaming solution, with $F_c = 4 p_c (v + v_A)$ only near the base of the wind prior to the shocks developing and in the vicinity of the shocks where the CR pressure changes significantly.

Figure \ref{fig:shocks} shows the properties of the shocks in detail.   From Figure  \ref{fig:vel_flux_stream} (left panel) there is a rough hierarchy of velocities at the shocks with $v_A \gg (p_c/\rho)^{1/2} \sim v \gtrsim c_i$.    The gas density jump at the shock is roughly consistent with an isothermal shock in which the density increases by a factor of $(V_s/c_i)^2$ where $V_s$ is the speed of the shock.   This is apriori somewhat surprising given that $p_c \sim \rho V_s^2$ so one might expect  CR pressure to limit how compressive the shocks are.   Indeed, if the CRs were fully coupled, $p_c \propto \rho^{2/3}$ would imply that the density could only jump by order unity for a shock in which $p_c \sim \rho V_s^2$.   The reason this is not realized in our numerical solutions is that the CRs are in fact not well-coupled throughout most of the shock.    Figure  \ref{fig:shocks} shows this explicitly:   the CR pressure is flat in much of the region where the density changes appreciably and $p_c/\rho^{2/3}$ varies strongly as a result.   Another way to see this is that only in a very small sub-region of the shock is the CR energy flux equal to the coupled one-moment value of $4 p_c(v_A+v)$.   

We interpret the fact that the CRs are not well coupled throughout the shock as a consequence of a global constraint on the CR pressure profile. This constraint regulates the change in the CR pressure at each shock.   Note that per Figure \ref{fig:vel_flux_stream} most of the flow has a velocity less than the CR sound speed and so is all in causal contact.    As we show in \S \ref{sec:timeavg} below, the time-averaged CR pressure profile roughly satisfies $p_c \propto \rho^{1/2}$ so that on average the CR pressure scale-height is only a bit larger than the gas scale-height.   The CR pressure only changes significantly at the shocks so this constraint on the CR pressure scale-height implies that on average $N_{shock} \Delta p_c/p_c \sim 1/2$ where $N_{shock}$ is the number of shocks per density scale-height and $\Delta p_c/p_c$ is the fractional change in CR pressure at each shock.   The instabilities that seed the shock initially grow on a wide range of length-scales but the longest wavelength modes saturate at the highest amplitudes, and some shocks merger with each other, leading to $N_{shock} \sim 1$, i.e., modes with wavelengths comparable to the local density scale-height dominate.  This prediction of $\Delta p_c/p_c \sim 1/2$ is consistent with the modest change in CR pressure per shock in the numerical simulations (e.g., Fig. \ref{fig:rho_stream}).

\subsubsection{Physical Origin of the Instability}
\label{sec:lin}

\citet{Begelman1994} showed that sound waves are driven unstable by CR streaming when the magnetic energy density is larger than the gas energy density.   This result only applies, however, in the absence of strong radiative cooling.   For an isothermal gas equation of state as we use here, sound waves are stable given the physics included in \citet{Begelman1994}'s analysis.   To understand the origin of the instability in our simulations, in Appendix \ref{sec:appendixA} we calculate the local linear stability of an isothermal hydrostatic atmosphere with CRs.  We identify two new instabilities in the presence of CR streaming.  The first is present in one-moment CR models and is driven by background gradients in density and CR pressure.   This instability is a streaming analogue of the acoustic instability with CR diffusion and a background CR pressure gradient studied by \citet{Drury1986}.    It leads to sound waves being  amplified after propagating a few scale-heights.  We believe that this is the dominant instability in most of our simulations since the numerical and analytic growth rates are reasonably consistent with each other (Fig. \ref{fig:growth}) and the numerical growth rates are largely independent of the reduced speed of light, consistent with an instability in the well-coupled one-moment regime.   We also identify a second instability driven by CR streaming in the two-moment formulation that is not present in one-moment CR transport models, i.e., the instability relies on the finite speed of light.   The unstable mode is  a sound wave at high $c_i/v_A$ but the entropy mode at low $c_i/v_A$; the latter is the most relevant to CR-driven wind models which tend to be magnetically dominated.     Physically, the finite speed of light introduces a phase shift between the CR flux and the CR pressure (eq. \ref{eq:CR2mom}).   This phase shift effectively produces a negative diffusion coefficient that amplifies linear waves (eq. \ref{eq:negdiff}).  We believe that this instability causes the very rapid short wavelength modes to grow in our highest resolution simulation in Figure \ref{fig:rho_stream}.

Galactic winds are likely subject to other CR-driven instabilities that might affect their dynamics in a way qualitatively akin to that found here.  We return to this point in \S \ref{sec:summary}.

\subsubsection{Quasi-Steady Structure of Streaming Solutions}
\label{sec:timeavg}
\begin{figure}
\centering
\includegraphics[width=84mm]{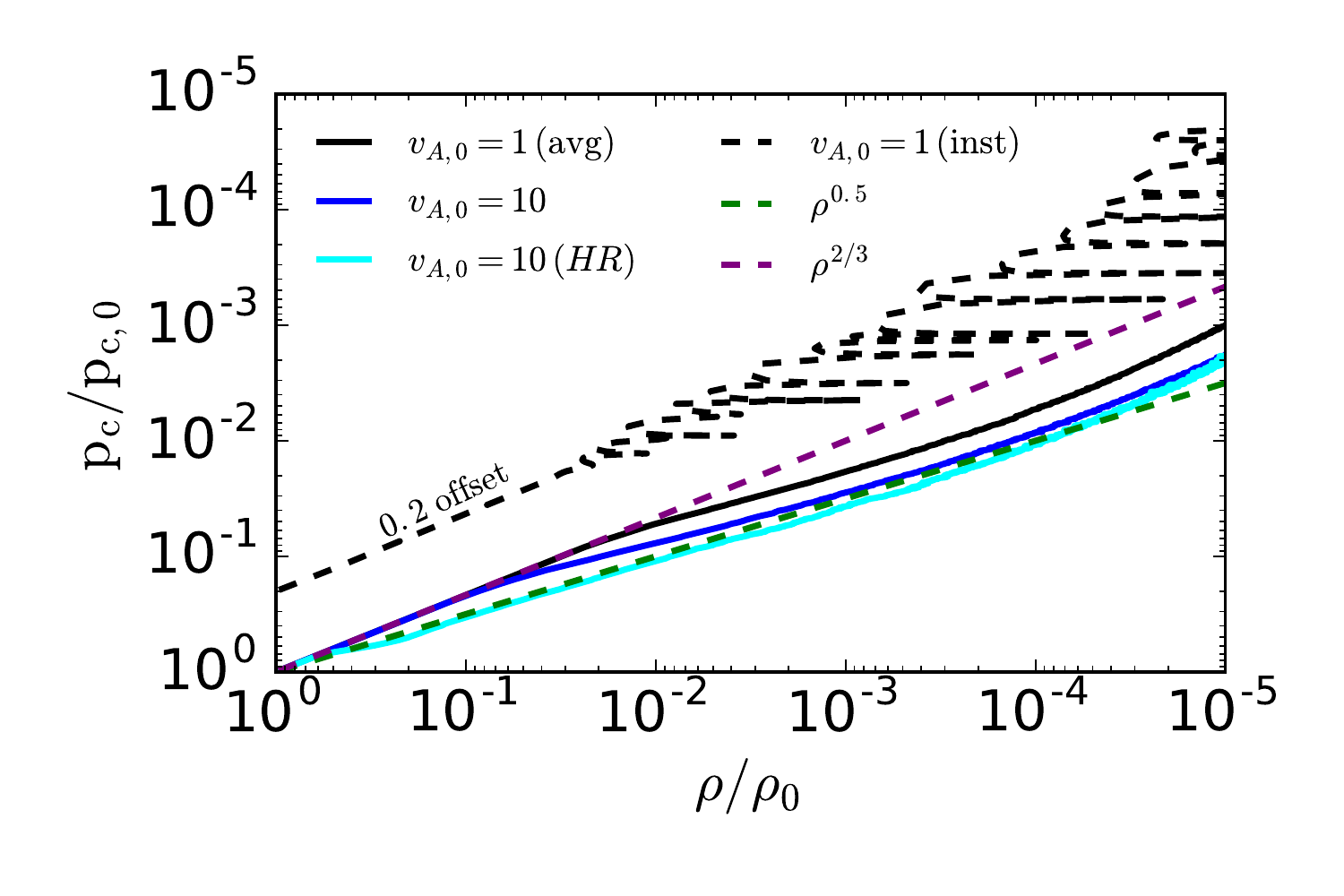}
\vspace{-0.2cm}
\caption{Time-averaged $V_g=10$ solutions in the $p_c-\rho$ plane, along with one instantaneous snapshot. At high densities near the base of the wind, $p_c \propto \rho^{2/3}$. However, once the shocks set in, the effective equation of state is closer to $p_c \propto \rho^{1/2}$.  This occurs at smaller radii for higher $v_{A,0}$ and/or higher resolution (Fig. \ref{fig:rho_stream}). }
\label{fig:pc-rho_stream}
\end{figure}

In this section we show that despite the time-dependent nature of our galactic wind simulations with CR streaming, the time-averaged properties of the winds can be  understood using  a modified version of CR-driven wind theory.    The key modification is that the inhomogeneous nature of the flow leads to an average relation between CR pressure and gas density that differs from the standard $p_c \propto \rho^{2/3}$ result derived for CR streaming at high $v_A$ (eq. \ref{simple_crenergy}).   To see this, Figure \ref{fig:pc-rho_stream} shows the time-averaged CR pressure vs. time-averaged CR density for three of our simulations.   For comparison, we also show the instantaneous $p_c(\rho)$ profile for the  $v_{A,0}=1$ simulation.   Near the base of the wind $p_c \propto \rho^{2/3}$.    Once the shocks develop, however, the time-averaged relation between $p_c-\rho$ flattens to be closer to an effective adiabatic index $\gamma_{\rm eff} \sim 1/2$, i.e., $p_c \propto \rho^{1/2}$.   This transition happens closer to the base of the wind at higher $v_{A,0}$ and higher resolution, consistent with where the shocks first develop in these simulations in Figure \ref{fig:rho_stream}.

\begin{figure*}
\centering
\includegraphics[width=175mm]{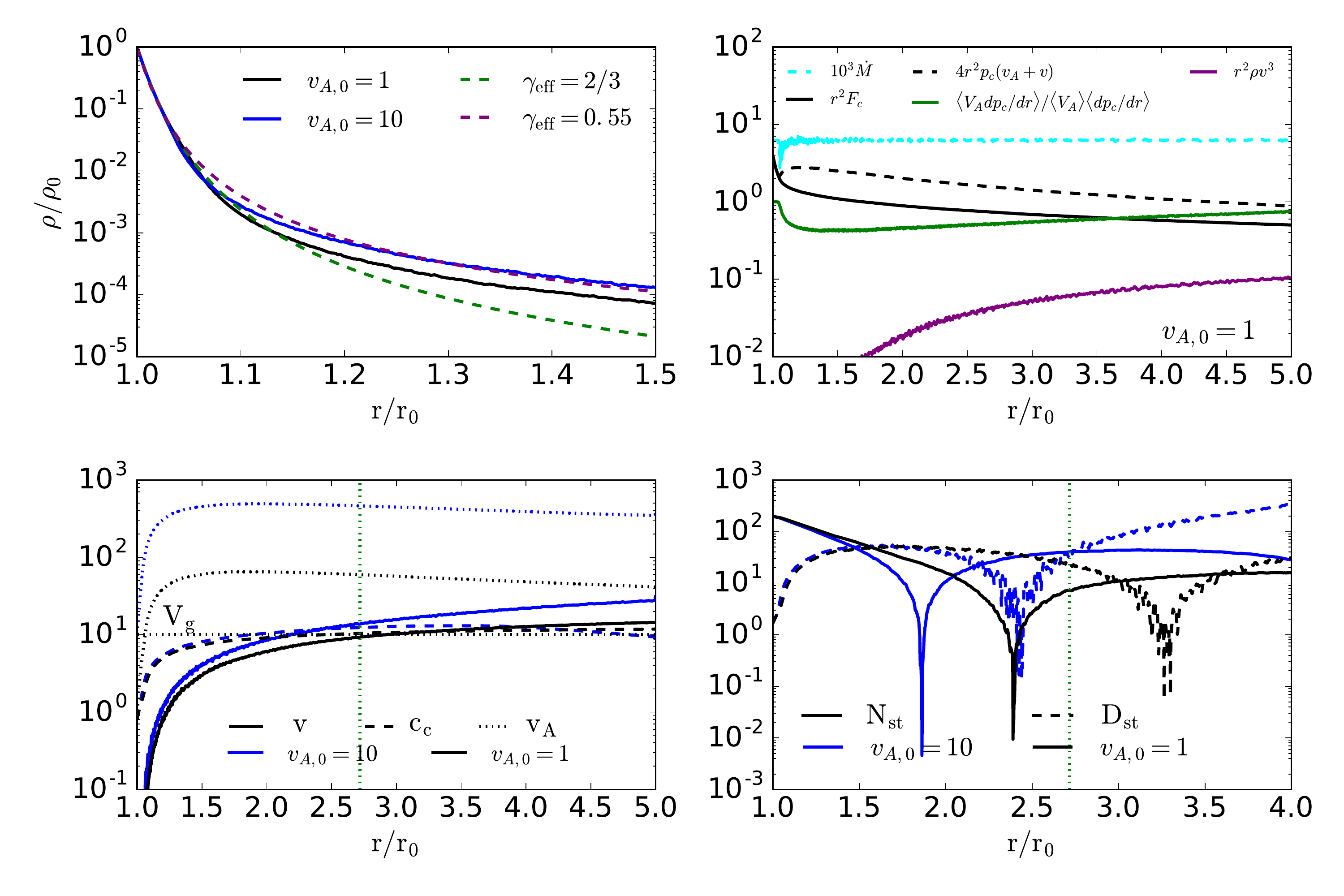}
\vspace{-0.2cm}
\caption{Time-averaged properties of galactic winds with CR streaming for $V_g=10$ and $v_{A,0}=1, 10$.   {\em Upper Left:}.   Time-averaged density profiles compared to our analytics; standard high $v_A$ ($\geff=2/3$) models underestimate the density at larger radii and thus the mass-loss rate while the lower $\geff$ models from \S \ref{sec:strmod} do much better.   {\em Upper Right:}. Time-averaged mass and energy loss-rates for $V_g=10$ \& $v_{A,0}=1$.   The total energy flux is dominated by streaming even at large radii; the gas kinetic energy flux is subdominant.   The green solid line quantifies how the inhomogeneous flow, with strong shocks surrounded by bottleneck regions in which $p_c=$ constant (Fig. \ref{fig:rho_stream}), minimizes streaming losses by concentrating the regions with a large CR pressure gradient where the density is high and the Alfv\'en speed is low.  This suppression of streaming losses makes $\geff \sim 0.5$ a better approximation than the canonical $\geff=2/3$ solution (see Fig. \ref{fig:pc-rho_stream}).   {\em Lower Left:}  Time-averaged velocity profiles.   The green vertical dotted line is the analytic prediction of the sonic point from eq. \ref{eq:son_geff0.5}.   {\em Lower Right:}  Numerator and Denominator of the steady state wind equation (eq. \ref{sonic}).  Because the flow is time dependent it does not strictly satisfy the steady state wind equation.  Nonetheless the analytic prediction of the sonic point for $\geff=0.5$ (vertical dotted line) is a reasonable estimate of the radius at which $v \sim c_c \sim V_g$ and $N_{st} \sim D_{st} \sim 0$.}
\label{fig:steady_stream}
\end{figure*}

To understand why $p_c(\rho)$ becomes shallower than $p_c \propto \rho^{2/3}$, it is helpful to remember the origin of the latter:  for a steady solution with $v_A \gg v$, the CR energy equation becomes $r^{-2} d (r^2 F_c)/dr = v_A dp_c/dr$ with $F_c = 4 p_c v_A$, which implies $p_c \propto \rho^{2/3}$ for a split-monopole.      Physically, the change in CR power $\sim r^2 F_c$ is due to energy transfer from the CRs to the gas at a rate  $v_A dp_c/dr$.     In the time dependent simulations, however, over most of the volume $dp_c/dr=0$ and the CRs and gas are not well-coupled and do not exchange energy with each other.  With $dp_c/dr =0$, $r^2 F_c \sim$ const so that $F_c \sim p_c v_A$ implies $p_c \propto \rho^{1/2}$.  In more detail, the exchange of energy between CRs and the gas, i.e., $d p_c/dr \ne 0$,  occurs only in the vicinity of the shocks which are regions of higher than average density and thus lower than average $v_A$ (see Fig. \ref{fig:vel_flux_stream} and \ref{fig:shocks}).   This inhomogeneity causes $\langle v_A dp_c/dr \rangle < \langle v_A \rangle \langle dp_c/dr \rangle$, where $\langle \rangle$ denotes a time average, as we show explicitly in the upper right panel of Figure \ref{fig:steady_stream} (green line).   As a result, the time-averaged energy equation for the CRs is better modeled as $r^{-2} d (r^2 F_c)/dr = f v_A dp_c/dr$ with $f \sim 1/3$ or so.  In the limit $f \ll 1$ and $F_c \sim p_c v_A$, the time-averaged CR energy equation again reduces simply to $r^2 F_c \sim$ constant, i.e., $p_c \propto \rho^{1/2}$.    This is roughly consistent with the time-averaged $p_c(\rho)$ in Figure \ref{fig:pc-rho_stream} and the time-averaged CR luminosity shown in Figure \ref{fig:steady_stream} (solid line in the upper right panel), which does not change  much with radius away from the base of the wind.

Figure \ref{fig:steady_stream} shows additional properties of our time-averaged galactic wind solutions with CR streaming.   In the upper left panel, we show that the density profiles near the base of the wind, where the solution is roughly hydrostatic, are  well approximated by the analytic theory developed in \S \ref{sec:strmod}, in which the CRs are modeled as a fluid with an effective adiabatic index $\gamma_{\rm eff} < 2/3$, consistent with the modified CR energetics summarized in Figure \ref{fig:pc-rho_stream}. The change from $\geff = 2/3$ to $\geff=1/2$ leads to significantly higher gas densities (factor of $\sim 10$) in the nearly hydrostatic portion of the wind.   As a result, the mass-loss rate in the wind is larger than standard wind theory would predict:   in \S \ref{sec:strmod} we show analytically that the seemingly modest of change of $\gamma_{\rm eff} = 2/3 \rightarrow 0.5$ in fact can change the mass-loss rate in CR-driven winds by a factor of a few-100.  This is driven primarily by the change in the gas density profile seen in Figure \ref{fig:steady_stream}.   

The lower left panel of Figure \ref{fig:steady_stream} shows the time-averaged velocity profiles of the wind for $v_{A,0} = 1, 10$ (in this plot, and those that follow, we define the time-averaged velocity as $\langle \rho v \rangle/\langle \rho \rangle$).  The gas accelerates to a speed somewhat larger than both the CR sound speed and the circular velocity of the potential. The lower right panel of Figure \ref{fig:steady_stream} shows the numerator and denominator of the steady state wind equation (eq. \ref{wind_streaming}).  Both pass through 0 as expected for a steady wind.   However, the numerator and denominator do not vanish at the same radius, as they would for a true steady state wind (e.g., Fig. 3 of Paper I).   The reason is that the wind equation does not formally hold for the time average of a time dependent solution.\footnote{For example, interpreting eq. \ref{wind_streaming} in terms of time-averaged variables requires assuming $\langle \rho v \rangle = \langle \rho \rangle \langle v \rangle$, which is not in general true for inhomogeneous flows like those found here.}  The vertical dotted  line in the lower panels of Figure \ref{fig:steady_stream} is the analytic prediction of the sonic point for our modified solutions from \S \ref{sec:strmod}, $r_s \simeq e r_0 \simeq 2.7 r_0$ (see eq. \ref{eq:sonic_mod}); the standard CR-wind theory in \S \ref{section:streaming} predicts a larger sonic point radius of $r_s=e^{3/2}r_0 \simeq 4.5 r_0$ (eq.~\ref{sonic}). Despite the significant time dependence of the solutions, the analytic estimate does a reasonable job of capturing the location where the flow reaches $v \sim c_c \sim V_g$, even though the solution does not rigorously satisfy the time-steady critical point conditions.

Finally, the upper right panel of Figure \ref{fig:steady_stream} quantifies the energy fluxes in the steady state solution.    For the steady wind problem with CR streaming and isothermal gas there is no conserved total energy flux because streaming losses from the CRs to the gas are assumed to be instantaneously radiated away.   Nor is there formally a separately conserved gas or CR energy flux.  As noted in our discussion of the origin of $p_c \propto \rho^{1/2}$, however, CR streaming losses are less in our inhomogeneous wind solutions than in standard  wind theory, so that $\propto r^2 F_c$ is only a weak function of radius, particularly away from the base of the wind.    Figure \ref{fig:steady_stream}  also shows that the steady state CR energy flux is a factor of few lower than the canonical value of $4 (v+v_A) p_c$, consistent with the individual snapshot in Figure \ref{fig:vel_flux_stream}.  Finally, the gas kinetic energy flux is a factor of $\sim 10$ less than the CR energy flux, even at large radii where the gas is supersonic.   

\subsection{Solutions For Different $V_g$}

\label{sec:Vg}

\begin{figure*}
\centering
\includegraphics[width=175mm]{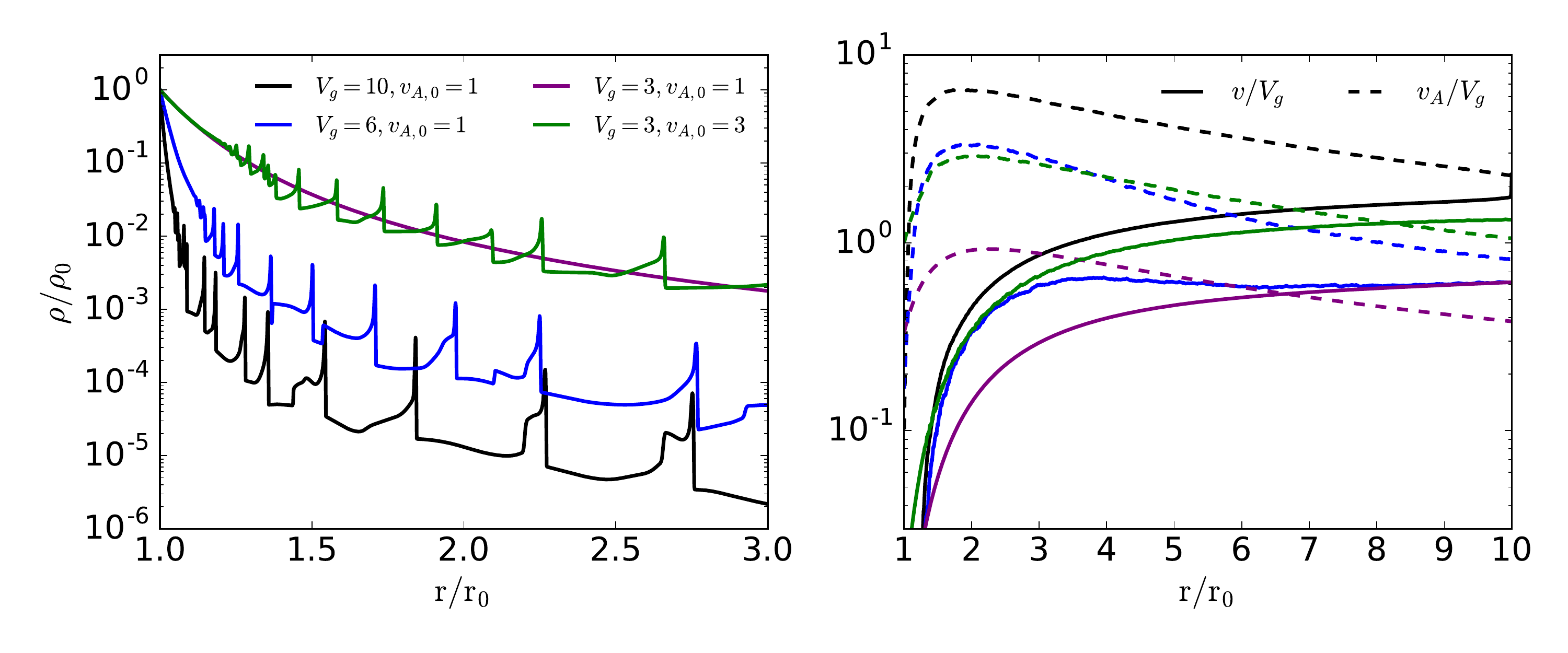}
\vspace{-0.2cm}
\caption{{\em Left:} Instantaneous density profiles for different $V_g$, with two different $v_{A,0}$ for $V_g=3$.  The instability and shocks that significantly modify the wind dynamics are present in all cases except for $V_g=3,v_{A,0}=1$ (see \S \ref{sec:Vg} for a physical explanation).  {\em Right:}. Time-averaged radial velocity and Alfv\'en velocity for the same four simulations.  The kinematics of the four different simulations are similar scaled by $V_g$ as we have done here.}
\label{fig:Vg}
\end{figure*}

In this section we present wind solutions for different values of $V_g/c_i = 3, 6, 10$.  Decreasing $V_g/c_i$ corresponds to a shallower potential for fixed gas thermodynamics or hotter gas for a given escape speed.   The latter, i.e., varying $c_i$, can physically be thought of as describing cosmic-rays coupling primarily to different phases of the ISM (e.g., warm ionized medium for lower $c_i$ vs. the hot ISM for higher $c_i$).   For these three values of $V_g/c_i$, the left panel of Figure \ref{fig:Vg} shows instantaneous density profiles (at randomly chosen times in the statistical steady state) while the right panel shows time-averaged velocity and Alfv\'en speed profiles.   We normalize the velocity profile using $v/V_g$ rather than $v/c_i$ because this highlights the similarity of the kinematics across a range of $V_g$.

\begin{figure}
\centering
\includegraphics[width=87mm]{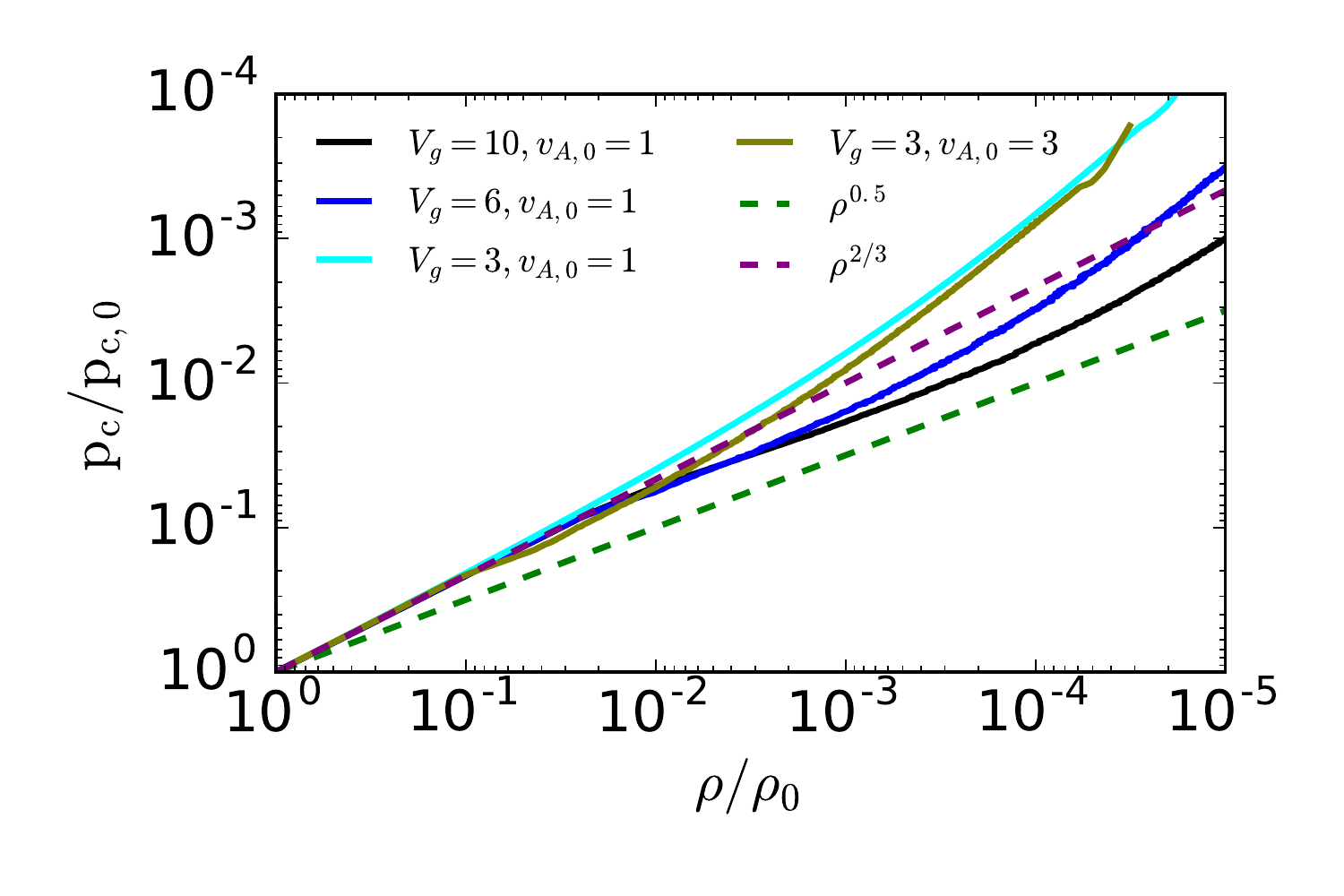}
\vspace{-0.2cm}
\caption{Time-averaged solutions for different $V_g$ in the $p_c-\rho$ plane.  Solutions with $V_g=6$ show the transition to $p_c \propto \rho^{1/2}$ just as the $V_g=10$ solutions do in Figure \ref{fig:pc-rho_stream}.   By contrast, solutions with $V_g = 3$ are more consistent with $p_c \propto \rho^{2/3}$ (until at lower densities $v > v_A$ and $p_c \propto \rho^{4/3}$).   This is due to the smaller density fluctuations introduced by shocks when the gas temperature is effectively larger (smaller $V_g/c_i$); see Fig. \ref{fig:Vg}.}
\label{fig:pc-rhoVg}
\end{figure}

The left panel of Figure \ref{fig:Vg} shows that for our fiducial $v_{A,0}=c_i$ magnetic field, the strong shocks are present for $V_g= 6$ but absent for $V_g = 3$. The reason for the latter is that the sound wave instability identified in Appendix \ref{sec:appendixA} operates most effectively when $v_A \gg v$; in the opposite regime of $v < v_A$ there is no instability because the gas is nearly adiabatic.    Smaller values of $V_g$ correspond to significantly larger density scale-heights (left panel of Fig. \ref{fig:Vg}) and thus smaller values of $v_A$. This  inhibits the acoustic instability that is driven by rapid streaming.    Additional evidence for this interpretation is that a simulation with $V_g = 3$ and a larger base Alfv\'en speed, $v_{A,0}=3$ (also shown in Figure \ref{fig:Vg}), does show the instability and shocks with  similar properties to the higher $V_g$ simulations.

A second striking result of Figure \ref{fig:Vg} is that simulations with different values of $V_g$ have similar kinematics, with terminal velocities that are within a factor of $\lesssim 2$ of $V_g$.   We compare this to our analytic predictions in \S \ref{sec:synthesis}.   

Figure \ref{fig:pc-rhoVg} shows the time-averaged solutions from Figure \ref{fig:Vg}  in the $p_c-\rho$ plane (analogous to Figure \ref{fig:pc-rho_stream}).   The $V_g = 6$ simulation is similar to the $V_g = 10$ simulation, with $p_c \propto \rho^{1/2}$ a better approximation to the effective equation of state at intermediate densities $10^{-2}-10^{-3.5} \rho_0$ where the shocks are present and $v_A/v$ is large.   This is not really the case for the $V_g = 3$ simulations, including the $v_{A,0}=3$ simulation in which strong shocks are present just as in the $V_g = 6$ and 10 simulations.  Our interpretation is that this is because for higher $c_i/V_g$, the density jump at the shocks is smaller, as is evident in the left panel of Figure \ref{fig:Vg}.   The smaller density contrast between the shocks and the rest of the solution minimizes the inhomogeneous nature of the flow, and thus the differences relative to canonical streaming solutions.   A smaller value of $v_A/v$ also increases the effective adiabatic index of the gas, with $\geff \rightarrow 4/3$ for $v \ll v_A$ (this indeed applies at the lowest densities in Fig. \ref{fig:pc-rhoVg}).  However, the $V_g = 6, v_{A,0}=1$ and $V_g = 3, v_{A,0}=3$ solutions in Figure \ref{fig:Vg} have very similar values of $v_A/v$ as a function of radius, and yet different $\geff$ in Figure \ref{fig:pc-rhoVg}. This points to the magnitude of the density jump at the shocks as the primary reason that the standard $\geff=2/3$ solution is a better approximation for $V_g = 3$ (and both $v_{A,0}$) than it is for $V_g=6$ and higher.  

\subsection{Solutions For Different $v_{A,0}$}
\label{sec:vA}

\begin{figure}
\centering
\includegraphics[width=87mm]{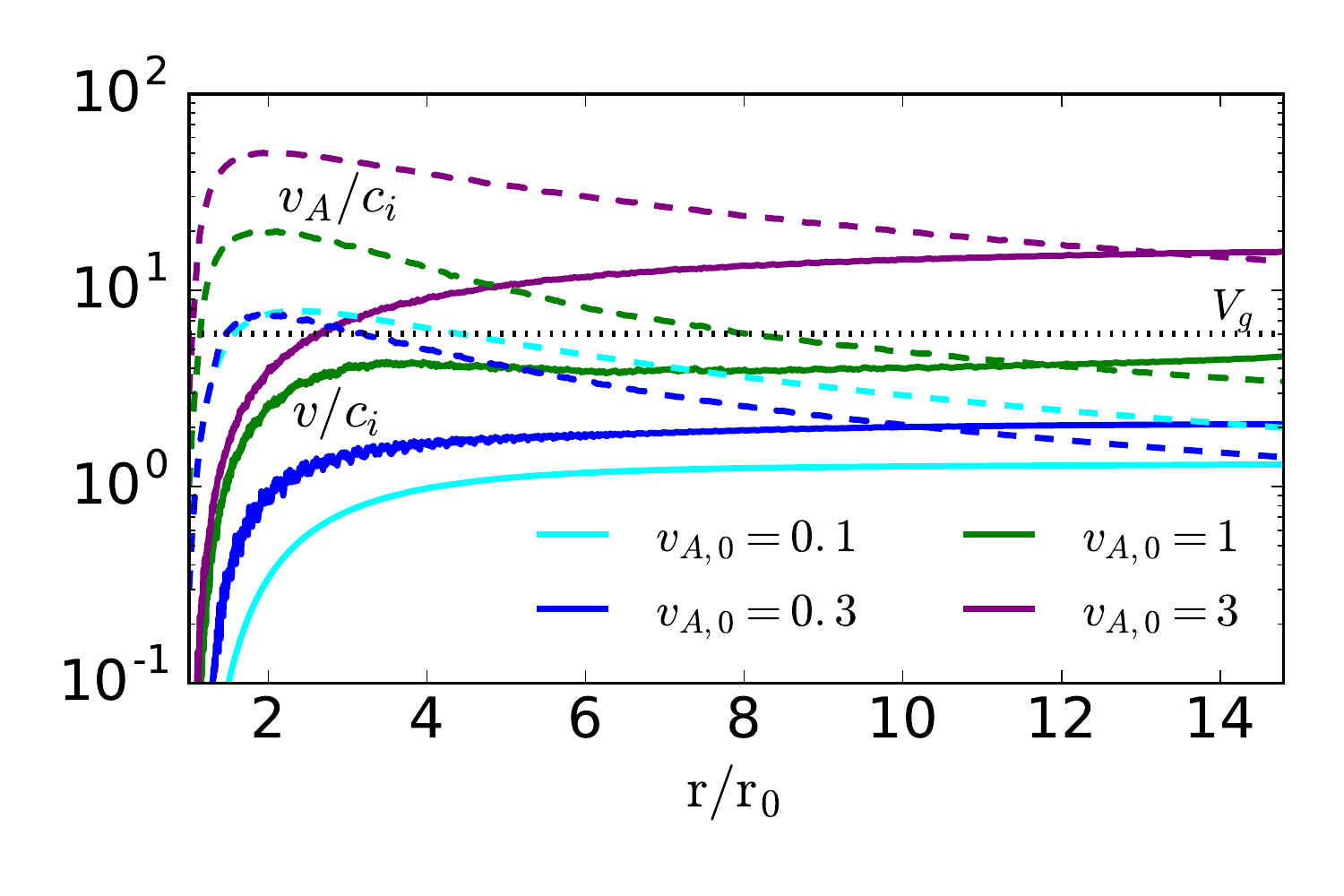}
\vspace{-0.2cm}
\caption{Time-averaged velocity (solid) and Alfv\'en velocity (dotted) profiles for $V_g =6$, $p_{c,0}=\rho_0 c_i^2$ and four different values of the base Alfv\'en velocity.   The solutions with lower base Alfv\'en speed, $v_{A,0}=0.1,0.3$, never reach $v \sim V_g$;  these are outflows but `breezes' rather than transonic winds.}
\label{fig:vA}
\end{figure}

Figure \ref{fig:vA} shows time-averaged radial velocity and Alfv\'en velocity profiles for different values of $v_{A,0} = 0.1, 0.3, 1$, and 3, all for $V_g = 6$ and $p_{c,0} = \rho_0 c_i^2$.   The most striking result in Figure \ref{fig:vA} is that the two low base Alfv\'en velocity solutions ($v_{A,0}=0.1, 0.3$) never reach $v \sim V_g$, i.e., they never reach the sonic point.  These solutions are formally `breezes' rather than transonic winds.   Note also that since $v_A \lesssim v$ at large radii for these solutions, the effective adiabatic index of the gas $\geff$ is increasing towards the adiabatic value of $4/3$ (eq. \ref{ceffstr}).  Once $\geff \gtrsim 1$, such that the CR sound speed $\propto \rho^{\geff-1}$ decreases with decreasing density, the flow can no long accelerate by tapping into the increasing CR sound speed.  Thus the low $v_{A,0}$ solutions in Figure \ref{fig:vA} will not be able to accelerate to $v \sim V_g$ at yet larger radii.    By contrast, the $v_{A,0} = 3$ solution in Figure \ref{fig:vA} is a robust transonic wind and $v_{A,0}=1$ solution is on the border with $v \sim V_g$.

The transition between transonic winds and breezes in Figure \ref{fig:vA} at $v_{A,0} \sim 1$ is reasonably well captured by our analytic estimate of $v_{A,crit}$, the base Alfv\'en velocity required to maintain $v_A > v$ out to the sonic point (\S \ref{sec:highvA}).   For the parameters of Figure \ref{fig:vA}, equation \ref{eq:vAlarge} yields $v_{A,crit} \simeq 0.5$ while equation \ref{eq:vAlargemod} (derived in the next section for the modified CR energetics found in our simulations) yields $v_{A,crit} \simeq 0.6$.  This supports our conjecture in \S \ref{sec:highvA} that $v_{A,0} \gtrsim v_{A,crit}$ is an approximate condition for a supersonic wind.

It is also instructive to compare the mass-loss rate for the solutions with varying $v_{A,0}$ to the maximum mass-loss rate allowed by energy conservation (\S \ref{sec:maxmdot}).  For our simulations with a $\ln(r)$ potential, we can define the escape speed as the speed at $r_0$ needed to just reach the outer radius of the box, i.e., $v_{esc} = 2 \sqrt{\ln(r_{out})} V_g$. Given the energy flux in the wind at $r_{out}$ in Table \ref{tab:compare}, we find that the simulations in Figure \ref{fig:vA} with $v_{A,0}=0.1, 0.3, 1, \& \, 3$ have $\dot M_w/\dot M_{max} = 0.78, 0.74, 0.51, \& \, 0.33$, respectively.   The $V_g = 6$ simulation with $v_{A,0}=10$ has $\dot M_w/\dot M_{max}=0.15$.  These results confirm the conjecture in \S \ref{sec:highvA} \& \ref{sec:maxmdot} that solutions with smaller values of $v_{A,0}$ (relative to $v_{A,crit}$; eq. \ref{eq:vAlarge}) have mass-loss rates approaching the maximum possible given the CR energy available in the wind.  $\dot M_w \simeq \dot M_{\rm max}$ is also consistent with the low speeds $\lesssim V_g$ of the lower $v_{A,0}$ models at all radii in Figure \ref{fig:vA}.

\subsection{Solutions with Streaming and Diffusion}
\label{sec:str_diff}
\begin{figure*}
\centering
\includegraphics[width=175mm]{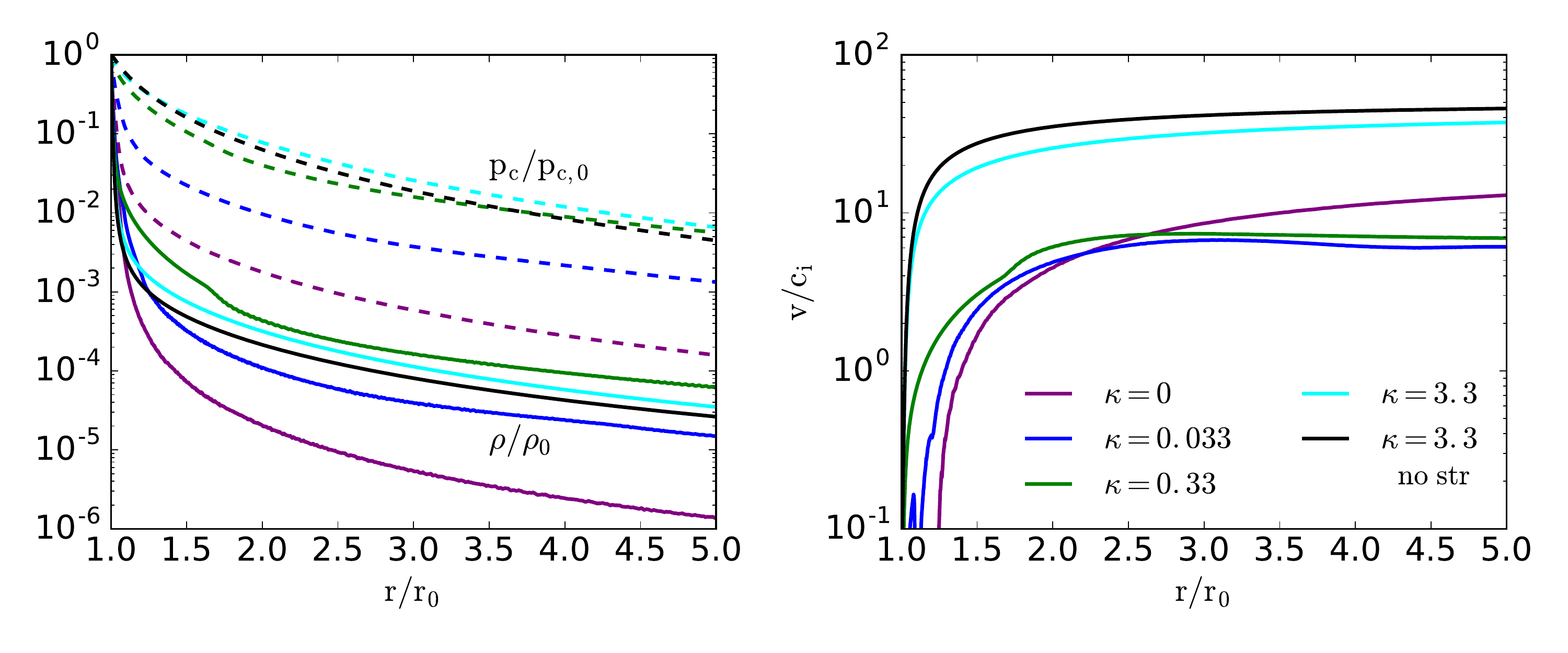}
\vspace{-0.2cm}
\caption{Time-averaged properties of galactic winds compared for CR streaming alone  ($v_{A,0}=1$),  diffusion alone ($\kappa = 3.3$) and both streaming and diffusion for several diffusion coefficients; all simulations are for $V_g=10$.   With increasing $\kappa$, the solutions with streaming and diffusion approach the pure diffusion solutions; even small diffusion coefficients $\kappa \sim 0.033-0.33$ (which on their own do not drive fast winds), significantly modify the streaming solution.  The kinks in the time-averaged density and velocity profile for $\kappa=3.3$ at $r \sim 1.7$ are associated with the onset of shocks at this radius (see Fig. \ref{fig:rho-both}).   }
\label{fig:rho-p-both}
\end{figure*}

\begin{figure}
\centering
\includegraphics[width=86mm]{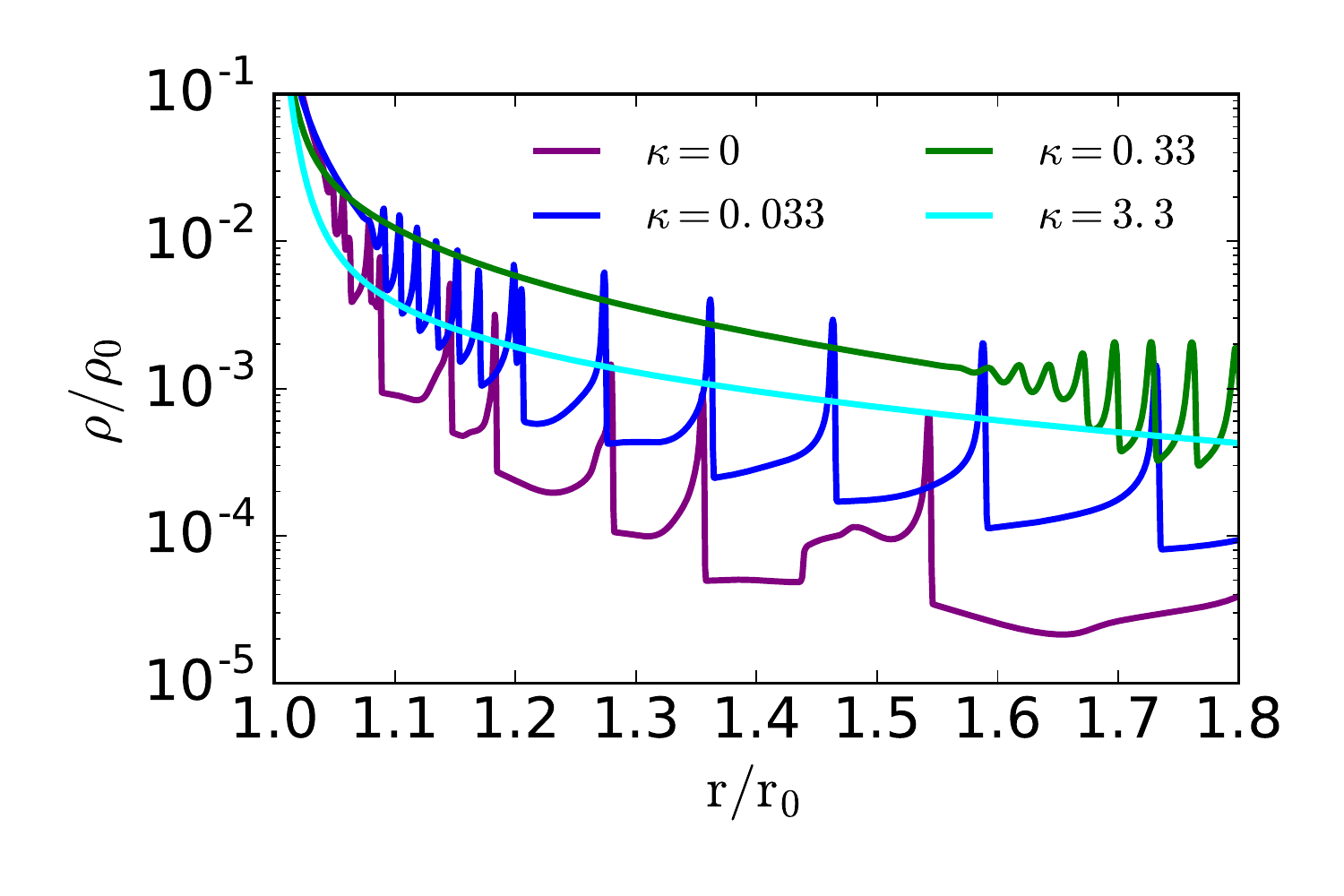}
\vspace{-0.2cm}
\caption{Instantaneous density profiles for solutions with streaming ($v_{A,0}=1$; $\kappa=0$) and diffusion; all simulations take $V_g=10$.  For increasing diffusion coefficient $\kappa$, the instability giving rise to shocks is suppressed and shows up at progressively larger radii.  For $\kappa = 3.3$, the solution is stable throughout.}
\label{fig:rho-both}
\end{figure}

CR transport is unlikely to be well-modeled as either pure streaming or pure diffusion:  even in models of self-confinement of CRs by the streaming instability, there is in general a diffusive-like correction to streaming whose magnitude is larger when damping processes counteract the streaming instability and suppress the amplitude of the resulting Alfv\'enic fluctuations (e.g., \citealt{Wiener2013}).   To understand the implications of this property of CR transport for galactic winds, we briefly consider the properties of winds with both streaming and diffusion. We find that even a modest diffusion coefficient significantly modifies the properties of streaming-only solutions.

Figure \ref{fig:rho-p-both} compares the steady state density, pressure, and velocity profiles of winds with pure streaming (a fiducial $v_{A,0}=1$ simulation), pure diffusion ($\kappa=3.3 r_0 c_i$) and streaming plus diffusion for a range of diffusion coefficients (all in units of $r_0 c_i$).   The mass loss rates for these simulations are given in Table \ref{tab:compare}.  Even a  small diffusion coefficient of $\kappa \sim 0.033-0.33$, which absent streaming does not produce a high speed wind (see Paper I), changes the mass-loss rate of the wind by a factor of $\sim 10-20$ relative to streaming alone.  Indeed, by $\kappa = 1/3$, the mass-loss rate in the simulation with streaming and diffusion is within a factor of a few of the pure diffusion solutions.   Likewise Figure \ref{fig:rho-p-both} shows that the density and CR pressure profiles in the simulations with both streaming and diffusion are within a factor of a few of the pure diffusion result by $\kappa \sim 1/3$.   The velocity profiles of the simulations with both transport processes are somewhat less sensitive to $\kappa$ and only accelerate as rapidly as the pure diffusion solution once $\kappa \sim 3$ (Fig. \ref{fig:rho-p-both}).

The strong sensitivity of the galactic wind solutions driven by CR streaming to a modest diffusion coefficient can be understood as follows.  We can estimate the magnitude of the diffusive CR flux in a calculation with streaming alone.   The CR pressure gradient is then set by the density scale height which is $H \simeq r_0 (c_i/V_g)^2$ for our spherical model.   Thus \be
\frac{F_{\rm diff}}{F_{\rm str}} \simeq \frac{\kappa}{r_0 c_i} \frac{V_g^2}{v_{A,0} c_i}
\simeq 100 \frac{\kappa}{r_0 c_i} \left(\frac{V_g}{100 \kms}\right)^2 \left(\frac{(10 \kms)^2}{v_{A,0}c_i}\right)
\label{eq:Fratio}
\ee
 As a result if $\kappa/(r_0 c_i) \gtrsim v_{A,0} c_i/V_g^2$ the diffusive CR flux will be larger than the streaming flux near the base of the wind. For the simulation in Figure \ref{fig:rho-p-both} this condition becomes $\kappa/(r_0 c_i) \gtrsim 0.01$, which is satisfied by all of the simulations in Figure \ref{fig:rho-p-both}.  Physically, solutions with streaming alone have smaller density and CR pressure scale-heights than diffusion solutions, so that only a small diffusion coefficient is needed for diffusive CR transport to be energetically important.  The latter tends to establish the much flatter CR pressure profiles seen in Figure \ref{fig:rho-p-both}, which in turn accelerates the gas more efficiently and increases the mass-loss rate (see Paper I).

Figure \ref{fig:rho-both} highlights an additional feature of the simulations with both CR streaming and diffusion: even modest diffusion coefficients suppress the linear instability and shocks described in \S \ref{sec:streamnum}.   In particular, with increasing  diffusion coefficient, the strong shocks in the outflow only show up at larger radii and by $\kappa = 3.3$ the solution is laminar and steady state with no shocks.  

\section{Analytic Approximations with  Modified Cosmic Ray Energetics}
\label{sec:strmod}

The  results in \S \ref{sec:streamnum} show that the standard assumptions of models of galactic winds driven by streaming CRs, in particular equations \ref{eq:ceff} \& \ref{simple_crenergy}, do not in fact apply to the time-averaged properties of many of our numerical wind solutions.  
A  self-consistent generalization of CR wind theory with streaming that accounts for this would require developing in detail a new closure model for the time-averaged CR energy equation.  Appendix \ref{sec:AppB} briefly discusses some of the subtleties of doing so.  Here we restrict ourselves to the simpler task of using an effective equation of state in which $p_c \propto \rho^\geff$ with $\geff < 2/3$.   This generalizes the $v_A \gg v$ analytics of \S \ref{section:streaming} to account for the different CR energetics found in the simulations (Fig. \ref{fig:pc-rho_stream}).  We first consider general $\geff$ and then specialize to $\geff = 1/2$.
\subsection{General $\geff$}
\label{sec:geffgen}
If we model the CRs as a fluid with $p_c \propto \rho^\geff$, the effective CR sound speed is 
\beq
c_{\rm eff}^2 = \geff \frac{p_c}{\rho} 
\label{eq:ceffgen}
\eeq
and the  steady state sonic point equations become
\be
\frac{1}{v}\frac{dv}{dr} = \frac{N_{\rm eff}}{D_{\rm eff}}
\ee
with
\be
D_{\rm eff} = v^2 - c_i^2 - c_{\rm eff}^2 \ \ \ \ \ \ N_{\rm eff} = \frac{2}{r}\left(c_i^2 + c_{\rm eff}^2 - V_g^2\right)
\label{eq:sonic_mod}
\ee

We proceed as in \S \ref{section:streaming} by considering the nearly hydrostatic portion of the flow interior to the sonic point, for which 
\beq
c_i^2\frac{d\ln \rho}{dr} + c_{\rm eff,0}^2 \left(\frac{\rho_0}{\rho}\right)^{1-\geff}\frac{d \ln \rho}{dr} = - \frac{2 V_g^2}{r}.
\label{eq:CRHEeff}
\eeq
Equation \ref{eq:CRHEeff} can be solved numerically for $\rho(r)$ in the hydrostatic portion of the wind.  Figure \ref{fig:steady_stream} (top left panel) shows that these solutions reproduce the density profiles in our time dependent simulations much better than standard $\geff=2/3$ CR wind theory.

\begin{figure*}
\centering
\includegraphics[width=170mm]{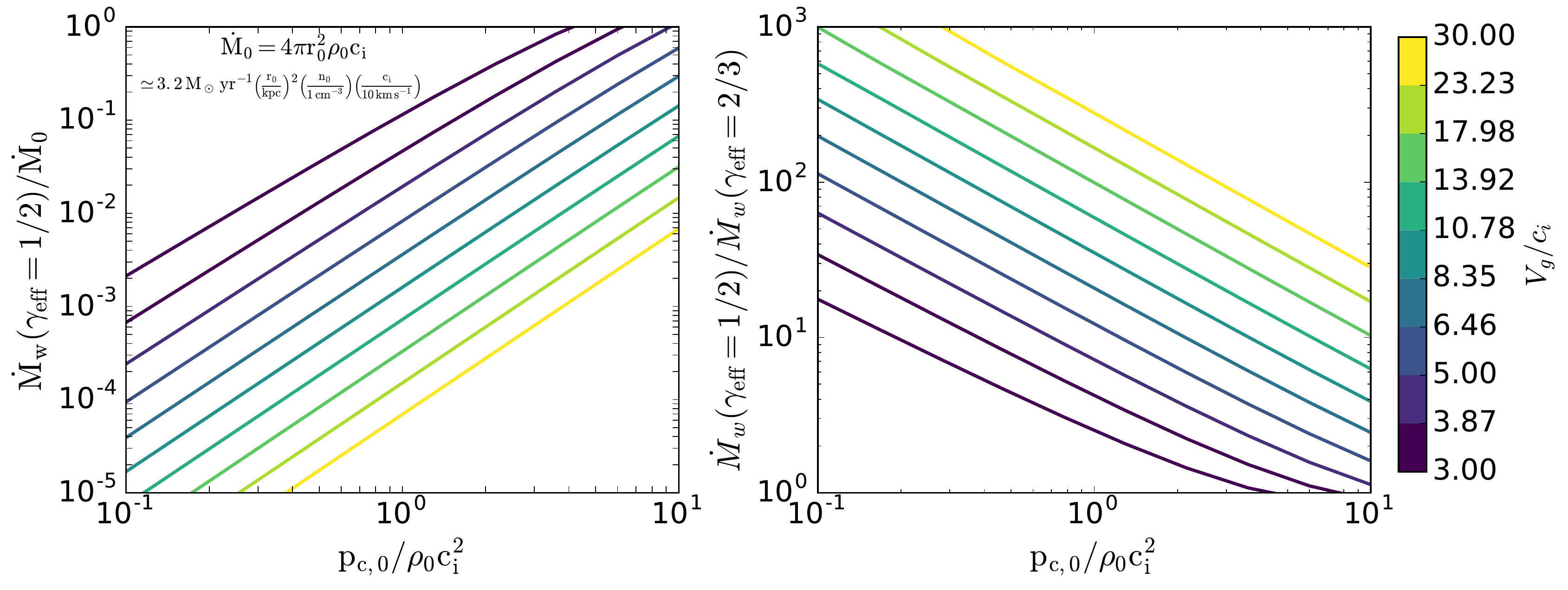}
\vspace{-0.2cm}
\caption{{\em Left:}. Analytic mass-loss rate for CR driven galactic winds in the limit of rapid streaming (large $v_A$) for $\geff=1/2$ (see eq. \ref{eq:Mdotdef} and associated text), as a function of the strength of gravity relative to the gas sound speed in the disk ($V_g/c_i$) and the base CR pressure (${\rm p_{c,0}/\rho_0 c_i^2}$).  The mass-loss rate is normalized by eq. \ref{eq:mdot0}.   {\em Right:}.  Ratio of the predicted mass-loss rate for $\geff=1/2$ to that for $\geff = 2/3$ (again using eq. \ref{eq:Mdotdef} with the appropriate $\geff$ in eqs. \ref{eq:CRHEeff} and \ref{eq:sonic_sol_mod}).  The modified CR thermodynamics due to strong shocks leading to $\geff=1/2$ can significantly increase the mass-loss rate, particularly for larger $V_g/c_i$ or smaller $p_{c,0}/\rho_0 c_i^2$.   For both panels, labeled values of $V_g/c_i$ on the color bar are logarithmically distributed and correspond to the curves on the plot.}  
\label{fig:mdotstrgam0.5}
\end{figure*}

The critical point conditions (eqs \ref{eq:sonic_mod}) determine the density and flow velocity at the sonic point, namely
\be
v(r_s) = V_g \ \ \ \ {\rm and} \ \ \ \  \rho(r_s) = \rho_0 \left(\frac{c_{\rm eff,0}^2}{\Veffsq}\right)^{\frac{1}{1-\geff}}
\label{eq:sonic_sol_mod}
\ee
The mass-loss rate in the wind can then be estimated as 
\be
\dot M_w = 4 \pi r_s^2 \rho(r_s) V_g
\label{eq:Mdotdef}
\ee
Note that $\rho(r_s)$ in equation \ref{eq:sonic_sol_mod} is analytic in terms of the base properties of the wind, but the location of the sonic point $r_s$ can in general only be determined numerically given the solution to equation \ref{eq:CRHEeff}.  Figure \ref{fig:mdotstrgam0.5} shows the resulting mass-loss rate (in units of $\dot M_0$; eq.~\ref{eq:mdot0}) for $\geff=1/2$ as a function of $p_{c,0}/\rho_0 c_i^2$ and $V_g/c_i$, as well as the ratio of the mass-loss rate predicted by $\geff=1/2$ to that predicted by $\geff=2/3$.   The larger CR pressure predicted by $\geff=1/2$ significantly increases the mass-loss rate, particularly for larger $V_g/c_{\rm eff,0}$.

As in \S \ref{section:streaming} we can analytically approximate the density profile in Figure \ref{fig:steady_stream} and the mass-loss rate in Figure \ref{fig:mdotstrgam0.5} if we focus on either of the limits $c_{\rm eff,0} \gg c_i$ or $c_{\rm eff,0} \ll c_i$. Consider first the case of massive galaxies with $V_g \gg c_{\rm eff,0} \gtrsim c_i$.   In this limit, the solution of equation \ref{eq:CRHEeff} is
\be
\label{eq:rhoapprox}
\rho(r) \simeq \rho_0 \left(1 + 2(1-\geff)\frac{V_g^2}{c_{\rm eff,0}^2} \ln[r/r_0]\right)^{-\frac{1}{1-\geff}},
\ee
the sonic point is located at
\be
r_s \simeq r_0 e^{\frac{1}{2(1-\geff)}}
\label{eq:rsapprox}
\ee
and the mass loss rate is 
\be
\dot M_w \simeq 4 \pi \rho_0 r_0^2 V_g e^{\frac{1}{1-\geff}} \left(\frac{c_{\rm eff,0}^2}{V_g^2}\right)^{\frac{1}{1-\geff}}.
\label{eq:mdotstrgeff}
\ee

In the opposite limit of weak CR pressure compared to gas pressure at the base in the galactic disk, i.e., $c_{\rm eff,0} \ll c_i$, the gas density profile is initially set by gas pressure, and is given by equation \ref{eq:weakCR}.   As the density drops, the CR pressure increases in importance relative to the gas pressure.   As in \S \ref{section:streaming}, the sonic point condition (eq. \ref{eq:sonic_mod}) requires that the pressure be CR dominated at the sonic point (assuming $c_i < V_g$). The transition between gas pressure and CR pressure support happens at a radius 
\be
r_{tr} \simeq r_0 \left(\frac{\rho_0 c_i^2}{p_{c,0}}\right)^\frac{c_i^2}{2 V_g^2 (1-\geff)}.
\label{eq:rtrmod}
\ee 
The density profile exterior to this transition radius is then like equation \ref{eq:rhoapprox} but with a different boundary condition set by continuity at $r_{tr}$.   This yields
\be
\frac{\rho}{\rho_0}\simeq \left(\frac{p_{c,0}}{\rho_0 c_i^2}\right)^{\frac{1}{1-\geff}}\left[1+ \frac{2(1-\geff)}{\geff}\frac{V_g^2}{c_i^2}\,\ln\left(\frac{r}{r_{tr}}\right)\right]^{\frac{-1}{1-\geff}} \ \ (r > r_{tr})
\label{eq:rhostrweakmod}
\ee
The sonic point is then located at $r_s \simeq e^{0.5/(1-\geff)} r_{tr}$ and the mass-loss rate is
\be
\dot{M}_w \simeq 4\pi r_0^2\,\rho_0\,V_g \, e^{\frac{1}{1-\geff}} \,
\left(\frac{\rho_0 c_i^2}{p_{c,0}}\right)^\frac{c_i^2}{V_g^2(1-\geff)} \left(\frac{c_{\rm eff,0}^2}{V_g^2}\right)^{\frac{1}{1-\geff}}.
\label{mdotstrgamweak}
\ee

\subsection{Specialization to $\geff=1/2$, i.e., $p_c \propto \rho^{1/2}$}
\label{sec:geff0.5}

To obtain somewhat simpler expressions, we now focus on the case of $\geff=1/2$ motivated by the simulations in \S \ref{sec:streamnum}.  In this case, the expressions for the sonic point  
and mass-loss rate for massive galaxies with $V_g \gg c_{\rm eff,0} \gtrsim c_i$ become
\be
r_s \simeq e \, r_0
\label{eq:son_geff0.5}
\ee
and
\beq
\begin{split}
\dot{M}_w\simeq & \, 4\pi r_0^2\,\rho_0\,V_g \, e^2\,\left(\frac{\,c_{\rm eff,0}}{V_g}\right)^4 \simeq 0.023 \mspy \left(\frac{r_0}{1 \kpc}\right)^2 \\ & \times \, \left(\frac{n_0}{1 \, {\rm cm^{-3}}}\right)   \left(\frac{c_{\rm eff,0}}{10 \, \kms}\right)^4 \left(\frac{V_g}{100 \, \kms}\right)^{-3}.
\label{eq:mdotstrmod}
\end{split}
\eeq
while for $c_{\rm eff,0} \lesssim c_i \ll V_g$ they become
\be
r_s \simeq e \, r_0 \left(\frac{\rho_0 c_i^2}{p_{c,0}}\right)^\frac{c_i^2}{V_g^2}.
\label{eq:son_geff0.5lowpc}
\ee
and
\beq
\dot{M}_w\simeq \, 4\pi r_0^2\,\rho_0\,V_g \, e^2\,\left(\frac{\,c_{\rm eff,0}}{V_g}\right)^4 \left(\frac{\rho_0 c_i^2}{p_{c,0}}\right)^\frac{2 c_i^2}{V_g^2}
\label{eq:mdotstrmodlowpc}
\eeq
Note that equation \ref{eq:mdotstrmodlowpc} for high base CR sound speed differs from equation \ref{eq:mdotstrmod} for low base CR sound speed only by a factor of $\left(\rho_0 c_i^2/p_{c,0}\right)^{2 c_i^2/V_g^2} > 1$ (this is similar to the standard theory in \S \ref{section:streaming}; see the discussion around eq. \ref{mdotstrgam0.6weak}).  As a result, the two can be approximately combined to yield
\beq
\dot{M}_w\simeq \, 4\pi r_0^2\,\rho_0\,V_g \, e^2\,\left(\frac{\,c_{\rm eff,0}}{V_g}\right)^4 {\rm Max}\left[1,  \left(\frac{\rho_0 c_i^2}{p_{c,0}}\right)^\frac{2 c_i^2}{V_g^2}\right]
\label{eq:mdotstrmodgen}
\eeq

Perhaps most importantly, the mass-loss rate in equation \ref{eq:mdotstrmod} is a factor of $\sim (V_g/c_{\rm eff,0})^2 \gg 1$ {\it larger} than the standard CR streaming mass-loss rate (eq. \ref{mdotstrgam0.6}).   This highlights how the seemingly modest modification to the CR energetics introduced by the strong shocks in our numerical simulations ($\geff < 2/3$) in fact significantly modifies the properties of the resulting winds.  Indeed, if we scale for parameters appropriate to the `average' ISM of the Milky Way with $V_g\simeq150$\,km s$^{-1}$ and $c_{\rm eff,0}\simeq10$\,km s$^{-1}$, the CR driven mass loss rate estimated in equation \ref{eq:mdotstrgeff} is a factor of $\sim 100$ larger than the standard theory.   

The expression for the mass-loss rate in equation \ref{eq:mdotstrmod} can be rewritten as $\dot M_w \propto p_{c,0} \, c_{\rm eff,0}^2/V_g^4$.   This shows that, to the (uncertain) extent that the CR pressure is roughly similar in different phases of the ISM, the mass-loss rate will be dominated by the hotter ISM phases, which have larger values of $c_{\rm eff,0}$.   As an example, if we take $p_{c,0} \simeq 10^{-12} \, {\rm dyne \, cm^{-2}}$, $V_g \simeq 150 \kms$, and $r_0 \sim 3$ kpc as representative of the MW, equation \ref{eq:mdotstrmod} implies $\dot M_w \simeq 0.06$, 0.5, and  1.5\,$\mspy$ for $c_{\rm eff,0} \simeq 10$, 30, and 50\,$\kms$, respectively.   The latter two estimates of the mass-loss rate, representing CR-driven winds from the warm-hot ISM, are dynamically important in the sense of being of order the star formation rate, while the direct mass loss from the colder ISM is negligible.  The same qualitative conclusion was true for standard CR streaming wind theory in \S \ref{section:streaming}.

It is also instructive to generalize other properties of CR-driven wind theory to $\geff=1/2$.   In particular, the condition for the validity of the high Alfv\'en speed limit becomes 
\be
v_{A,0} \gtrsim v_{A,crit} \equiv {c_{\rm eff,0}} \, \frac{V_g \, c_{\rm eff,0}}{V^2_{\rm g, eff}}\left(\frac{r_s}{r_0}\right)^2 \simeq c_{\rm eff,0} \, e^2 \frac{c_{\rm eff,0}}{V_g}
\label{eq:vAlargemod} 
\ee
where the second equality is for massive galaxies with $V_g \gg c_{\rm eff,0}$.   As before, the high $v_A$ approximation is most applicable for massive galaxies with large $V_g/c_{\rm eff, 0}$ (physically, larger $V_g/c_{\rm eff,0}$ corresponds to lower mass-loss rates and thus lower densities and larger $v_A$ for a given base magnetic field strength).

The energetics of the wind can be understood by noting that so long as $p_c \propto \rho^{1/2}$, the CR pressure gradient in equation \ref{gasenstr} can again be rewritten as a CR enthalpy, leading to a conserved Bernoulli-like constant 
\be
\frac{d}{dr}\left(\frac{1}{2} v^2 + c_i^2 \ln \rho + 2 V_g^2 \ln r - 2 c_{\rm eff,0}^2  \left[\frac{\rho}{\rho_0}\right]^{-1/2}\right)  = 0.
\label{gasenstrmod}
\ee

The wind terminal velocity, power, and momentum loss-rate can then be estimated using arguments that parallel those given in \S \ref{section:streaming}.    The generalization of equations \ref{eq:vinfstr}, \ref{eq:vinfstr2},  \ref{eq:Edotstr}, and \ref{eq:pdotw} to $\geff=1/2$ are
\be
v_\infty^2 \simeq V_g^2 + 4 \Veffsq\left(\left[\frac{\rho(r_A)}{\rho(r_s)}\right]^{-1/2} - 1 \right),
\label{eq:vinfstrmod}
\ee

\beq
\begin{split}
    v_\infty \simeq &  \ \frac{2 V_g}{e} \frac{\sqrt{v_{A,0} V_g}}{c_{\rm eff,0}}   \simeq 230 \kms  \, \times \\ &  \, \left(\frac{V_g}{100 \kms}\right)^{3/2} \left(\frac{v_{A,0}}{10 \kms}\right)^{1/2} \left(\frac{c_{\rm eff,0}}{10 \kms}\right)^{-1}
    \label{eq:vinfstrmod2}
    \end{split}
\eeq
\be
\frac{0.5 \dot M_w v_\infty^2}{\dot E_c(r_0)} \sim 0.25,
\label{eq:Edotstrmod}
\ee
and
\be
\begin{split}
\frac{\dot p_w}{\dot p_{ph}} & \simeq \frac{\dot M_w}{\dot M_*} \frac{v_\infty}{\epsilon_{ph} c} \\ & \simeq  2.4 \, \epsilon_{ph,\,-3.3}^{-1} \, \left(\frac{c_{\rm eff,0}}{10 \, \kms}\right) \left(\frac{p_*/m_*}{3000 \, \kms}\right) \\ & \times \left(\frac{V_g}{100 \, \kms}\right)^{-3/2} \left(\frac{v_{A,0}}{10 \, \kms}\right)^{1/2}  \frac{p_{c,0}}{\pi G \Sigma_g^2}
\end{split}
\label{eq:pdotwmod}
\ee
Finally, the maximum mass-loss rate in CR-driven winds set by energy conservation (\S \ref{sec:maxmdot}) is
\be
\frac{\dot M_{\rm max}}{\dot M_*} \simeq \, 2.2 \, \zeta_{0.25} \, \epsilon_{c,-6.3} \, \left(\frac{100 \, \kms}{v_{\rm esc}(r_0)}\right)^2,
\label{eq:mdotmaxvsmdotstarmod}
\ee
where we have assumed that a fraction $0.25 \, \zeta_{0.25}$ of the base CR power is available to drive gas to larger radii. This normalization of $\zeta$ is reasonably consistent with eq. \ref{eq:Edotstrmod} and Figure \ref{fig:strfin} (bottom panel) discussed in the next section.

Equation \ref{eq:pdotwmod} shows that the momentum loss-rate in CR-driven winds with $\geff \sim 1/2$ is substantially larger than the standard theory (eq. \ref{eq:pdotw}), with values that are comparable to or exceed the momentum carried by the stellar radiation field.   Equation \ref{eq:Edotstrmod} is particularly striking compared to the standard theoretical result in equation \ref{eq:Edotstr} and highlights how the modified CR energetics found in our simulations implies a much larger fraction of the CR energy being transferred to kinetic energy of the gas at large radii.   Equation \ref{eq:Edotstrmod} can be understood physically by noting that $\geff=1/2$ is motivated in the first place by the suppression of streaming losses due to the effects of the strong shocks in the simulations (e.g., Fig \ref{fig:steady_stream}).    Absent streaming losses $r^2 p_c v_A \sim {\rm const}$ so that a significant fraction of the CR energy at the base of the wind is available to convert into wind energy at large radii.

\section{Discussion}
\label{sec:disc}

In this section we first synthesize the results of our streaming analytics and simulations (\S \ref{sec:synthesis}) and then compare the properties of CR-driven winds with streaming and diffusive CR transport (\S \ref{sec:strvsdiff}).   \S \ref{sec:iso} discusses the validity of the isothermal gas approximation used in this work.   We conclude this section by summarizing some of the observational implications of our results (\S \ref{sec:obs}).

\subsection{Synthesis of Streaming Results} 
\label{sec:synthesis}

\begin{figure}
\centering
\includegraphics[width=84mm]{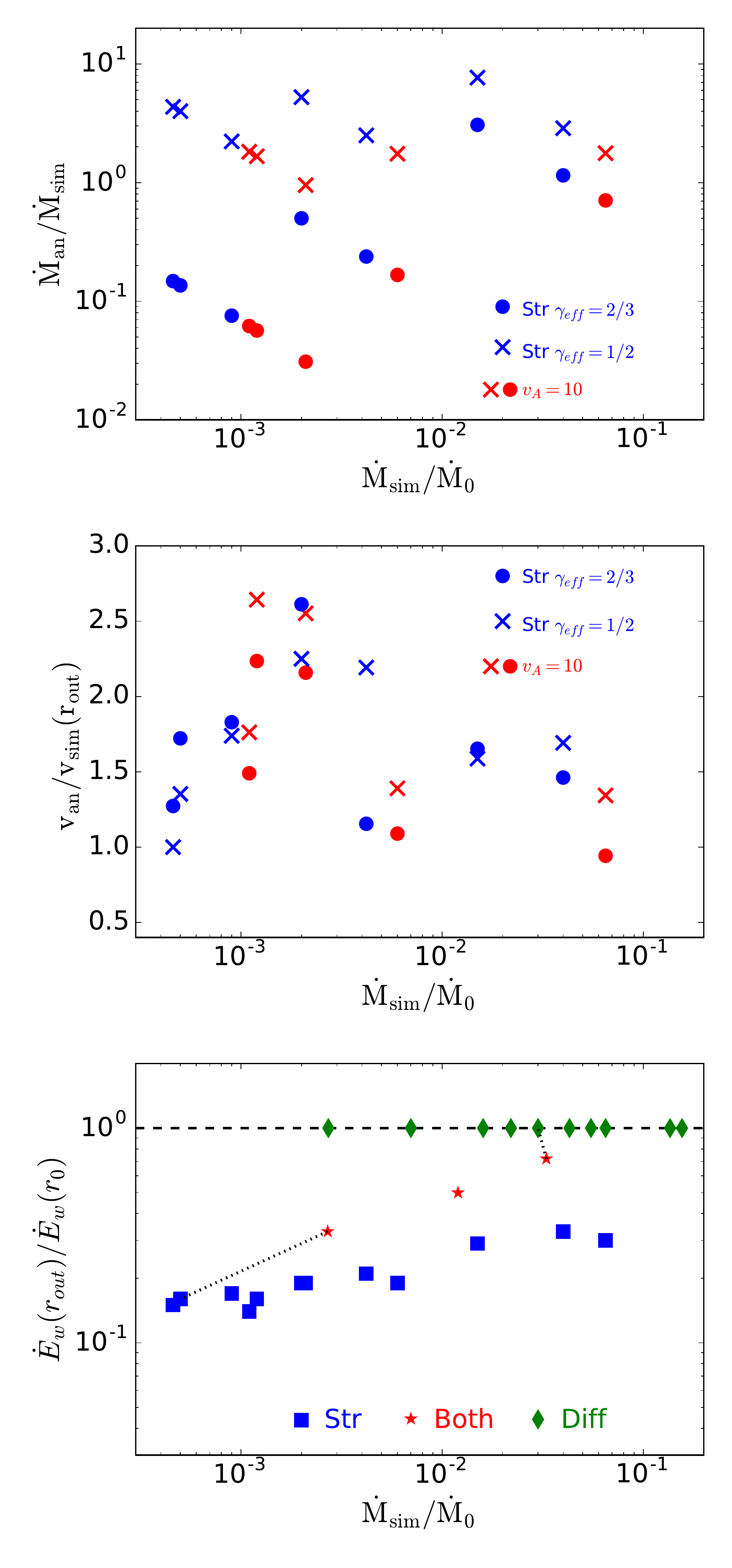} 
\caption{{\em Top:}  Ratio of the analytically predicted-mass loss rate in the limit of rapid streaming, for both the canonical $\geff=2/3$ and our modified $\geff=1/2$, to the simulated mass-loss rate for all of our streaming simulations; $\geff = 1/2$ (eq. \ref{eq:mdotstrmod}) is the better predictor of the simulated mass-loss rates.  {\em Middle:} Ratio of the analytically predicted terminal velocity in the limit of rapid streaming, for both the canonical $\geff=2/3$ and our modified $\geff=1/2$, to the simulated velocity at the edge of the box for all of our streaming simulations.  The analytics somewhat overestimate the terminal speed, largely because the acceleration in streaming solutions is slow (Fig. \ref{fig:steady_stream}) and the flow has often not reached the terminal velocity on the domain.   {\em Bottom:}  Energy flux at the top of the domain relative to that input in CR at the base, for streaming simulations and the diffusion simulations from Paper I.  CR transport via diffusion is energy-conserving while with streaming, the terminal energy flux is only $\sim 20 \%$ of the input energy flux because of transfer of energy to the gas and our assumed isothermal equation of state.  Dotted lines connect simulations with streaming and diffusion to their pure streaming or pure diffusion counterparts.}
\label{fig:strfin}
\end{figure}

Figure \ref{fig:strfin} brings together our various numerical and analytic solutions for the properties of CR-driven galactic winds with streaming at the Alfv\'en speed.  In the top panel we compare our analytic estimates of the mass-loss rate for streaming CRs with $\geff=0.5$ and $2/3$ (in particular, eqs. \ref{eq:mdotstrmod} \& \ref{eq:mdotstr2}, respectively) to the numerical simulations.  Perhaps most strikingly, the high $v_A$ simulations (red circles and $\times$) agree very well with the $\geff=1/2$ analytics over a factor of $\sim 100$ in mass-loss rate while the standard CR wind theory is significantly in error at low mass-loss rates (which are the higher $V_g/c_i$ simulations in the lower left of the plot).  For lower $v_A$, the modified $\geff=1/2$ analytics overestimates $\dot M_w$ by a factor of few; this is not so surprising since our high $v_A$ analytics is then somewhat less applicable.   Overall, we find that equation \ref{eq:mdotstrmod} with a numerical pre-factor of $f \sim 0.3 f_{0.3}$ captures the numerical mass-loss rates reasonably well.

The middle panel of Figure \ref{fig:strfin} compares the analytic estimate of the wind terminal velocity  to the speed in the simulations at the top of the box.  There is reasonably good agreement, though the analytic speeds tend to be somewhat larger than the simulations.   This is primarily because some of the energy remains in the CRs in the simulations given the finite outer radius of the computational domain.  This is more true of the streaming simulations in this paper than the diffusion simulations in Paper I because the acceleration of the flow is significantly slower with CR streaming than diffusion (see \S \ref{sec:strvsdiff} below).    Finally,  the bottom panel of Figure \ref{fig:strfin} shows the energy flux at large radii in our simulated winds relative to the energy flux in CRs at the base.    For the case of CR diffusion from Paper I, energy is globally conserved so that the energy flux at large radii is essentially identical to that supplied by the CRs at the base of the wind.   For streaming on the other hand, energy is lost because of the $v_A dp_c/dr$ work done by the CRs on the (isothermal) gas:   the net energy flux at large radii is $\sim 0.1-0.3$ of that supplied by the CRs at the base of the wind, roughly consistent with equation \ref{eq:Edotstrmod}.

\subsection{Comparison of Winds Driven by CRs in the Streaming and Diffusive Regimes}
\label{sec:strvsdiff}

\begin{figure}
\centering
\includegraphics[width=84mm]{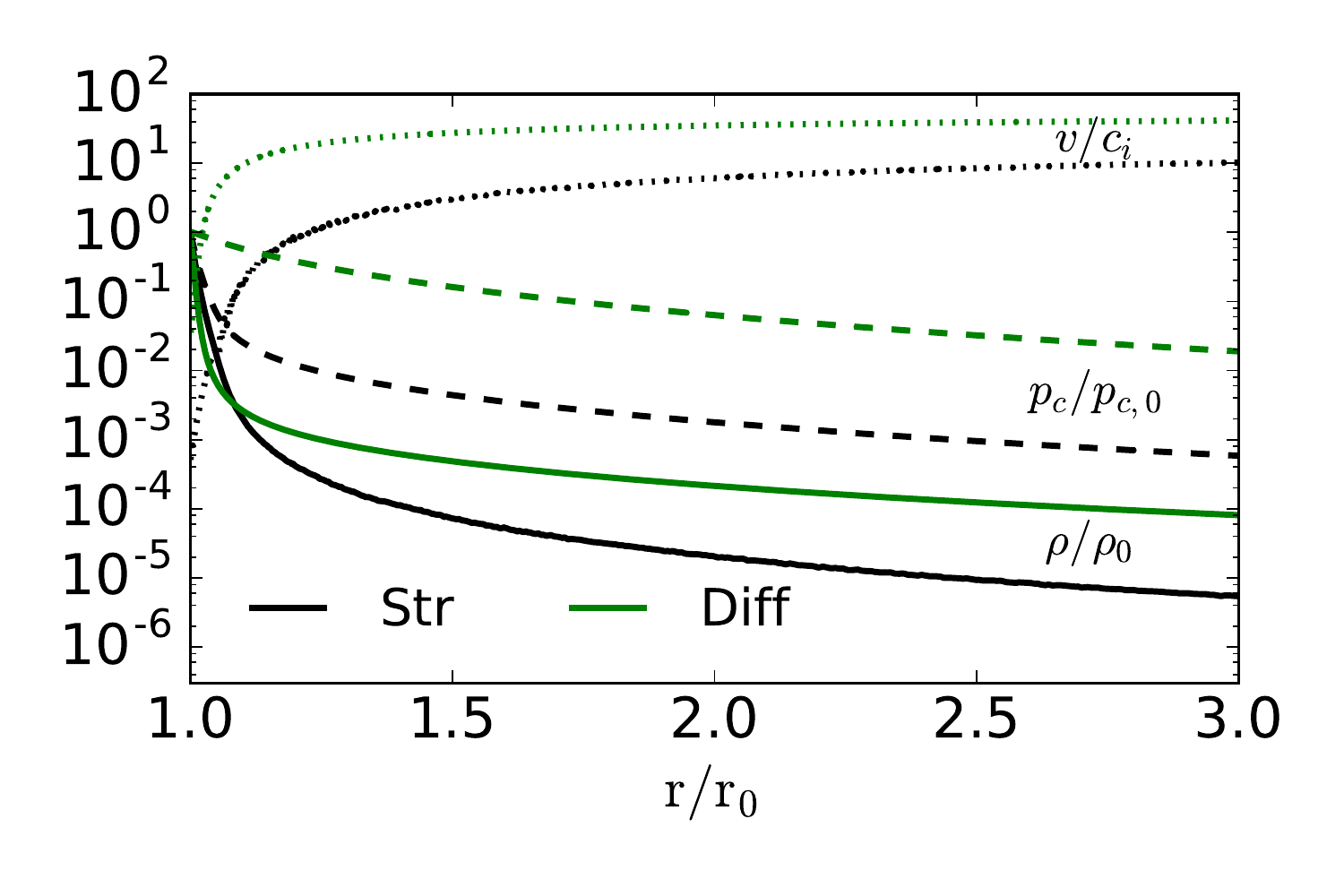} 
\includegraphics[width=84mm]{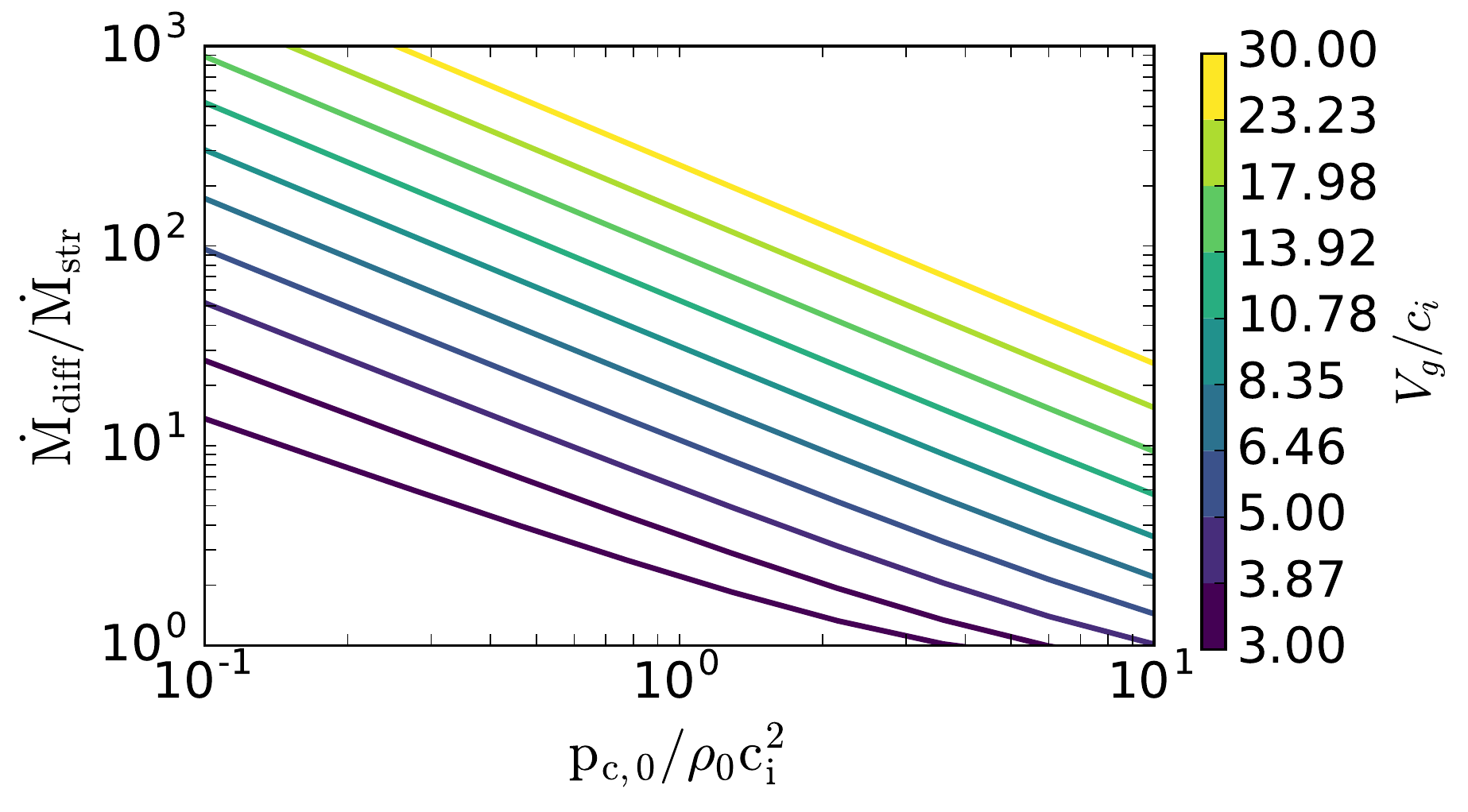} 
\caption{{\em Top:}  Comparison of the time-averaged velocity, CR pressure, and gas density profiles for streaming ($v_{A,0}=c_i$) and diffusion ($\kappa = 3.3 r_0 c_i$) solutions for CR-driven galactic winds with $V_g=10 c_i$ and $p_{c,0}=\rho_0 c_i^2$.   With CR diffusion, the CR pressure declines significantly more slowly near the base of the wind, increasing the mass-loss rate and causing the flow to accelerate much more rapidly than with streaming transport.  {\em Bottom:}    Mass-loss rate predicted by CR diffusion relative to CR streaming as a function of the base CR pressure $p_{c,0}$ and the depth of the gravitational potential $V_g$; mass-loss rates are analytic estimates validated by our simulations (see \S \ref{sec:strvsdiff} for details). For most parameters, diffusive transport predicts a mass-loss rate significantly larger than streaming transport. This is a consequence of the larger CR pressure (and pressure gradient) in the subsonic portion of the wind at small radii (top panel).}
\label{fig:StrvsDiff}
\end{figure}

In this section we briefly compare some of the properties of galactic winds driven by streaming CRs (this paper) with those produced by diffusing CRs (Paper I).   The top panel of Figure \ref{fig:StrvsDiff} shows the density, velocity, and CR pressure profiles for our $V_g=10 c_i$, $p_{c,0}=\rho_0 c_i^2$ simulations with streaming transport ($v_{A,0}=c_i$) and diffusive transport ($\kappa=3.3 r_0 c_i$).  The latter was discussed in detail in Paper I and is included here in Table \ref{tab:compare} as well.   Note that the mass-loss rate is a factor of $\simeq 60$ times larger for the diffusion solution than for the streaming solution.

The key difference between the streaming and diffusion solutions is the CR pressure profile.  With diffusion, the CR scale-height is large and the solution approaches $p_c \propto 1/r$  in the limit of rapid diffusion.   By contrast, with CR streaming and $v_A \gg v$, $p_c \propto \rho^{\geff}$ with $\geff = 2/3$ for canonical CR-wind theory (eq. \ref{eq:ceffstr}) and $\geff \sim 0.5$ for our modified theory accounting for strong shocks due to instabilities in the flow (Fig. \ref{fig:pc-rho_stream}).   In either of the latter two regimes, the CR pressure scale-height is thus of order the gas scale-height.  In general, this implies a smaller CR scale-height for CR streaming than for CR diffusion.  The larger CR pressure (gradient) in the presence of CR diffusion leads to the other key features of the solution shown in Figure \ref{fig:StrvsDiff}:  winds with CR diffusion in general have much larger mass-loss rates, larger terminal speeds, and retain a larger fraction of the CR energy supplied at the base of the wind (a factor of $\simeq 60$, 2.5, and 6 larger, respectively, for the comparison in Figure \ref{fig:StrvsDiff}).

To quantify the properties of winds with CR diffusion vs. streaming over a wide parameter regime, we turn to our analytic estimates validated by simulations.  Figure 6 of Paper I shows that for $\kappa \gtrsim r_0 c_i$, winds driven by CR diffusion have mass-loss rates
\be
\dot M_{\rm diff} \simeq \frac{2 \pi r_0^2 \, p_{c,0} c_i}{h_c \Veffsq}
\label{eq:Mdotdiff}
\ee
where 
\be
h_c \sim \min\bigg(1,\frac{c_i}{V_g}\sqrt{\frac{\kappa}{r_0 c_i}}\bigg)
\label{eq:hc}
\ee
is the base CR scale-height in units of the base radius of the flow.  

The bottom panel of Figure \ref{fig:StrvsDiff} compares the diffusion and streaming estimates of the mass-loss rate for a wide range of base CR pressures $p_{c,0}$ and strength of gravity relative to the isothermal gas sound speed $V_g/c_i$.  For streaming we estimate the mass-loss rate by numerically solving for the location of the sonic point as described below equation \ref{eq:sonic_mod}.  We multiply the resulting $\dot M_{\rm stream}$ by a factor of $0.3$ motivated by Fig. \ref{fig:strfin} (top panel).  The key result of Figure \ref{fig:StrvsDiff} (bottom panel) is that mass-loss rates due to diffusing CRs are a factor of a few-1000 larger than those produced by streaming CRs for the same base conditions.  It is worth stressing that this is true even though we have taken into account that our simulations (and the modified analytics in \S \ref{sec:strmod}) have larger mass-loss rates driven by streaming CRs than canonical CR wind theory (i.e., the diffusion vs. streaming mass-loss rates would have been yet more different using the standard theory of \S \ref{section:streaming}).

A useful analytic expression for the relative mass-loss rates in the diffusive and streaming limits can be derived comparing equations \ref{eq:mdotstrmodgen} \& \ref{eq:Mdotdiff}, which yields
\be
\frac{\dot M_{\rm diff}}{\dot M_{\rm str}} \simeq \frac{0.5}{f_{0.3}} \frac{V_g}{c_{\rm eff,0}}\frac{{\rm Max}\left(\frac{c_i}{c_{\rm eff,0}}, \frac{V_g}{c_{\rm eff,0}}\sqrt{\frac{r_0 c_i}{\kappa}}\right)}{{\rm Max}\left[1,  \left(\frac{\rho_0 c_i^2}{p_{c,0}}\right)^{2 c_i^2/V_g^2}\right]}
\label{eq:mdotdiffvsstr}
\ee
where $0.3 f_{0.3}$ is the dimensionless number we multiply equation \ref{eq:mdotstrmodgen} by, motivated by Figure \ref{fig:strfin}.  
Equation \ref{eq:mdotdiffvsstr} reproduces the results of Figure \ref{fig:StrvsDiff} to better than a factor of 2, with differences of this magnitude only occurring for the smallest $V_g/c_i$.   

\subsection{Validity of the Isothermal Gas Approximation}
\label{sec:iso}

\begin{figure}
\centering
\includegraphics[width=84mm]{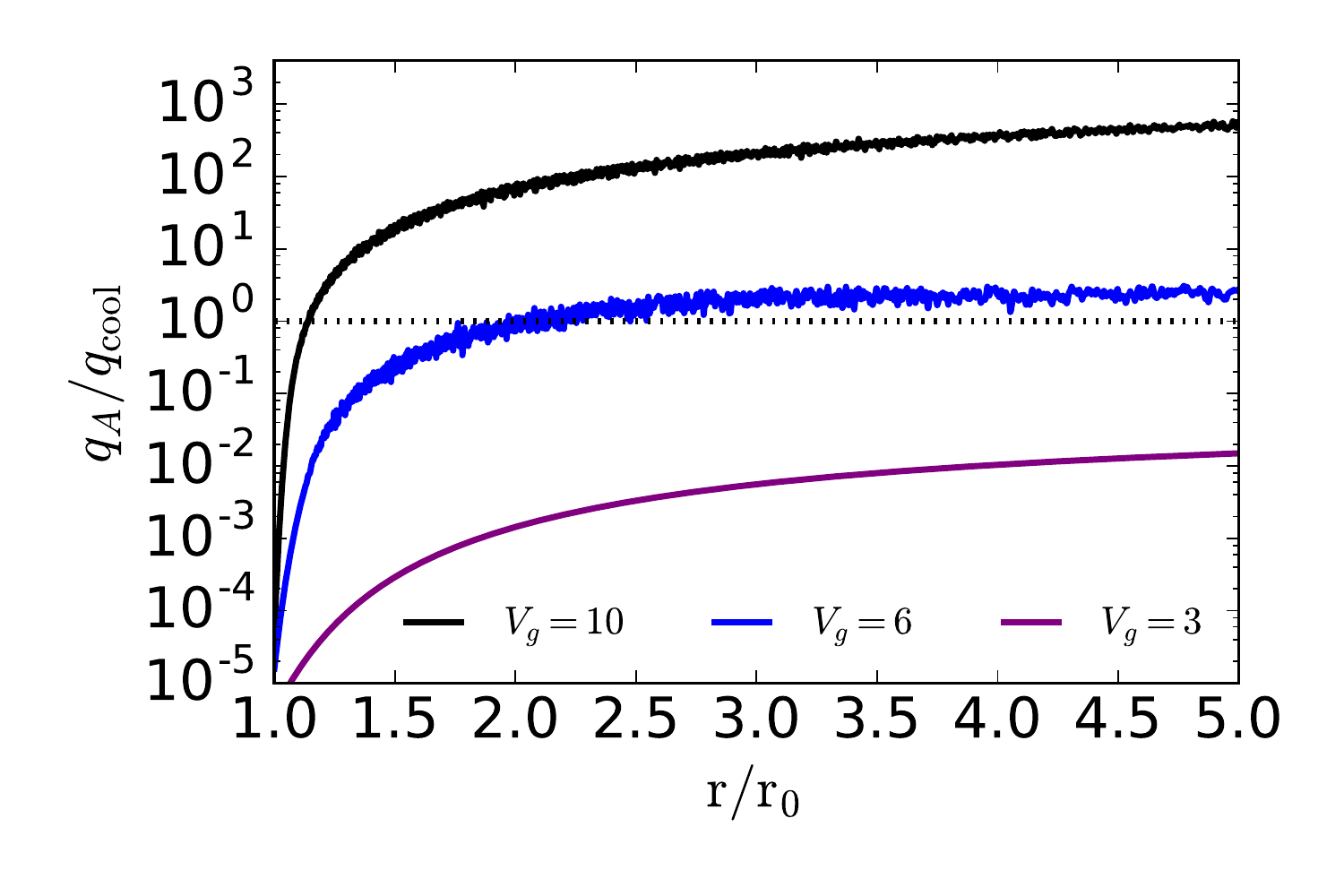} 
\caption{Ratio of the time-averaged gas heating rate due to CR streaming $q_A$ to the optically thin cooling rate $\qc$ for our solutions with $v_{A,0}=c_i$ and $p_{c,0}=\rho_0 c_i^2$. Cooling is calculated post facto and is not included in the simulations.   Self-consistency of our isothermal solutions requires rapid cooling, i.e., $\qc > q_A$.  We assign physical units to our simulations by assuming $v_{A,0}=10 \kms$, $p_{c,0} = 10^{-12}$ erg cm$^{-3}$, $r_0=1 \kpc$, $n_0=1$ cm$^{-3}$, and $\Lambda=10^{-21}$ erg s$^{-1}$ cm$^3$; for other parameter values the curves in this Figure can be scaled up or down using equation \ref{eq:qratio}.}
\label{fig:cool}
\end{figure}

In his original treatment of galactic winds driven by streaming CRs, \citet{Ipavich1975} separately studied solutions that included Alfv\'enic heating of the gas ($\sim v_A dp_c/dr$), but neglected radiative losses, and "zero-temperature" solutions in which the heating of the gas is assumed to be rapidly radiated away.   The subsequent literature has considered variants of one or both of these limits.  For example, \citet{Breitschwerdt1991} neglected Alfv\'enic heating of the gas but also evolved the gas energy equation without radiative cooling.\footnote{ \citet{Breitschwerdt1991} effectively assumed that energy is transferred from CRs to Alfv\'en waves by the streaming instability, but that the waves are not damped, and hence that there is no heating of the gas.   However, all physical damping mechanisms for Alfv\'en waves excited by the streaming instability (e.g., ion-neutral, turbulent damping, etc.; see \citealt{Squire2021} for a recent discussion) lead to the Alfv\'en waves rapidly heating the gas, so we view this model variant as relatively unphysical.}   \citet{Mao2018} assumed an isothermal equation of state, as we have in this paper (our analysis in  \S\ref{section:streaming} is most analogous to their work).

In this section we provide estimates of when the isothermal gas approximation is plausibly self-consistent in the sense that the cooling time of the gas in the wind is shorter than the heating timescale associated with energy transfer from the CRs to the gas, mediated by the Alfv\'en waves.    This is a necessary but not sufficient condition for the plausibility of the isothermal gas approximation -- necessary because the Alfv\'enic heating term is necessarily present in models of CR transport mediated by the streaming instability, but not sufficient because it focuses only on Alfv\'enic heating neglecting other heating mechanisms (e.g., shocks).

The optically thin cooling rate per unit volume given a cooling function $\Lambda(T)$ is $\qc = n^2 \Lambda(T)$ while the Alfv\'enic heating rate per unit volume is given by $q_{A} = v_A |dp_c/dr|$.  The ratio of the heating and cooling rates can thus be written as
\be
\frac{q_A}{\qc} = \left(\frac{v_{A,0} p_{c,0}}{n_0^2 r_0 \Lambda}\right) \left(\frac{\rho_0}{\rho}\right)^2 \frac{v_A}{v_{A,0}} \frac{r_0}{p_{c,0}}\bigg|\frac{dp_c}{dr}\bigg|
\label{eq:qratio}
\ee
where we have written quantities in terms of dimensionless variables such that the pre-factor in front is roughly the ratio of the heating time to the cooling time at the base of the wind. 

To evaluate equation \ref{eq:qratio} in our simulations, we need to assume physical values for the quantities in the first parentheses. We take $v_{A,0}=10 \kms$, $p_{c,0} = 10^{-12}$\,erg cm$^{-3}$, $r_0=1 \kpc$, $n_0=1$ cm$^{-3}$, and $\Lambda=10^{-21}$ erg s$^{-1}$ cm$^3$, in which case the first term in parentheses in equation \ref{eq:qratio} is $\simeq 3 \times 10^{-7}$.  For these parameters, Figure \ref{fig:cool} shows the ratio of the time-averaged heating and cooling rates $q_A/\qc$ for our $v_{A,0}=c_i$, $p_{c,0}=\rho_0 c_i^2$, and $V_g/c_i = 3$, 6,  $10$ simulations.  In evaluating $q_A$ and $\qc$, we explicitly calculate $\langle v_A dp_c/dr \rangle$ and $\langle \rho^2 \rangle$ to best capture the volume averaged heating and cooling rates in the inhomogeneous flow, though this only makes a factor of few difference for the results in Figure \ref{fig:cool}.   The curves in Figure \ref{fig:cool} can be trivially scaled to other choices of $c_i$, $r_0$, $\Lambda$, \& $n_0$ using equation \ref{eq:qratio}.   For example, a higher base density of $n_0 \simeq 10-1000$ cm$^{-3}$ and a somewhat smaller physical scale of the system, $r_0 \sim 0.1-0.3$ kpc, would be appropriate for starbursts, which would decrease $q_A/\qc$ in Figure \ref{fig:cool} by a factor of few-100s.

The sonic point in our analytic models is $r_s \simeq 2.7 r_0$ (eq.~\ref{eq:sonic_mod}) and indeed this is roughly the radius at which the flow reaches $v \sim V_g \sim v_\infty$ in the simulations (Fig.~\ref{fig:steady_stream}).  As a result, the isothermal approximation is  well-motivated if $q_A \lesssim \qc$ out to $r/r_0 \sim $ few, so that cooling can regulate the temperature and maintain a roughly constant value of $c_i$ at the radii within which the mass-loss rate and wind velocity are determined.\footnote{In our isothermal solutions, gas pressure is only important at much smaller radii than $r \sim r_s \sim $ few $r_0$, namely $r \lesssim r_{tr}$ (eq. \ref{eq:rtr}).   One might thus be tempted to evaluate the importance of cooling only at those radii.  This is not correct, however, because if $q_A/\qc \gtrsim 1$ at $r_{tr} \lesssim r \lesssim r_s$, the gas will heat up appreciably, significantly increasing the importance of gas pressure at radii interior to the sonic point in the isothermal solutions, thus fundamentally modifying the structure of the flow.}    A second comparison that is important is the ratio of the heating/cooling timescale and the flow time in the wind $\sim H/v$.  For our CR-driven wind solutions, this ratio is $t_{\rm adv}/t_{\rm heat} \simeq (v_A/v)(p_c/p_{gas})$.   Since $v \lesssim v_A$ and $p_c \gtrsim p_{gas}$ near the sonic point, $t_{\rm heat} \lesssim t_{\rm adv}$: the short heating timescale implies that it is indeed the comparison of the heating and cooling rates (captured by equation \ref{eq:qratio} and Fig. \ref{fig:cool}) that sets the thermodynamics of the gas.   

Figure \ref{fig:cool} shows that near the base of the wind $\qc \gg q_A$ for all of our models so that cooling is important at small radii.\footnote{The exception to this is if the gas is very hot to begin with, e.g., if $c_i \simeq 300 \kms$, $\Lambda \simeq 10^{-22.5}$ erg s$^{-1}$ cm$^3$, in which case the curves in Figure \ref{fig:cool} would all be scaled up by $\sim 10^6$.   This is essentially a thermally-driven wind in which CRs do not play a dominant role.} By $r \sim 2 r_0$, however the importance of cooling depends strongly on the depth of the gravitational potential $V_g$, with cooling dominant for $V_g = 3 c_i$, marginally important for $V_g = 6 c_i$, and negligible $V_g = 10 c_i$.   This implies that the $V_g = 10 c_i$ isothermal solutions are only plausibly self-consistent if the pre-factor in equation \ref{eq:qratio} is significantly smaller than assumed in Figure \ref{fig:cool}, e.g., if the gas density at the base is $n_0 \gg 1$ cm$^{-3}$, as might be the case in a starburst or at higher redshift.  For $V_g = 3 c_i$, the mass-loss rate is sufficiently larger that cooling of the gas is dynamically important at all radii and so the isothermal approximation is reasonable.  $V_g = 6 c_i$ lies in between the $V_g = 3$ \& $10 c_i$ cases, with comparable heating and cooling rates for $r \sim 2-5 r_0$.

The strong dependence of $q_A/\qc$ on $V_g$ in Figure \ref{fig:cool} reflects the strong dependence of the mass-loss rate on $V_g$ (e.g., Fig. \ref{fig:mdotstrgam0.5}).   These results can be understood analytically as follows.  We evaluate the importance of cooling at the analytic sonic point using our modified streaming models with general $\geff$.   The density at the sonic point is then given by equation \ref{eq:sonic_sol_mod} and the location of the sonic point by equation \ref{eq:son_geff0.5}.   
Assuming a split monopole magnetic field and taking $|d \ln p/d \ln r|_{r_s} \sim 1$, we find
\be
\label{eq:qratioan}
\begin{split}
& \frac{q_A}{\qc}\bigg|_{r_s}  \simeq   \frac{v_{A,0} p_{c,0}}{n_0^2 r_0 \Lambda}   \, e^{-3/(2-2 \geff)} \, \left(\frac{c_{\rm eff,0}}{\Veff}\right)^{-\frac{2(2.5-\geff)}{1-\geff}}
\\ & \simeq  \frac{5}{n_0 \Lambda_{-21}} \, \frac{v_{A,0} \, c_{\rm eff,0}^2}{(10 \kms)^3} \frac{\rm kpc}{r_0} \, \left(\frac{c_{\rm eff,0}}{0.1 \, \Veff}\right)^{-8} \ (\geff=1/2)
\end{split}
\ee
where $\Lambda = 10^{-21} \Lambda_{-21}$ erg cm$^3$ s$^{-1}$ and we have assumed $\geff = 1/2$ in the last equality. Recall that optically thin atomic cooling of solar metallicity gas has $\Lambda \sim 10^{-22}-10^{-21}$ erg s$^{-1}$ cm$^3$ for $T \sim 10^{4-6}$K, with slower cooling (lower $\Lambda$) at yet higher T.   For comparison, for $\geff = 2/3$ the result is similar to equation \ref{eq:qratioan} but $\propto (c_{\rm eff,0}/\Veff)^{-11}$.  

Equation \ref{eq:qratioan} describes well the numerical results in Figure \ref{fig:cool}, both the normalization of $q_A/\qc$ at $r \sim {\rm few} \, r_0$ and the strong dependence on $V_g$.  Overall, we conclude that the validity of the isothermal gas approximation is determined primarily by the ratio of the base CR sound speed in the wind to the galaxy escape speed, via the strong dependence of $q_A/\qc$ on $c_{\rm eff,0}/V_g$ in equation \ref{eq:qratioan}.   Indeed, the constraint $q_c \gtrsim q_A$ at $r \sim r_s$ requires 
\be
\frac{c_{\rm eff,0}}{\Veff} \gtrsim 0.13 \left(\frac{v_{A,0}}{10 \kms} \frac{\rm kpc}{r_0} n_0^{-1} \Lambda_{-21}^{-1} \right)^{1/6} \left(\frac{\Veff}{100 \kms}\right)^{1/3}.
\label{eq:cool_cond}
\ee
As noted in \S \ref{section:streaming} \& \ref{sec:strmod}, the mass-loss rate increases significantly with increasing $c_{\rm eff,0}$ such that warmer phases of the ISM, which likely have larger $c_{\rm eff,0}$, will probably dominate the mass-loss rate.   Equation \ref{eq:cool_cond} shows that precisely because of the larger mass-loss rate, these phases will also have stronger cooling and are more likely to be in the (roughly) isothermal limit studied in this paper.   That being said, many of the isothermal CR-streaming-driven solutions in the literature with lower values of $c_{\rm eff,0}$, including several in this paper, are not self-consistent because CR heating of the gas is likely to overwhelm radiative cooling, invalidating the isothermal gas approximation.

One important caveat to our analysis in this section is that we have compared the time-averaged CR heating and radiative cooling rates in the wind.   Figure \ref{fig:rho_stream} shows, however, that CR heating is negligible over most of the volume because $dp_c/dr \simeq 0$.  Cooling will still be important in those regions, driving the gas towards photoionization equilibrium.  Likewise, CR heating and radiative cooling are both enhanced in the vicinity of the strong shocks that permeate the flow (Fig. \ref{fig:shocks}).  This is not accounted for in our analysis of time-averaged heating and cooling rates in Figure \ref{fig:cool}.   Time-dependent simulations with both streaming heating and cooling are needed to fully understand when (and, indeed, if) the isothermal approximation used here and in previous work is appropriate.  

\subsection{Observational Implications}

\label{sec:obs}

In Paper I, we highlighted the importance of empirically constraining CR transport models in other galaxies given the theoretical uncertainties in the mechanism(s) of CR transport.  We argued that GeV gamma-ray observations by FERMI  \citep{Fermi2010,Fermi2010b,Fermi2012}, though modest in number, are particularly valuable because pion decay directly constrains the energetics and transport of the GeV protons that dominate the CR energy density (e.g., \citealt{Lacki2010,Lacki2011,Crocker2021a}).  We defer to a future paper detailed models of streaming-driven galactic winds calibrated to gamma-ray observations.  It is worth stressing, however, that the gamma-ray calibrated diffusion coefficients in Paper I of $\kappa \sim 10^{29}$ cm$^2$ s$^{-1}$ correspond to CR escape times of $H_c^2/\kappa \sim 30$ Myrs for $H_c \sim 3$ kpc.  For streaming transport to produce a similar escape timescale with $H_c \sim 3$ kpc, we require an Alfv\'en speed of $v_{A,0} \sim 100 \kms$.   This requires strong magnetic fields in the warm ionized or hot phases of the ISM.   

Our general expression for the mass-loss rate validated by the simulations (eq.~\ref{eq:mdotstrmod}) can be written as $\dot M_w \propto p_{c,0} \, c_{\rm eff,0}^2/V_g^4$, where $p_{c,0}$ is the base CR pressure in the wind.   This shows that the mass-loss rate will in general be dominated by the hotter ISM phases, which have larger values of $c_{\rm eff,0}$; see also Fig.~\ref{fig:mdotstrgam0.5}. Importantly, streaming CRs are thus ineffective at directly driving cold gas out of galaxies, though CR-driven winds in hotter phases of the ISM may entrain cooler gas along the way (e.g., \citealt{Wiener2019,Bruggen2020}).   The mass-loss rate in the warm-hot ISM phases can in principle approach the maximum mass-loss rate allowed by energy conservation, $\dot M_{\rm max}/\dot M_* \simeq \, 2 \,  \epsilon_{c,-6.3} \, (100 \, \kms/v_{\rm esc}(r_0))^2$ (eq.~\ref{eq:mdotmaxvsmdotstarmod}).  This maximum mass-loss rate is, however, itself already modest compared to the mass-loss rates typically required in cosmological models to reconcile the stellar mass and dark matter halo mass-functions (e.g., \citealt{Somerville2015}).   The reason is that isothermal winds with CR streaming can only supply at most $\sim 2.5$\% of the energy of SNe to galactic winds:   $\sim10$\% of the SNe energy goes into CRs but the CRs only retain $\sim 1/4$ of this energy in the wind when cooling is efficient (Fig.~\ref{fig:strfin}).  This suggests that streaming CRs are an energetically inefficient source of feedback for driving galactic winds.   Pionic losses, which are likely important in luminous starbursts given their gamma-ray luminosities (e.g., \citealt{Lacki2010}), would  strengthen this conclusion by further decreasing the CR energy available to drive galactic winds.

Our simulations show that the terminal speed in CR streaming-driven galactic winds is typically within a factor of few of the galaxy escape speed set by $V_g$ (see Table \ref{tab:compare}).   This property of our wind models is intriguingly similar to some observational inferences (e.g., \citealt{Martin2005,Weiner2009}) and to the class of `momentum-conserving' wind models \citep{Murray2005} that have been successful at reproducing the galaxy stellar mass function and the galaxy mass-metallicity relationship (e.g., \citealt{Oppenheimer2006, Oppenheimer2010}).   However, as noted in the previous paragraph, the somewhat low (maximum) mass-loss rates in our models are in tension with what is required in cosmological models of galaxy formation. The wind momentum fluxes we find are similar to those expected from supernova-heated hot gas wind models and those driven by radiation pressure (eq.~\ref{eq:pdotwmod}; \citealt{Lochhaas2020}).

One of the most important new results presented in this paper vis-a-vis observations of galactic winds is the presence of strong shocks that permeate the flow (Fig.~\ref{fig:rho_stream} \& Fig.~\ref{fig:Vg}).  These shocks significantly increase the mass-loss rate relative to previous steady state calculations (Fig.~\ref{fig:mdotstrgam0.5}).     It is likely that such shocks would also produce a wide range of observational signatures.  In particular, shocks are likely to give rise to multi-phase gas with hot post-shock gas cooling rapidly and producing X-ray emission, optical line emission, and even (re-)forming molecules and dust in the post shock region if there is ambient dense gas in the flow \citep{Hollenbach1989}.  Strong in-situ shocks could also accelerate electrons producing  synchrotron emission off the plane of the galactic disk.   This could be important for interpreting gradients in synchrotron emission and synchrotron spectral indices above galaxy disks and/or in galaxy outflows  (e.g., \citealt{Dumke1995,Buckman2020}), since models typically focus on the interplay between cooling and escape (e.g., \citealt{Seaquist1991}), rather than re-acceleration.  Future work extending our results to models with heating, cooling, and magnetic fields would be able to directly predict the observational signatures of the strong shocks found here, for comparison to observations.    

\section{Summary}
\label{sec:summary}

In this paper we have used analytic estimates and idealized time-dependent spherically symmetric simulations to study the properties of isothermal galactic winds driven by CRs streaming at the Alfv\'en velocity.  This complements Paper I \citep{Quataert2021} in which we carried out a similar analysis for diffusive CR transport. These two mechanisms of CR transport differ dramatically in their predictions for how the CR pressure decreases away from a galaxy (see the top panel of Fig. \ref{fig:StrvsDiff}):   in the limit of rapid CR streaming, $p_c \propto \rho^{2/3}$ according to standard CR-driven wind theory (eq. \ref{simple_crenergy}) and so the CR pressure scale-height is tied to that of the gas, while for rapid CR diffusion, $p_c \propto r^{-1}$, i.e., the CR pressure scale-height is of order the size of the system.   One aim of this paper and its companion is to understand how these differences in the CR pressure profiles translate into differences in the predicted properties of galactic winds.

To start, we analytically estimated galactic wind properties using standard theoretical assumptions for CR streaming (\S \ref{section:streaming}).  In particular, when the Alfv\'en velocity is large compared to the flow velocity, the standard treatment of CR streaming implies $p_c \propto \rho^{2/3}$ (\citealt{Ipavich1975}; see eq. \ref{simple_crenergy}).   This effective equation of state applies at small radii in the wind, before the flow has accelerated to large velocities $\gtrsim v_A$.  The assumption $p_c \propto \rho^{2/3}$ is sufficient to allow analytic estimates of the wind properties: the mass-loss rate $\dot M_w$ (eq.~\ref{eq:mdotstr2}),  terminal speed $v_\infty$ (eq.~\ref{eq:vinfstr2}),  asymptotic wind power $0.5 \dot M_w v_\infty^2$ (eq.~\ref{eq:Edotstr}), and wind momentum loss-rate $\dot p_w$ (eq.~\ref{eq:pdotw}).

To test these models, we carried out time-dependent simulations for the same idealized spherically-symmetric galactic wind problem that we studied analytically (isothermal gas and an isothermal gravitational potential).  The simulations use the two-moment numerical method for CR transport developed by \citet{JiangOh2018}.  The two-moment method is particularly important for this study because we require an accurate treatment of CR streaming. For our spherically-symmetric problem, the magnetic field is not evolved dynamically but is only present via the  Alfv\'en velocity, which sets the CR streaming speed.  

For most of the parameter space we surveyed (Table \ref{tab:compare}), the simulations yield solutions that are intrinsically time-dependent, invalidating previous steady-state wind calculations. Small amplitude waves grow near the base of the wind and then steepen into strong shocks that permeate the flow (Fig. \ref{fig:rho_stream}-\ref{fig:shocks}).  To interpret the instabilities found in the simulations, we identified two new linear instabilities driven by CR streaming (\S \ref{sec:lin} and Appendix \ref{sec:appendixA}).   The first is an acoustic instability at high Alfv\'en speed due to the finite background CR pressure gradient present in hydrostatic equilibrium; this instability, which dominates in most of our simulations, is a streaming analogue of the acoustic instability discussed by \citet{Drury1986}.  The second instability is a short-wavelength instability of the CR entropy mode due to the finite speed of light retained in the two-moment CR equations (eqs. \ref{eq:CR2mom}).  We note that the sound wave instability driven by CR streaming discovered by \citet{Begelman1994} is not present in our simulations because of our assumption of isothermal gas, but it may be present in winds where cooling is inefficient (e.g., \citealt{Huang2021}).     

The shocks seeded by the streaming-generated linear instabilities have at least two particularly important dynamical consequences. (1) As the waves grow in amplitude they eventually reverse the sign of the local pressure gradient $dp_c/dr$.  Since CRs cannot stream up the pressure gradient, this leads to the CR pressure profile having a 'staircase' like structure (Fig. \ref{fig:rho_stream}):  most of the volume is filled  with $d p_c/dr=0$ while the CR pressure changes appreciably only in small regions near the shocks.   (2)  The latter implies that the CR pressure changes appreciably only in denser than average regions and thus regions of smaller than average Alfv\'en velocity (Fig.~\ref{fig:shocks}).  This leads to the streaming-induced transfer of energy from the CRs to the gas $|{\bf v_A \cdot \nabla} p_c|$ being less efficient than would be the case in a laminar solution.  This is particularly true for lower gas sound speeds $c_i$ (i.e., larger $V_g/c_i$ where the circular velocity of the potential $\sqrt{2} V_g$ sets the velocity scale for shocks), because shock-induced density compressions are then larger.   The net result is that the time-averaged relation between CR pressure and gas density can  be shallower than the canonical $p_c \propto \rho^{2/3}$, with $\langle p_c \rangle \propto \langle \rho \rangle ^{\geff}$ and $\geff \simeq 1/2$ rather than $\geff = 2/3$ (Figs.~\ref{fig:pc-rho_stream} \& \ref{fig:pc-rhoVg}).  The scaling $p_c \propto \rho^{1/2}$ is equivalent to ${\bf \nabla} \cdot (p_c {\bf v_A)} \sim$\,constant, i.e., the CR power is roughly independent of radius.  This reflects the fact that the inhomogeneous time-dependent solutions minimize the energetic importance of streaming losses which otherwise sap energy out of the CRs.    It is worth noting that our fluid model for CRs neglects the fact that CRs can be efficiently accelerated at shocks, which would produce an additional source of CRs in the vicinity of shocks.  This could modify how CRs behave in shocks from what we find in our calculations (Fig.~\ref{fig:shocks}).

Motivated by the simulation results we carried out analytic estimates of the wind mass-loss rate, kinetic power, terminal velocity, and momentum loss-rate for general $\geff$ (\S \ref{sec:strmod}).  The case of $\geff \simeq 1/2$ does a good job of reproducing our simulation results (see \S \ref{sec:synthesis} and Fig.~\ref{fig:strfin}), including the density profiles in the inner hydrostatic portion of the flow (top left panel of Fig. \ref{fig:steady_stream}).  In particular, the scaling of the simulation mass-loss rates with $V_g/c_{\rm eff,0}$ (where $c_{\rm eff,0}$ is the base CR sound speed) is much better predicted by our $\geff=1/2$ theoretical results than by standard CR wind theory.   This reflects the fact that the mass-loss rate for $\geff=1/2$ (eq. \ref{eq:mdotstrmod}) is a factor of $\sim (V_g/c_{\rm eff,0})^2 \gg 1$ {larger} than the standard CR streaming mass-loss rate (eq. \ref{mdotstrgam0.6}).  The seemingly modest modification to the CR equation of state introduced by the strong shocks in our numerical simulations in fact significantly modifies the properties of the resulting winds.  

Our expression for the mass-loss rate accounting for the effects of strong shocks in the flow is $\dot M_w \propto p_{c,0} \, c_{\rm eff,0}^2/V_g^4$, where $p_{c,0}$ is the base CR pressure in the wind (eq.~\ref{eq:mdotstrmod}).   This shows that streaming CRs do not efficiently drive outflows from the cold, dense phases of the ISM, which have smaller values of the base CR sound speed  $c_{\rm eff,0}$.   Instead, the CR-driven mass-loss-rate will be dominated by the hotter phases, which have larger values of $c_{\rm eff,0}$; see Fig.~\ref{fig:mdotstrgam0.5}.   The warm and hot ISM also dominate the volume of the ISM and are likely the phases that set the rate at which CRs escape a galaxy.  

Our analytic derivations explicitly assume that the Alfv\'en speed is large compared to the flow velocity so that the effective equation of state is given by $p_c \propto \rho^\geff$ with $\geff \sim 1/2-2/3$.  Once $v > v_A$ the effective adiabatic index instead approaches $4/3$ (eq.~\ref{eq:ceff}; \citealt{Ipavich1975}) and the CRs no longer efficiently accelerate the gas.   This implies that for $v_A \rightarrow 0$ CR streaming cannot drive a wind.   We analytically derived a critical base Alfv\'en velocity required for efficient acceleration of a wind to terminal velocities $\gtrsim V_g$.  For  $\geff=1/2$, the result is $v_{A,0} \gtrsim v_{A, crit} \simeq 7 c_{\rm eff, 0}^2/V_g \simeq 7 \kms (c_{\rm eff,0}/10 \kms)^2 (V_g/100 \kms)^{-1}$ (eq.~\ref{eq:vAlargemod}).  For hotter phases of the ISM which perhaps have $c_{\rm eff,0} \sim 30-50 \kms$, $v_{A, crit} \simeq 40-100 \kms$ for a MW mass galaxy with $V_g \simeq 150 \kms$; this corresponds to $B \gtrsim 1 \mu G$ (for $n \sim 10^{-3} {\rm cm^{-3}}$) for efficient acceleration of a wind in the hot ISM, which is plausible given synchrotron estimates of magnetic field strengths \citep{Beck2015}.

What happens for $v_{A,0} < v_{A, crit}$?   The simulations show that in this case the gas is slowly lifted out to large radii with velocities below the escape velocity  (Fig.~\ref{fig:vA}).  These are formally `breezes' rather than transonic winds in the sense that the time-averaged flow never reaches the sonic point.   The mass-loss rate in this regime approaches the maximum allowed by energy conservation given the base CR power (see eq. \ref{eq:mdotmaxvsmdotstarmod} and \S \ref{sec:maxmdot} \& \ref{sec:vA}).   These low $v_{A,0}$ solutions are thus analogous to the low diffusion coefficient $\kappa$ solutions in Paper I, both of which are CR analogues of photon-tired stellar winds \citep{Owocki1997}.   One difference between the low $v_{A,0}$ solutions here and the low $\kappa$ solutions in Paper I is that for CR streaming, the total energy available to lift gas out to large radii is only a fraction $\sim 0.25$ of the initial CR power because of streaming-induced losses.  This implies that the maximum mass-loss for isothermal streaming solutions is smaller than for diffusive CRs by the same factor of $\sim 4$ (compare eq.~\ref{eq:mdotmaxvsmdotstarmod} vs.~eq.~39 of Paper I).

In most regions of parameter space, we find that the Alfv\'en speed becomes larger than both the  flow speed and the gas and CR sound speeds somewhat exterior to the base of the wind (Fig.~\ref{fig:vA}).   In our spherically symmetric simulations, the magnetic field is assumed to be a split-monopole and exerts no forces on the gas.   This is unlikely to be true in a more realistic multi-dimensional calculation.   Instead, there would be a combination of closed field lines in which magnetic forces dominate and open field lines with outflow.   Multi-dimensional simulations to study this would be valuable.   Including a realistic ambient circumgalactic medium in such calculations would also be useful, since this may limit how magnetically dominated the solution can become exterior to the bulk of the galaxy.

Our combined analysis in this paper and Paper I highlights several   properties of galactic winds that depend on the mechanism of CR transport. Using our analytic estimates validated by the numerical simulations, we find that diffusive CR transport leads to larger mass-loss rates for a given set of galaxy conditions, by a factor of a few-$10^2$ (Fig. \ref{fig:StrvsDiff}, bottom panel, and eq. \ref{eq:mdotdiffvsstr}).  The differences are largest for cooler ISM phases (large $V_g/c_i$) and lower base CR pressures.   The origin of this difference is that the CR pressure falls off more slowly with distance from the galaxy for diffusive CR transport (Fig. \ref{fig:StrvsDiff}, top panel).  A corollary of this result is that winds driven by CRs with streaming transport accelerate significantly more slowly than those driven by CRs with diffusive transport (Fig.~\ref{fig:StrvsDiff}, top panel) and the critical point that determines the mass-loss rate in the wind is at a radius of several times the base radius $r_0$ (eq.~\ref{eq:son_geff0.5}).   The slow acceleration of winds with streaming transport implies that the wind properties are likely to be much more sensitive to spatial variation in CR transport in the halos of galaxies.   As an indication of this sensitivity, we find that including even a small diffusion coefficient in our simulations of galactic winds with CR streaming significantly modifies the resulting solutions.  Specifically, including a diffusion coefficient that by itself does not drive a strong wind (Paper I) is sufficient to increase the mass-loss rate by a factor of $\simeq 5$ relative to a streaming-only calculation (Table \ref{tab:compare}, \S \ref{sec:str_diff}, and Fig.~\ref{fig:rho-p-both}).   Streaming-only wind models are thus sensitive to including a small diffusive correction to the transport, which is inevitably present due to the finite mean free path in the frame of the self-excited Alfv\'en waves (e.g., \citealt{Skilling1971,Recchia2016, Thomas2019}).   The converse is not true: including streaming in our fiducial diffusion simulation from Paper I has little impact on the resulting wind properties (see the last two entries in Table \ref{tab:compare}).    

Another key difference between winds with CR diffusion and streaming is that in the isothermal streaming solutions the asymptotic wind power is only $\sim 0.25$ of the power supplied to CRs by supernovae and other stellar processes in the galactic disk (Fig.~\ref{fig:strfin}).   This is a consequence of streaming-induced transfer of energy from the CRs to the gas; in the isothermal gas approximation this energy is assumed to be rapidly radiated away.   By contrast, in winds with CR diffusion, energy is globally conserved.   Winds driven by CR diffusion thus deposit more energy into the surrounding circumgalactic medium than winds driven by CR streaming.  The combination of the higher wind power and higher mass-loss rates strongly suggests that galactic winds driven by diffusive CRs would be more dynamically  important for galaxy formation than winds driven by streaming CRs (\citealt{Wiener2017,Hopkins2020a} reached this same conclusion in their simulations).  Which mechanism is indeed the dominant one remains, however, uncertain and may well depend on the thermal, magnetic, and turbulent state of the plasma.   

An important approximation in our analysis in both Paper I and this paper is the assumption of isothermal gas.   This assumption is often made in wind models driven by mechanisms other than thermal gas pressure because it simplifies the solutions considerably.  For CR streaming, however, the isothermal gas approximation is only plausibly self-consistent if radiative losses are fast compared to the rate at which CRs heat the gas (mediated by the streaming-generated Alfv\'en waves) $|{\bf v_A \cdot \nabla} p_c|$, since the latter inevitably accompanies CR streaming \citep{Wentzel1971, Ipavich1975}.  In the absence of cooling, previous work has shown that CR heating can rapidly transfer CR energy to the gas, leading to what is effectively a strong thermally driven wind \citep{Ipavich1975}.   By estimating the importance of radiative losses post facto in our solutions, we find that the rapid cooling limit requires $c_{\rm eff, 0}/V_g \gtrsim 0.13$ (with a weak dependence on other parameters in the problem; see eq. \ref{eq:cool_cond} and Fig. \ref{fig:cool}).   Physically, a larger base CR sound speed $c_{\rm eff, 0}$ implies a larger mass-loss rate and hence stronger cooling.   Importantly, the phases of the ISM with larger $c_{\rm eff,0}$ likely dominate the CR-driven mass-loss rate.  Our post facto analysis of the importance of radiative cooling is based, however, on the time-averaged cooling and heating rates in our simulations.  The large spatial variation in gas density and the CR pressure gradient ($\propto$ the CR heating rate) in our time-dependent simulations (Fig. \ref{fig:rho_stream}) shows that the interplay between heating and cooling will be more subtle than captured by the time-averaged analysis.  The presence of self-consistent heating and cooling will also generate additional instabilities that may be dynamically important:  thermal instability in regions of strong cooling \citep{Kempski2020} and the acoustic instability of \citet{Begelman1994} in regions of high Alfv\'en velocity but slow cooling.   A careful study of time-dependent CR-driven winds with realistic cooling will clearly be necessary to fully understand the interplay between cooling, CR heating, and instabilities driven by CR streaming. This will be the focus of future work.  

The results in this paper have a number of  implications for observations of galactic winds, which we briefly sketch in \S\ref{sec:obs}. Important directions for future work include studying the observational signatures of the multiphase gas that is likely produced by the strong shocks found in our simulations, understanding the multi-dimensional magnetohydrodynamics of winds with large Alfv\'en velocity, and constructing self-consistent models of galaxies across the star-forming sequence with synchrotron and pionic losses to understand how radio and gamma-ray observations constrain the streaming-only wind models developed here.

\section*{Data Availability} The numerical simulation results used in this paper will be shared on request to the corresponding author.  
 \vspace{-0.3cm}
 \section*{Acknowledgments} We thank Andrea Antoni, Phil Hopkins, Philipp Kempski, S. Peng Oh, Eve Ostriker, and Jono Squire for useful conversations.  EQ thanks the Princeton Astrophysical Sciences department and the theoretical astrophysics group and Moore Distinguished Scholar program at Caltech for their hospitality and support.  EQ was supported in part by a Simons Investigator Award from the Simons Foundation and by NSF grant AST-1715070.   The Center for Computational Astrophysics at the Flatiron Institute is supported by the Simons Foundation. TAT is supported in part by NSF grant 1516967 and NASA grant 80NSSC18K0526. TAT acknowledges support from a Simons Foundation Fellowship and an IBM Einstein Fellowship from the Institute for Advanced Study, Princeton, while a portion of this work was completed.  This research made extensive use of Matplotlib \citep{Hunter:2007} and Astropy,\footnote{http://www.astropy.org} a community-developed core Python package for Astronomy \citep{Astropy1, Astropy2}.

\bibliography{ref}

\begin{thebibliography}{}
\makeatletter
\relax
\def\mn@urlcharsother{\let\do\@makeother \do\$\do\&\do\#\do\^\do\_\do\%\do\~}
\def\mn@doi{\begingroup\mn@urlcharsother \@ifnextchar [ {\mn@doi@}
  {\mn@doi@[]}}
\def\mn@doi@[#1]#2{\def\@tempa{#1}\ifx\@tempa\@empty \href
  {http://dx.doi.org/#2} {doi:#2}\else \href {http://dx.doi.org/#2} {#1}\fi
  \endgroup}
\def\mn@eprint#1#2{\mn@eprint@#1:#2::\@nil}
\def\mn@eprint@arXiv#1{\href {http://arxiv.org/abs/#1} {{\tt arXiv:#1}}}
\def\mn@eprint@dblp#1{\href {http://dblp.uni-trier.de/rec/bibtex/#1.xml}
  {dblp:#1}}
\def\mn@eprint@#1:#2:#3:#4\@nil{\def\@tempa {#1}\def\@tempb {#2}\def\@tempc
  {#3}\ifx \@tempc \@empty \let \@tempc \@tempb \let \@tempb \@tempa \fi \ifx
  \@tempb \@empty \def\@tempb {arXiv}\fi \@ifundefined
  {mn@eprint@\@tempb}{\@tempb:\@tempc}{\expandafter \expandafter \csname
  mn@eprint@\@tempb\endcsname \expandafter{\@tempc}}}

\bibitem[\protect\citeauthoryear{{Abdo} et~al.,}{{Abdo}
  et~al.}{2010a}]{Fermi2010b}
{Abdo} A.~A.,  et~al., 2010a, \mn@doi [\aap] {10.1051/0004-6361/201015759},
  \href {https://ui.adsabs.harvard.edu/abs/2010A&A...523L...2A} {523, L2}

\bibitem[\protect\citeauthoryear{{Abdo} et~al.,}{{Abdo}
  et~al.}{2010b}]{Fermi2010}
{Abdo} A.~A.,  et~al., 2010b, \mn@doi [\apjl] {10.1088/2041-8205/709/2/L152},
  \href {https://ui.adsabs.harvard.edu/abs/2010ApJ...709L.152A} {709, L152}

\bibitem[\protect\citeauthoryear{{Ackermann}, {Ajello}, {Allafort}, {Baldini},
  {Ballet}, {Bastieri}  \& {et al.}}{{Ackermann} et~al.}{2012}]{Fermi2012}
{Ackermann} M.,  {Ajello} M.,  {Allafort} A.,  {Baldini} L.,  {Ballet} J.,
  {Bastieri} D.,   {et al.} 2012, \mn@doi [\apj] {10.1088/0004-637X/755/2/164},
  \href {https://ui.adsabs.harvard.edu/abs/2012ApJ...755..164A} {755, 164}

\bibitem[\protect\citeauthoryear{{Amato} \& {Blasi}}{{Amato} \&
  {Blasi}}{2018}]{Amato2017}
{Amato} E.,  {Blasi} P.,  2018, \mn@doi [Advances in Space Research]
  {10.1016/j.asr.2017.04.019}, \href
  {https://ui.adsabs.harvard.edu/abs/2018AdSpR..62.2731A} {62, 2731}

\bibitem[\protect\citeauthoryear{{Astropy Collaboration} et~al.,}{{Astropy
  Collaboration} et~al.}{2013}]{Astropy1}
{Astropy Collaboration} et~al., 2013, \mn@doi [\aap]
  {10.1051/0004-6361/201322068}, \href
  {https://ui.adsabs.harvard.edu/abs/2013A&A...558A..33A} {558, A33}

\bibitem[\protect\citeauthoryear{{Astropy Collaboration} et~al.,}{{Astropy
  Collaboration} et~al.}{2018}]{Astropy2}
{Astropy Collaboration} et~al., 2018, \mn@doi [\aj] {10.3847/1538-3881/aabc4f},
  \href {https://ui.adsabs.harvard.edu/abs/2018AJ....156..123A} {156, 123}

\bibitem[\protect\citeauthoryear{{Bai}, {Ostriker}, {Plotnikov}  \&
  {Stone}}{{Bai} et~al.}{2019}]{Bai2019}
{Bai} X.-N.,  {Ostriker} E.~C.,  {Plotnikov} I.,   {Stone} J.~M.,  2019,
  \mn@doi [\apj] {10.3847/1538-4357/ab1648}, \href
  {https://ui.adsabs.harvard.edu/abs/2019ApJ...876...60B} {876, 60}

\bibitem[\protect\citeauthoryear{{Beck}}{{Beck}}{2015}]{Beck2015}
{Beck} R.,  2015, \mn@doi [\aapr] {10.1007/s00159-015-0084-4}, \href
  {https://ui.adsabs.harvard.edu/abs/2015A&ARv..24....4B} {24, 4}

\bibitem[\protect\citeauthoryear{{Begelman} \& {Zweibel}}{{Begelman} \&
  {Zweibel}}{1994}]{Begelman1994}
{Begelman} M.~C.,  {Zweibel} E.~G.,  1994, \mn@doi [\apj] {10.1086/174519},
  \href {https://ui.adsabs.harvard.edu/abs/1994ApJ...431..689B} {431, 689}

\bibitem[\protect\citeauthoryear{{Blandford} \& {Eichler}}{{Blandford} \&
  {Eichler}}{1987}]{Blandford1987}
{Blandford} R.,  {Eichler} D.,  1987, \mn@doi [\physrep]
  {10.1016/0370-1573(87)90134-7}, \href
  {https://ui.adsabs.harvard.edu/abs/1987PhR...154....1B} {154, 1}

\bibitem[\protect\citeauthoryear{{Booth}, {Agertz}, {Kravtsov}  \&
  {Gnedin}}{{Booth} et~al.}{2013}]{Booth2013}
{Booth} C.~M.,  {Agertz} O.,  {Kravtsov} A.~V.,   {Gnedin} N.~Y.,  2013,
  \mn@doi [\apjl] {10.1088/2041-8205/777/1/L16}, \href
  {https://ui.adsabs.harvard.edu/abs/2013ApJ...777L..16B} {777, L16}

\bibitem[\protect\citeauthoryear{{Breitschwerdt}, {McKenzie}  \&
  {Voelk}}{{Breitschwerdt} et~al.}{1991}]{Breitschwerdt1991}
{Breitschwerdt} D.,  {McKenzie} J.~F.,   {Voelk} H.~J.,  1991, \aap, \href
  {http://adsabs.harvard.edu/abs/1991A%26A...245...79B} {245, 79}

\bibitem[\protect\citeauthoryear{{Br{\"u}ggen} \& {Scannapieco}}{{Br{\"u}ggen}
  \& {Scannapieco}}{2020}]{Bruggen2020}
{Br{\"u}ggen} M.,  {Scannapieco} E.,  2020, \mn@doi [\apj]
  {10.3847/1538-4357/abc00f}, \href
  {https://ui.adsabs.harvard.edu/abs/2020ApJ...905...19B} {905, 19}

\bibitem[\protect\citeauthoryear{{Buckman}, {Linden}  \& {Thompson}}{{Buckman}
  et~al.}{2020}]{Buckman2020}
{Buckman} B.~J.,  {Linden} T.,   {Thompson} T.~A.,  2020, \mn@doi [\mnras]
  {10.1093/mnras/staa875}, \href
  {https://ui.adsabs.harvard.edu/abs/2020MNRAS.494.2679B} {494, 2679}

\bibitem[\protect\citeauthoryear{{Butsky} \& {Quinn}}{{Butsky} \&
  {Quinn}}{2018}]{Butsky2018}
{Butsky} I.~S.,  {Quinn} T.~R.,  2018, \mn@doi [\apj]
  {10.3847/1538-4357/aaeac2}, \href
  {https://ui.adsabs.harvard.edu/abs/2018ApJ...868..108B} {868, 108}

\bibitem[\protect\citeauthoryear{{Chan}, {Kere{\v{s}}}, {Hopkins}, {Quataert},
  {Su}, {Hayward}  \& {Faucher-Gigu{\`e}re}}{{Chan} et~al.}{2019}]{Chan2019}
{Chan} T.~K.,  {Kere{\v{s}}} D.,  {Hopkins} P.~F.,  {Quataert} E.,  {Su} K.~Y.,
   {Hayward} C.~C.,   {Faucher-Gigu{\`e}re} C.~A.,  2019, \mn@doi [\mnras]
  {10.1093/mnras/stz1895}, \href
  {https://ui.adsabs.harvard.edu/abs/2019MNRAS.488.3716C} {488, 3716}

\bibitem[\protect\citeauthoryear{{Crocker}, {Krumholz}  \&
  {Thompson}}{{Crocker} et~al.}{2021}]{Crocker2021a}
{Crocker} R.~M.,  {Krumholz} M.~R.,   {Thompson} T.~A.,  2021, \mn@doi [\mnras]
  {10.1093/mnras/stab148}, \href
  {https://ui.adsabs.harvard.edu/abs/2021MNRAS.tmp..195C} {}

\bibitem[\protect\citeauthoryear{{Dalgarno}}{{Dalgarno}}{2006}]{Dalgarno2006}
{Dalgarno} A.,  2006, \mn@doi [Proceedings of the National Academy of Science]
  {10.1073/pnas.0602117103}, \href
  {https://ui.adsabs.harvard.edu/abs/2006PNAS..10312269D} {103, 12269}

\bibitem[\protect\citeauthoryear{{Drury} \& {Falle}}{{Drury} \&
  {Falle}}{1986}]{Drury1986}
{Drury} L.~O.,  {Falle} S.~A.~E.~G.,  1986, \mn@doi [\mnras]
  {10.1093/mnras/223.2.353}, \href
  {https://ui.adsabs.harvard.edu/abs/1986MNRAS.223..353D} {223, 353}

\bibitem[\protect\citeauthoryear{{Dumke}, {Krause}, {Wielebinski}  \&
  {Klein}}{{Dumke} et~al.}{1995}]{Dumke1995}
{Dumke} M.,  {Krause} M.,  {Wielebinski} R.,   {Klein} U.,  1995, \aap, \href
  {https://ui.adsabs.harvard.edu/abs/1995A&A...302..691D} {302, 691}

\bibitem[\protect\citeauthoryear{{Everett}, {Zweibel}, {Benjamin}, {McCammon},
  {Rocks}  \& {Gallagher}}{{Everett} et~al.}{2008}]{Everett2008}
{Everett} J.~E.,  {Zweibel} E.~G.,  {Benjamin} R.~A.,  {McCammon} D.,  {Rocks}
  L.,   {Gallagher} III J.~S.,  2008, \mn@doi [\apj] {10.1086/524766}, \href
  {http://adsabs.harvard.edu/abs/2008ApJ...674..258E} {674, 258}

\bibitem[\protect\citeauthoryear{{Guo} \& {Oh}}{{Guo} \& {Oh}}{2008}]{Guo2008}
{Guo} F.,  {Oh} S.~P.,  2008, \mn@doi [\mnras]
  {10.1111/j.1365-2966.2007.12692.x}, \href
  {https://ui.adsabs.harvard.edu/abs/2008MNRAS.384..251G} {384, 251}

\bibitem[\protect\citeauthoryear{{Hollenbach} \& {McKee}}{{Hollenbach} \&
  {McKee}}{1989}]{Hollenbach1989}
{Hollenbach} D.,  {McKee} C.~F.,  1989, \mn@doi [\apj] {10.1086/167595}, \href
  {https://ui.adsabs.harvard.edu/abs/1989ApJ...342..306H} {342, 306}

\bibitem[\protect\citeauthoryear{{Hopkins} et~al.,}{{Hopkins}
  et~al.}{2020}]{Hopkins2020a}
{Hopkins} P.~F.,  et~al., 2020, \mn@doi [\mnras] {10.1093/mnras/stz3321}, \href
  {https://ui.adsabs.harvard.edu/abs/2020MNRAS.492.3465H} {492, 3465}

\bibitem[\protect\citeauthoryear{{Hopkins}, {Squire}, {Chan}, {Quataert}, {Ji},
  {Kere{\v{s}}}  \& {Faucher-Gigu{\`e}re}}{{Hopkins}
  et~al.}{2021}]{Hopkins2020b}
{Hopkins} P.~F.,  {Squire} J.,  {Chan} T.~K.,  {Quataert} E.,  {Ji} S.,
  {Kere{\v{s}}} D.,   {Faucher-Gigu{\`e}re} C.-A.,  2021, \mn@doi [\mnras]
  {10.1093/mnras/staa3691}, \href
  {https://ui.adsabs.harvard.edu/abs/2021MNRAS.501.4184H} {501, 4184}

\bibitem[\protect\citeauthoryear{{Huang} \& {Davis}}{{Huang} \&
  {Davis}}{2021}]{Huang2021}
{Huang} X.,  {Davis} S.~W.,  2021, arXiv e-prints, \href
  {https://ui.adsabs.harvard.edu/abs/2021arXiv210511506H} {p. arXiv:2105.11506}

\bibitem[\protect\citeauthoryear{Hunter}{Hunter}{2007}]{Hunter:2007}
Hunter J.~D.,  2007, Computing In Science \& Engineering, 9, 90

\bibitem[\protect\citeauthoryear{{Ipavich}}{{Ipavich}}{1975}]{Ipavich1975}
{Ipavich} F.~M.,  1975, \mn@doi [\apj] {10.1086/153397}, \href
  {http://adsabs.harvard.edu/abs/1975ApJ...196..107I} {196, 107}

\bibitem[\protect\citeauthoryear{{Jiang} \& {Oh}}{{Jiang} \&
  {Oh}}{2018}]{JiangOh2018}
{Jiang} Y.-F.,  {Oh} S.~P.,  2018, \mn@doi [\apj] {10.3847/1538-4357/aaa6ce},
  \href {https://ui.adsabs.harvard.edu/abs/2018ApJ...854....5J} {854, 5}

\bibitem[\protect\citeauthoryear{{Kempski} \& {Quataert}}{{Kempski} \&
  {Quataert}}{2020}]{Kempski2020}
{Kempski} P.,  {Quataert} E.,  2020, \mn@doi [\mnras] {10.1093/mnras/staa385},
  \href {https://ui.adsabs.harvard.edu/abs/2020MNRAS.493.1801K} {493, 1801}

\bibitem[\protect\citeauthoryear{{Kulsrud} \& {Pearce}}{{Kulsrud} \&
  {Pearce}}{1969}]{Kulsrud1969}
{Kulsrud} R.,  {Pearce} W.~P.,  1969, \mn@doi [\apj] {10.1086/149981}, \href
  {https://ui.adsabs.harvard.edu/abs/1969ApJ...156..445K} {156, 445}

\bibitem[\protect\citeauthoryear{{Lacki}, {Thompson}  \& {Quataert}}{{Lacki}
  et~al.}{2010}]{Lacki2010}
{Lacki} B.~C.,  {Thompson} T.~A.,   {Quataert} E.,  2010, \mn@doi [\apj]
  {10.1088/0004-637X/717/1/1}, \href
  {http://adsabs.harvard.edu/abs/2010ApJ...717....1L} {717, 1}

\bibitem[\protect\citeauthoryear{{Lacki}, {Thompson}, {Quataert}, {Loeb}  \&
  {Waxman}}{{Lacki} et~al.}{2011}]{Lacki2011}
{Lacki} B.~C.,  {Thompson} T.~A.,  {Quataert} E.,  {Loeb} A.,   {Waxman} E.,
  2011, \mn@doi [\apj] {10.1088/0004-637X/734/2/107}, \href
  {http://adsabs.harvard.edu/abs/2011ApJ...734..107L} {734, 107}

\bibitem[\protect\citeauthoryear{{Lerche}}{{Lerche}}{1967}]{Lerche1967}
{Lerche} I.,  1967, \mn@doi [\apj] {10.1086/149045}, \href
  {https://ui.adsabs.harvard.edu/abs/1967ApJ...147..689L} {147, 689}

\bibitem[\protect\citeauthoryear{{Lochhaas}, {Thompson}  \&
  {Schneider}}{{Lochhaas} et~al.}{2020}]{Lochhaas2020}
{Lochhaas} C.,  {Thompson} T.~A.,   {Schneider} E.~E.,  2020, arXiv e-prints,
  \href {https://ui.adsabs.harvard.edu/abs/2020arXiv201106004L} {p.
  arXiv:2011.06004}

\bibitem[\protect\citeauthoryear{{Mao} \& {Ostriker}}{{Mao} \&
  {Ostriker}}{2018}]{Mao2018}
{Mao} S.~A.,  {Ostriker} E.~C.,  2018, \mn@doi [\apj]
  {10.3847/1538-4357/aaa88e}, \href
  {https://ui.adsabs.harvard.edu/abs/2018ApJ...854...89M} {854, 89}

\bibitem[\protect\citeauthoryear{{Martin}}{{Martin}}{2005}]{Martin2005}
{Martin} C.~L.,  2005, \mn@doi [\apj] {10.1086/427277}, \href
  {http://adsabs.harvard.edu/abs/2005ApJ...621..227M} {621, 227}

\bibitem[\protect\citeauthoryear{{Murray}, {Quataert}  \& {Thompson}}{{Murray}
  et~al.}{2005}]{Murray2005}
{Murray} N.,  {Quataert} E.,   {Thompson} T.~A.,  2005, \mn@doi [\apj]
  {10.1086/426067}, \href {http://adsabs.harvard.edu/abs/2005ApJ...618..569M}
  {618, 569}

\bibitem[\protect\citeauthoryear{{Oppenheimer} \& {Dav{\'e}}}{{Oppenheimer} \&
  {Dav{\'e}}}{2006}]{Oppenheimer2006}
{Oppenheimer} B.~D.,  {Dav{\'e}} R.,  2006, \mn@doi [\mnras]
  {10.1111/j.1365-2966.2006.10989.x}, \href
  {http://adsabs.harvard.edu/abs/2006MNRAS.373.1265O} {373, 1265}

\bibitem[\protect\citeauthoryear{{Oppenheimer}, {Dav{\'e}}, {Kere{\v s}},
  {Fardal}, {Katz}, {Kollmeier}  \& {Weinberg}}{{Oppenheimer}
  et~al.}{2010}]{Oppenheimer2010}
{Oppenheimer} B.~D.,  {Dav{\'e}} R.,  {Kere{\v s}} D.,  {Fardal} M.,  {Katz}
  N.,  {Kollmeier} J.~A.,   {Weinberg} D.~H.,  2010, \mn@doi [\mnras]
  {10.1111/j.1365-2966.2010.16872.x}, \href
  {http://adsabs.harvard.edu/abs/2010MNRAS.406.2325O} {406, 2325}

\bibitem[\protect\citeauthoryear{{Ostriker} \& {Shetty}}{{Ostriker} \&
  {Shetty}}{2011}]{Ostriker2011}
{Ostriker} E.~C.,  {Shetty} R.,  2011, \mn@doi [\apj]
  {10.1088/0004-637X/731/1/41}, \href
  {https://ui.adsabs.harvard.edu/abs/2011ApJ...731...41O} {731, 41}

\bibitem[\protect\citeauthoryear{{Owocki} \& {Gayley}}{{Owocki} \&
  {Gayley}}{1997}]{Owocki1997}
{Owocki} S.~P.,  {Gayley} K.~G.,  1997, {ThePhysics of Stellar Winds Near the
  Eddington Limit}.
p.~121

\bibitem[\protect\citeauthoryear{{Quataert}, {Thompson}  \& {Jiang}}{{Quataert}
  et~al.}{2021}]{Quataert2021}
{Quataert} E.,  {Thompson} T.~A.,   {Jiang} Y.-F.,  2021, arXiv e-prints, \href
  {https://ui.adsabs.harvard.edu/abs/2021arXiv210205696Q} {p. arXiv:2102.05696}

\bibitem[\protect\citeauthoryear{{Recchia}, {Blasi}  \& {Morlino}}{{Recchia}
  et~al.}{2016}]{Recchia2016}
{Recchia} S.,  {Blasi} P.,   {Morlino} G.,  2016, \mn@doi [\mnras]
  {10.1093/mnras/stw1966}, \href
  {https://ui.adsabs.harvard.edu/abs/2016MNRAS.462.4227R} {462, 4227}

\bibitem[\protect\citeauthoryear{{Seaquist} \& {Odegard}}{{Seaquist} \&
  {Odegard}}{1991}]{Seaquist1991}
{Seaquist} E.~R.,  {Odegard} N.,  1991, \mn@doi [\apj] {10.1086/169764}, \href
  {https://ui.adsabs.harvard.edu/abs/1991ApJ...369..320S} {369, 320}

\bibitem[\protect\citeauthoryear{{Skilling}}{{Skilling}}{1971}]{Skilling1971}
{Skilling} J.,  1971, \mn@doi [\apj] {10.1086/151210}, \href
  {https://ui.adsabs.harvard.edu/abs/1971ApJ...170..265S} {170, 265}

\bibitem[\protect\citeauthoryear{{Somerville} \& {Dav{\'e}}}{{Somerville} \&
  {Dav{\'e}}}{2015}]{Somerville2015}
{Somerville} R.~S.,  {Dav{\'e}} R.,  2015, \mn@doi [\araa]
  {10.1146/annurev-astro-082812-140951}, \href
  {https://ui.adsabs.harvard.edu/abs/2015ARA&A..53...51S} {53, 51}

\bibitem[\protect\citeauthoryear{{Squire}, {Hopkins}, {Quataert}  \&
  {Kempski}}{{Squire} et~al.}{2021}]{Squire2021}
{Squire} J.,  {Hopkins} P.~F.,  {Quataert} E.,   {Kempski} P.,  2021, \mn@doi
  [\mnras] {10.1093/mnras/stab179}, \href
  {https://ui.adsabs.harvard.edu/abs/2021MNRAS.502.2630S} {502, 2630}

\bibitem[\protect\citeauthoryear{{Stone}, {Tomida}, {White}  \&
  {Felker}}{{Stone} et~al.}{2020}]{Stone2020}
{Stone} J.~M.,  {Tomida} K.,  {White} C.~J.,   {Felker} K.~G.,  2020, \mn@doi
  [\apjs] {10.3847/1538-4365/ab929b}, \href
  {https://ui.adsabs.harvard.edu/abs/2020ApJS..249....4S} {249, 4}

\bibitem[\protect\citeauthoryear{{Thomas} \& {Pfrommer}}{{Thomas} \&
  {Pfrommer}}{2019}]{Thomas2019}
{Thomas} T.,  {Pfrommer} C.,  2019, \mn@doi [\mnras] {10.1093/mnras/stz263},
  \href {https://ui.adsabs.harvard.edu/abs/2019MNRAS.485.2977T} {485, 2977}

\bibitem[\protect\citeauthoryear{{Thompson}, {Quataert}  \&
  {Murray}}{{Thompson} et~al.}{2005}]{Thompson2005}
{Thompson} T.~A.,  {Quataert} E.,   {Murray} N.,  2005, \mn@doi [\apj]
  {10.1086/431923}, \href {http://adsabs.harvard.edu/abs/2005ApJ...630..167T}
  {630, 167}

\bibitem[\protect\citeauthoryear{{Weiner} et~al.,}{{Weiner}
  et~al.}{2009}]{Weiner2009}
{Weiner} B.~J.,  et~al., 2009, \mn@doi [\apj] {10.1088/0004-637X/692/1/187},
  \href {http://adsabs.harvard.edu/abs/2009ApJ...692..187W} {692, 187}

\bibitem[\protect\citeauthoryear{{Wentzel}}{{Wentzel}}{1971}]{Wentzel1971}
{Wentzel} D.~G.,  1971, \mn@doi [\apj] {10.1086/150794}, \href
  {https://ui.adsabs.harvard.edu/abs/1971ApJ...163..503W} {163, 503}

\bibitem[\protect\citeauthoryear{{Wiener}, {Oh}  \& {Guo}}{{Wiener}
  et~al.}{2013a}]{Wiener2013}
{Wiener} J.,  {Oh} S.~P.,   {Guo} F.,  2013a, \mn@doi [\mnras]
  {10.1093/mnras/stt1163}, \href
  {https://ui.adsabs.harvard.edu/abs/2013MNRAS.434.2209W} {434, 2209}

\bibitem[\protect\citeauthoryear{{Wiener}, {Zweibel}  \& {Oh}}{{Wiener}
  et~al.}{2013b}]{Wiener2013b}
{Wiener} J.,  {Zweibel} E.~G.,   {Oh} S.~P.,  2013b, \mn@doi [\apj]
  {10.1088/0004-637X/767/1/87}, \href
  {https://ui.adsabs.harvard.edu/abs/2013ApJ...767...87W} {767, 87}

\bibitem[\protect\citeauthoryear{{Wiener}, {Pfrommer}  \& {Oh}}{{Wiener}
  et~al.}{2017}]{Wiener2017}
{Wiener} J.,  {Pfrommer} C.,   {Oh} S.~P.,  2017, \mn@doi [\mnras]
  {10.1093/mnras/stx127}, \href
  {https://ui.adsabs.harvard.edu/abs/2017MNRAS.467..906W} {467, 906}

\bibitem[\protect\citeauthoryear{{Wiener}, {Zweibel}  \& {Ruszkowski}}{{Wiener}
  et~al.}{2019}]{Wiener2019}
{Wiener} J.,  {Zweibel} E.~G.,   {Ruszkowski} M.,  2019, \mn@doi [\mnras]
  {10.1093/mnras/stz2007}, \href
  {https://ui.adsabs.harvard.edu/abs/2019MNRAS.489..205W} {489, 205}

\bibitem[\protect\citeauthoryear{{Yan} \& {Lazarian}}{{Yan} \&
  {Lazarian}}{2002}]{Yan2002}
{Yan} H.,  {Lazarian} A.,  2002, \mn@doi [PRL] {10.1103/PhysRevLett.89.281102},
  \href {https://ui.adsabs.harvard.edu/abs/2002PhRvL..89B1102Y} {89, 281102}

\makeatother
\end{thebibliography}


\begin{appendix}
\section{Linear Stability}
\label{sec:appendixA}
In this Appendix, we study the linear stability of the CR magnetohydrodynamic equations with streaming transport.   Since the simulations are one-dimensional we restrict ourselves to one dimensional perturbations.   Physically, this system of equations admits longitudinal sound waves, in which both gas and CR pressure are the restoring force, as well as gas and CR entropy modes.  

In what follows, we identify two new (to the best of our knowledge) instabilities associated with CR transport by streaming, one that is present in the one-moment CR equations and is driven by background gradients in density and CR pressure (\S \ref{sec:lin1mom}) and one that is only present in the two-moment CR model and is driven by a phase shift introduced by the finite speed of light (\S \ref{sec:lin2mom}).   We show in \S \ref{sec:lin1mom} that the instability due to background gradients is the dominant instability in most of our streaming simulations.   As noted in \S \ref{sec:lin}, \citet{Begelman1994} identified an instability of sound waves when the Alfv\'en speed is large compared to the total (gas and CR) sound speed.  This instability relies on CR heating of the gas, which is not  present in our formulation of the problem because of our assumed isothermal equation of state.   Thus the \citet{Begelman1994} instability is not present in our analytic analysis in this Appendix (nor in our simulations).

\subsection{Instabilities of the Two-Moment CR Equations with Streaming Transport}
\label{sec:lin2mom}

We assume here that perturbations are $\propto \exp(-i \omega t + i k r)$ and that $k H \gg 1$ (where $H$ is a characteristic length-scale in the equilibrium state) so that a WKB analysis is appropriate.   We neglect background gradients in the problem, with the exception of requiring a background CR pressure gradient so that the streaming formulation of the CR equations is well-posed, as described below.

We carry out a linear perturbation analysis of equations \ref{eq:CR2mom}.   For CR transport due to streaming, the key frequencies in the problem are the isothermal gas sound wave frequency
\be
\omega_g = k c_i,
\label{eq:omg}
\ee
the adiabatic CR sound wave frequency
\be 
\omega_c = k c_{\rm eff} \equiv k \sqrt{4 p_c/3 \rho} 
\label{eq:omc}
\ee
and the Alfv\'en frequency
\be
\omega_A = k v_A .
\label{eq:omA}
\ee
Note that for this 1D purely longitudinal problem there is no perturbed magnetic field,  no magnetic tension, and hence no Alfv\'en waves.   The Alfv\'en frequency arises because the CRs are assumed to stream at $v_A$ relative to the gas. There is an entropy mode with frequency $\sim \omega_A$ associated with this streaming (see \citealt{Kempski2020}).

Finally, there is a frequency associated with the time dependent flux term in equation \ref{eq:CR2mom}, namely
\be
\omega_{m,s} =  \frac{3 V_m^2}{4 v_A H_c},
\label{eq:omM}
\ee
where $H_c = 1/|{\rm d\ln p_c/dr}|$ is the CR pressure scale-height.   In the WKB linear analysis for the streaming problem, $\omega_{m,s}$ is taken to be finite but all other terms related to background gradients are neglected.   Also note that in the usual one-moment CR equations, $V_m \rightarrow \infty$ and so $\omega_{m,s} \rightarrow \infty$.    

Under these approximations, it is straightforward to derive the linear longitudinal dispersion relation, which takes the form
\be
\begin{split}
& \left(\omega - \omega_A - i \frac{\omega(3 \omega + \omega_A)}{\omega_{m,s}} - \frac{4 v_A^2 \omega}{3 V_m^2} \right)\left(\omega^2 - \omega_g^2\right) \\ & = \left(\omega_c^2 - \frac{3 \omega^2 c_{\rm eff}^2}{V_m^2}\right)\left(\omega - \frac{\omega_A}{2}\right)
\end{split}
\label{eq:DRst}
\ee

For the one-moment system of CR equations, $\omega_{m,s} \propto V_m \rightarrow \infty$ and the dispersion relation reduces to $(\omega - \omega_A)(\omega^2 - \omega_g^2) - \omega_c^2(\omega - \omega_A/2) = 0$.   All solutions of this dispersion relation are stable.  Physically, this dispersion relation describes the linear coupling of sound waves to the gas and CR entropy modes.

The presence of a finite speed of light in the two-moment CR system leads to an instability.   This is most easily seen by assuming, as is physical, that $\omega_{m,s}$ is large compared to other frequencies in the problem.   In the limit $\omega_{m,s} \gg \omega_A \gg \omega_g, \omega_c$, the unstable solution is the CR entropy mode, with
\be 
\omega \simeq \omega_A + 4 i \frac{\omega_A^2}{\omega_{m,s}}
\label{eq:inst1}
\ee
By contrast, in the limit $\omega_{m,s} \gg \omega_g, \omega_c \gg \omega_A$ the instability is the sound wave, with
\be
\omega \simeq \pm \sqrt{\omega_g^2 + \omega_c^2} + \, i \, \frac{3}{2} \, \frac{\omega_c^2}{\omega_{m,s}}
\label{eq:inst2}
\ee

The high Alfv\'en speed limit is generally the most appropriate for our wind solutions (see \S \ref{sec:numerics}) so we focus on that in what follows.    Physically, the instability of the CR entropy mode in this limit arises due to the phase shift between the CR pressure and flux introduced by the time-dependent flux term in equation \ref{eq:CR2mom}, which effectively introduces a negative diffusion coefficient into the CR energy equation.    To see this, consider the limit $\omega \ll \omega_{m,s}$ in which case the equation for the linearly perturbed CR flux is
\be
\delta F_c \simeq 4 v_A \delta p_c - \frac{12 v_A}{\omega_{m,s}} \frac{\partial}{\partial t} \delta p_c
\label{eq:pertflux}
\ee
where we neglect terms proportional to perturbed gas quantities, which are negligible for the entropy mode at high $v_A$.  Substituting equation \ref{eq:pertflux} into the CR energy equation yields
\be
\frac{d \delta p_c}{dt} \simeq \frac{4 v_A}{\omega_{m,s}} \frac{\partial}{\partial r}\frac{\partial}{\partial t} \delta p_c
\label{eq:dpc}
\ee
where $d/dt = \partial/\partial t + v_A \partial/\partial r$.       For $\omega_{m,s} \rightarrow \infty$ the solution of equation \ref{eq:dpc} is just $\delta p_c = \delta p_c(\xi)$ with $\xi = r - v_A t$.   This describes  the adiabatic CR entropy mode that propagates along the field lines at $v_A$.   The right hand side of equation \ref{eq:dpc} represents the correction to the entropy mode due to the phase shift introduced by the frequency $\omega_{m,s}$.   To evaluate this term, we work in the quasi-adiabatic limit and use the leading order adiabatic solution $\delta p_c = \delta p_c(\xi)$ in the RHS of equation \ref{eq:dpc} to estimate the non-adiabatic correction.  This yields
\be
\frac{d \delta p_c}{dt} \simeq - \frac{4 v_A^2}{\omega_{m,s}} \frac{\partial^2}{\partial \xi^2} \delta p_c
\label{eq:negdiff}
\ee
Equation \ref{eq:negdiff} shows that the finite frequency $\omega_{m,s}$ gives rise to an effective negative diffusion coefficient for the otherwise  adiabatic entropy mode.   The growth rate from equation \ref{eq:negdiff} is exactly that in equation \ref{eq:inst1}.

To evaluate the importance of this instability in galactic winds, we can define the linear amplification a wave undergoes as it propagates out using $\rm A(r) \equiv \gamma dt = \gamma H_\rho/v_A$ where $\gamma$ is the growth rate and $dt = H_\rho/v_A$ is the time the wave spends in a region with roughly constant properties and $H_\rho$ is the density scale height.  The amplification factor $A(r)$ is essentially the number of e-foldings the wave undergoes before it propagates out to a region where the ambient  conditions have changed significantly.   Once $A(r) \gtrsim 1$, the wave will start to become nonlinear.    Using equation \ref{eq:inst1} and assuming $p_c \propto \rho^{2/3}$, we can write
\be
{\rm A(r)} = 8 \, k^2 H_\rho^2 \, \frac{v_{A,0}^2}{V_m^2} \, \frac{\rho_0 r_0^2}{\rho(r) r^2}
\label{eq:amp}
\ee

\begin{figure}
\centering
\includegraphics[width=87mm]{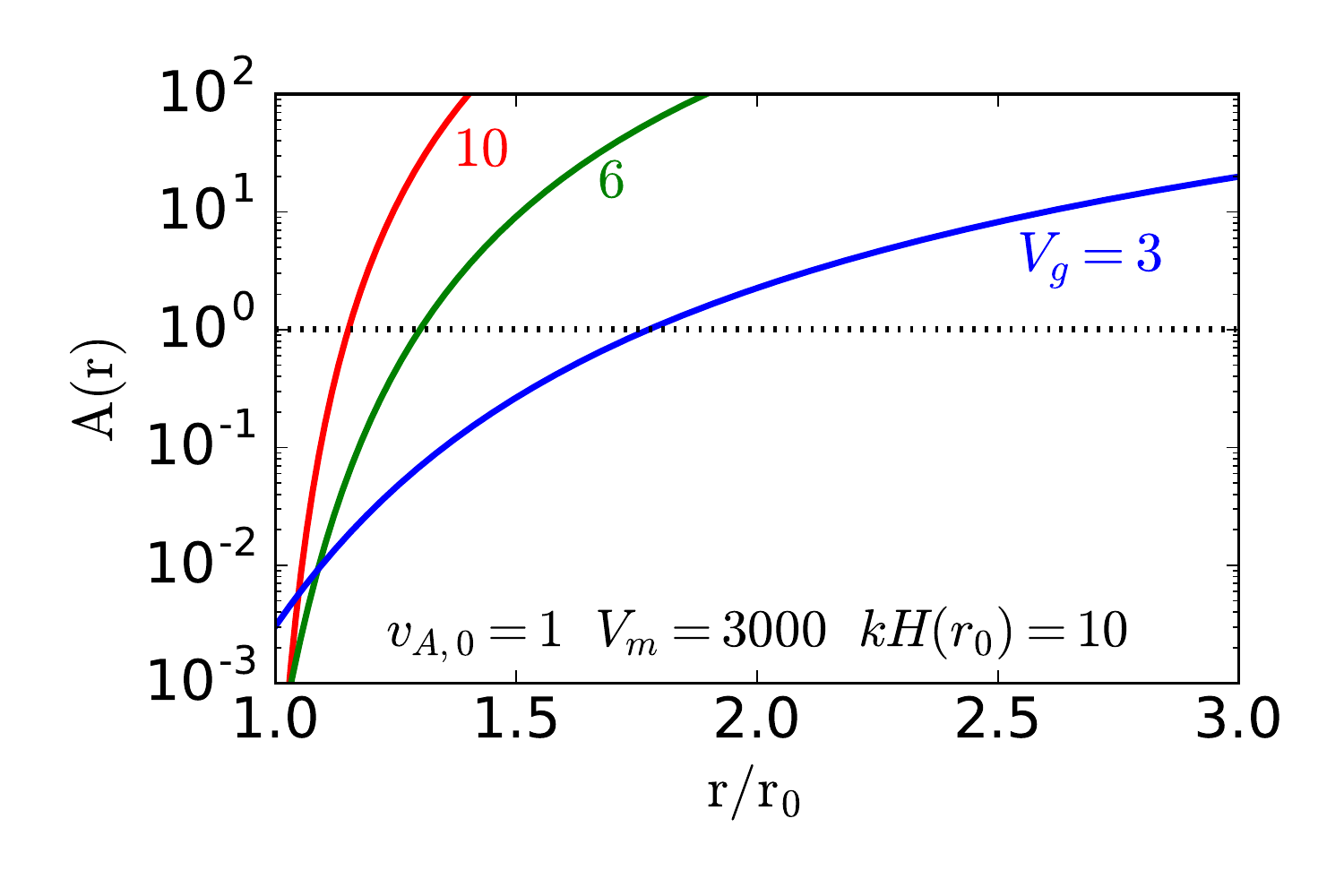}
\vspace{-0.2cm}
\caption{Linear amplification factor $A(r)$ for the high $v_A$ entropy mode instability of the two-moment CR equations (eq. \ref{eq:amp}). $A(r)$ is the number of e-foldings the wave undergoes before it propagates out to a region where the ambient conditions have changed significantly.  Parameters are similar to our fiducial streaming simulations in \S \ref{sec:streamnum}.   Although the entropy mode instability can grow and become nonlinear, we find that sound-wave instabilities driven by background gradients (and independent of the speed of light) grow faster and dominate in our simulations (Fig. \ref{fig:growth}). }
\label{fig:ampVg}
\end{figure}

Figure \ref{fig:ampVg} shows $A(r)$ for our semi-analytic streaming density profile from eq. \ref{density_he_streaming}  for parameters similar to the simulations in \S \ref{sec:streamnum}, namely $V_g/c_i = 3, 6, \& 10$, $p_c(r_0) = \rho(r_0) c_i^2$, $V_m/v_{A,0} = 3000$ and a wavelength $k H_\rho = 10$, i.e., $\lambda \simeq 0.006 r_0$, which is similar to the wavelength of the dominant mode in the simulations.   The strong dependence of the growth factor on radius is primarily driven by the rapid increase of $v_A$ as the density drops in the quasi-hydrostatic inner portion of the flow.    

Figure \ref{fig:ampVg} shows that for $V_g=10$, the instability is expected to become nonlinear by $r \sim 1.15 r_0$ with higher $k$ modes growing faster and becoming nonlinear at smaller radii (and vice-versa for lower $k$ modes).   For lower $V_g$ (weaker gravity), the instability only becomes nonlinear at larger radii. This is because of the larger gas scale-height and hence smaller $v_A$ when gravity is weaker.

In the fiducial $v_{A,0} = 1$ simulation (e.g., Fig \ref{fig:rho_stream}) we find that the modes become nonlinear at somewhat smaller radii $\sim 1.05 r_0$ than suggested by Figure \ref{fig:ampVg}.  We argue in \S \ref{sec:lin1mom} that this is because the dominant instability in most of our simulations is actually due to background gradients not kept in the analysis here.  The one exception is the high resolution $v_A=10$ simulation, which shows very rapid growth of short wavelength modes (Fig. \ref{fig:rho_stream}) that we believe are associated with the instability of the two-moment system identified here.

Equations \ref{eq:inst1}-\ref{eq:amp} predict that higher $k$ modes grow more quickly.   In the simulations the growth is limited by resolution so that the fastest growing modes cannot actually be resolved.    Physically, the wavelengths of the modes of interest cannot be smaller than the mean free path of the CRs that dominate the CR energy density, because the fluid approximation utilized here is only valid on scales above the CR mean free path.   The growth rates are also $\propto V_m^{-2}$ and so are smaller for a larger (and more realistic) speed of light.   Future work is required to fully understand the evolution and saturation of these instabilities for higher $k$ and more realistic $V_m$. To provide a guide to the relevant issues, consider $v_{A,0} = 10 \kms$.   Then $V_m = c$ should be $3 \times 10^4 v_{A,0}$, 10 times larger than our fiducial value.    Given that $\gamma \propto k^2/V_m^2$, to achieve the same growth rate would require a wavelength 10 times smaller, which we cannot resolve in most of our simulations.   Physically, the relevant wavelengths would then be $\sim 10^{-3} r_0$ which is $\sim $ parsec for typical galactic parameters.   This is comparable to the cosmic-ray mean free path in the Galaxy for GeV particles.    It is thus plausible that for more realistic parameters higher $k$ modes can grow fast enough to become dynamically important.

\subsubsection{Suppression of the Streaming Instability by Diffusion}
\label{sec:linear_w_str+diff}
The instability of the entropy mode described by equation \ref{eq:inst1} is suppressed by CR diffusion.    This can be derived simply by noting that the entropy mode instability in the limit $\omega_{m,s} \gg \omega_A \gg \omega_g, \omega_c$ is purely a consequence of the CR physics and can be derived from equation \ref{eq:CR2mom} setting all perturbed gas quantities to zero.  It is thus particularly easy to generalize equation \ref{eq:inst1} to include both streaming and diffusion.   Doing so, we find
\be
0 = \omega - \omega_A - \frac{4 \omega v_A^2}{3 V_m^2} - i \, \left(\frac{3 \omega^2}{\omega_{m,tot}} + \frac{\omega \omega_A}{\omega_{m,tot}} - \omega_d - \frac{\omega \omega_A \kappa}{V_m^2}\right)
\ee
where
\be
\omega_d = k^2 \kappa
\label{eq:omd}
\ee
is the CR diffusion frequency associated with the assumed constant diffusion coefficient $\kappa$ and 
\be
\omega_{m,tot} = \frac{3 V_m^2}{3 \kappa + 4 v_A H_c}
\ee
is the frequency associated with the finite speed of light including both streaming and diffusion.

In the limit $\omega_d \gg \omega_A$, diffusion fully suppresses the instability of the CR entropy mode.   In the limit, $\omega_A \gg \omega_d$, the solution of the dispersion relation of interest is
\be
\omega \simeq \omega_A + i \left(\frac{4 \omega_A^2}{\omega_{m,tot}} - \omega_d\right)
\label{eq:CRwdiff}
\ee
Instability requires that the term in () in equation \ref{eq:CRwdiff} be positive, which requires
\be
\kappa \lesssim \frac{16 v_A^2}{3 V_m^2} \, v_A H_c
\label{eq:kappac}
\ee
Given that $V_m \gg v_A$, this is a very stringent constraint on the diffusion coefficient, i.e., small amounts of diffusion on top of streaming stabilize the instability of the two-moment CR method found here.   

\subsection{Instabilities of the One-Moment CR System with Background Gradients}
\label{sec:lin1mom}
Instabilities of the one-moment CR system for a homogeneous background can be derived using the results in \S \ref{sec:lin2mom} by taking $V_m \rightarrow \infty$. As discussed there, the sound and entropy modes are both stable in this limit (for our assumption of isothermal gas, which eliminates the acoustic instability of \citealt{Begelman1994}).   We now show, however, that including background gradients in the calculation leads to instability of sound waves.  The instabilities are present in the one-moment CR system and so we restrict our analysis to this limit for ease of algebra.  The instability is similar to that derived by \citet{Drury1986} for diffusive CR transport.  

We consider an isothermal gas plus CR system that satisfies the following conservation laws 
\be
\frac{\partial \rho}{\partial t} + \frac{\partial \rho v}{\partial z}  = 0
\label{eq:massApp}
\ee
\be
\rho \frac{\partial v}{\partial t} + \rho v \frac{\partial v}{\partial z} = - c_i^2 \frac{\partial \rho}{\partial z} - \frac{\partial p_c}{\partial z} - \rho g
\label{eq:momApp}
\ee
\be
 \frac{\partial p_c}{\partial t} +  v \frac{\partial p_c}{\partial z} = -\frac{4}{3} p_c \frac{\partial v}{\partial z}  -\frac{4}{3} p_c \frac{\partial v_A}{\partial z} - v_A \frac{\partial p_c}{\partial z}
 \label{eq:CRApp}
 \ee
 We assume a constant gravitational acceleration $g$.   To simplify the algebra, we focus on the limit of $v_A \gg c_i, c_c$, where $c_c = \sqrt{p_c/\rho}$.   In this case, the CR energy equation for sound waves can be approximated by simply balancing the terms $\propto v_A$ in equation \ref{eq:CRApp}:
\be
v_A \frac{\partial p_c}{\partial z} = -\frac{4}{3} p_c \frac{\partial v_A}{\partial z} 
\label{eq:CRsimp}
\ee
In the equilibrium state, $p_c v_A^{4/3} = $ const, i.e., $p_c \propto \rho^{2/3}$.

We now linearize equations \ref{eq:massApp}, \ref{eq:momApp}, \& \ref{eq:CRsimp}.   To start we assume that all perturbations, labeled by $\delta$, are $\propto \exp(-i \omega t)$ but we do not Fourier transform in z.   We do the latter only at the end of the calculation to ensure that all background gradient terms are properly kept.    The linearly perturbed equations are then
\be 
i \omega \delta \rho = \frac{\partial (\rho \delta v)}{\partial z}
\label{eq:linmass}
\ee
\be
-i \omega \rho \delta v = -c_i^2 \frac{\partial \delta \rho}{\partial z} - \frac{\partial \delta p_c}{\partial z} - \delta \rho g
\label{eq:linmom}
\ee
\be
\frac{\partial \delta p_c}{\partial z} = \frac{2}{3} \delta p_c \frac{\partial \ln \rho}{\partial z} + \frac{2}{3} c_c^2\left(\frac{\partial \delta \rho}{\partial z} - \delta \rho \frac{\partial \ln \rho}{\partial z}\right)
\label{eq:CRlin}
\ee
where we have used the fact that $\partial \ln p_c/\partial z = 2/3 \partial \ln \rho/\partial z$ in the equilibrium state.   

Equations \ref{eq:linmass} and \ref{eq:linmom} can be combined to yield
\be
\omega^2 \delta \rho = -c_i^2 \frac{\partial^2  \delta \rho}{\partial^2  z} - \frac{\partial^2  \delta p_c}{\partial^2  z} - g \frac{\partial \delta \rho}{\partial z} 
\label{eq:comb}
\ee
We now calculate $\partial^2 \delta p_c/\partial^2 z$ to $O(1/kH)$ using equation \ref{eq:CRlin}, i.e., we neglect terms $O(1/H^2)$ where $H$ is the density scale-height (the CR pressure scale-height is of the same order).   This yields
\be
\frac{\partial^2 \delta p_c}{\partial^2 z} = \frac{2}{3} c_c^2  \frac{\partial^2 \delta \rho}{\partial^2 z} - \frac{2 c_c^2}{9}  \frac{\partial \delta \rho}{\partial z}  \frac{\partial \ln \rho}{\partial z}
\label{eq:eos}
\ee
Combining with equation \ref{eq:comb} and using
\be
c_i^2 \frac{\partial \ln \rho}{\partial z} + c_c^2 \frac{\partial \ln p_c}{\partial z}  = \left(c_i^2 + \frac{2}{3} c_c^2\right) \frac{\partial \ln \rho}{\partial z} = -g
\ee
for the background force balance yields a second order differential equation for the density perturbations, namely
\be
\omega^2 \delta \rho = -\left(c_i^2 + \frac{2}{3} c_c^2 \right) \frac{\partial^2 \delta \rho}{\partial^2 z}  +  \frac{d\delta \rho}{\partial z} \frac{\partial \ln \rho}{\partial z}  \left(c_i^2 + \frac{10}{9} c_c^2 \right)
\label{eq:fin}
\ee
We now look for the WKB solution to equation \ref{eq:fin} with $\delta \rho \propto \exp[ikz - z/(2H)]$.   The $\exp[-z/(2H)]$ factor is consistent with  conservation of energy flux (action)  absent wave dissipation/growth.    Equation \ref{eq:fin} then becomes
\be
\omega^2 = k^2\left( c_i^2 + \frac{2}{3} c_c^2 \right) - i \frac{4 k c_c^2}{9 H} + O(H^{-2})
\label{eq:WKB}
\ee
To leading order the solution is an adiabatic sound wave but the stratified background destabilizes the sound wave.    The instability is driven by the non-adiabatic phase shift between density and pressure introduced by CR streaming (eq. \ref{eq:eos}).   In particular, one might think that the CRs just behave like a $\gamma = 2/3$ gas from equation \ref{eq:CRsimp}.   This is not really correct, however.    In particular, as a sanity check it is straightforward to show that if $v_A = 0$, so that the CRs are  adiabatic  with $\gamma = 4/3$ (eq. \ref{eq:CRApp}), then the analogous calculation gives $\omega^2 = k^2 (c_i^2 + \gamma c_c^2) + O(H^{-2})$.   There is thus no instability for adiabatic CRs for any $\gamma > 0$, as makes physical sense.   The $O(H^{-2})$ correction here is what gives rise to the acoustic cutoff frequency.   The key difference between the CR magnetohydrodynamic equations with streaming and an adiabatic fluid is that for the latter, there is a simple relation between {Lagrangian} pressure and density perturbations, namely $\Delta p/p = \gamma \Delta \rho/\rho$ where $\Delta = \delta + v \cdot \nabla$ is a Lagrangian perturbation, $\delta$ is the Eulerian perturbation and $v$ is the fluid velocity.   The CR magnetohydrodynamic equations in the limit of high $v_A$ connect $p_c$ and $\rho$ via the Alfv\'en velocity, not the fluid velocity, leading to the phase shift in equation \ref{eq:eos} that is the origin of the instability found here.    

\begin{figure}
\centering
\includegraphics[width=87mm]{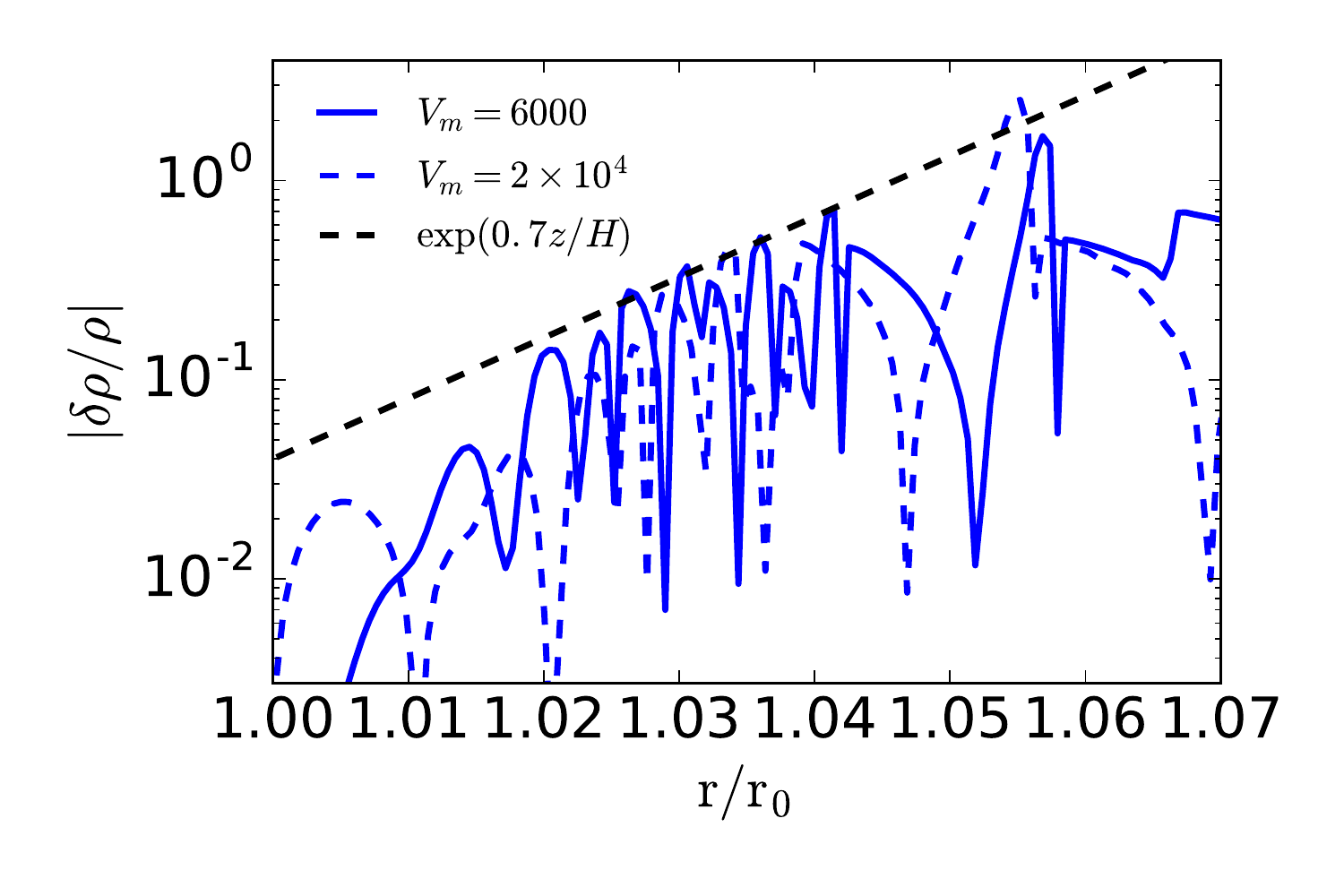}
\vspace{-0.2cm}
\caption{Comparison of analytic and numerical growth rates in our streaming simulations.   The numerical results are for our $v_{A,0}=10$ simulation for two values of the reduced speed of light $V_m$ and show $\delta \rho/\langle \rho \rangle$ with $\delta \rho = \rho(t)-\langle \rho \rangle$ where $\langle \rho \rangle$ is the time-averaged density profile.  Analytic growth rates are from eq. \ref{eq:WKBsol}.  The numerics and analytics agree well for both $V_m$, consistent with the sound wave instability in \S \ref{sec:lin1mom} being the dominant instability in the simulations.}
\label{fig:growth}
\end{figure}

Defining
\be
c_{tot}^2 = c_i^2 + \frac{2}{3} c_c^2
\ee
equation \ref{eq:WKB} becomes
\be
\omega \simeq |k| c_{tot} - i \frac{k}{|k|}\frac{2 c_c^2}{9 H c_{tot}}
\label{eq:WKBsol}
\ee
Equation \ref{eq:WKBsol} implies that backwards propagating ($k < 0$) sound waves are unstable.   Since it takes the waves a time $H/c_{tot}$ to travel a distance $H$, the growth rate in equation \ref{eq:WKBsol} implies that for $c_c \gtrsim c_i$, waves are amplified significantly once they propagate a few scale-heights.   More quantitatively, Figure \ref{fig:growth}  compares the growth rate predicted by equation \ref{eq:WKBsol} with the numerical growth in the $v_{A,0}=10$ simulation for two different $V_m$ and finds good agreement.  The numerical growth rates are calculated using $\delta \rho/\langle \rho \rangle$ with $\delta \rho = \rho(t)-\langle \rho \rangle$.  For the analytic growth rate we use the fact that the leading order solution is a sound wave with $\omega = k c_{tot}$ so that the wave crests/troughs are predicted to scale with height as $\delta \rho/\rho \propto \exp[z/H(0.5 + {\rm Im}(\omega)H/c_{tot}]$, which is $\sim \exp[0.7 z/H]$ for the parameters of the simulation.

The fact that the numerical growth rates in Figure \ref{fig:growth} are independent of $V_m$ is consistent with the fact that the background gradient instability identified here is the dominant instability in the simulations, not the finite speed of light instability identified in Appendix \ref{sec:lin2mom}.

\section{Time-averaged CR Energetics}
\label{sec:AppB}

The standard CR transport equations imply $p_c \propto \rho^{2/3}$ for high $v_A$, which is a consequence of $\nabla \cdot F_c = v_A dp_c/dr$ (\S \ref{section:streaming}).   This equation is not, however, well-satisfied by the time average of our simulations, because of the strong inhomogeneity introduced by shocks.   What is the appropriate time-averaged CR energy equation to use in its stead? 

Figure \ref{fig:steady_stream} shows that to decent accuracy we can approximate the time-averaged simulation over a range of radii as
\be
\langle F_c \rangle = \eta_1 \langle p_c \rangle \langle v+v_A \rangle  \ \ \ \  {\rm and} \ \ \ \  \langle v_A \frac{d p_c}{dr} \rangle = \eta_2 \langle v_A \rangle \langle \frac{d p_c}{dr} \rangle
\label{eq:avgs}
\ee
with $\eta_1 \sim 1.4$ and $\eta_2 \sim 0.3$.    To the extent that $\nabla \cdot \langle F_c \rangle = \eta_2 \langle v_A \rangle \langle \frac{d p_c}{dr} \rangle$, one could re-derive all of standard CR wind theory with the modifications to the CR energetics due to inhomogeneity encapsulated by $\eta_1$ and $\eta_2$.   For example, it is easy to show that in this approximation, the steady state wind equations for the sonic point in CR-driven winds (eq. \ref{wind_streaming}) are modified with the CR adiabatic index $4/3 \rightarrow \eta_1/(\eta_1-\eta_2)$.

However, in our simulations $\langle \nabla \cdot F_c \rangle \ne \nabla \cdot \langle F_c \rangle$.  This is not so surprising given the very large gradients near shocks that are not present in $\nabla \cdot \langle F_c \rangle$, but are in $\langle \nabla \cdot F_c \rangle$.   This shows that a new closure model for the CR energy equation to replace equation \ref{cr_energy} must be formulated in terms of an expression for $\langle \nabla \cdot F_c \rangle$ in terms of other time-averaged CR quantities.  We leave this to future work.  Instead, in \S \ref{sec:strmod} we develop a  phenomenological extension of CR wind theory in which the modified CR energetics found in our simulations is accounted for by a constant $\geff < 2/3$ in the limit $v_A \gg v$.

\end{appendix}
\label{lastpage}
\end{document}